\documentclass[a4paper,11pt]{book}
\usepackage[english]{babel}
\usepackage{amssymb}
\usepackage{amsmath}
\usepackage{tabularx}
\usepackage{titlesec}
\usepackage[dvipdf]{graphicx}
\usepackage{nomencl}
\usepackage[body={6.0in, 8.2in},left=1.20in,right=1.80in]{geometry} %before fancyhdr !! 
\usepackage{fancyhdr}
\usepackage{setspace}

\titleformat{\chapter}[display]{\huge\bfseries}{\huge Chapter \Huge\thechapter}{0.1em}{\titlerule\vspace{1em}}[\vspace{0.1em}]
\makeatletter
\def\cleardoublepage{\clearpage\if@twoside \ifodd\c@page\else%
\hbox{}%
\thispagestyle{empty}%              % Empty header styles
\newpage%
\if@twocolumn\hbox{}\newpage\fi\fi\fi}
\begin{document}
\frontmatter
\thispagestyle{empty}
\begin{center}
\begin{minipage}{0.2\textwidth}
\includegraphics[width=\textwidth]{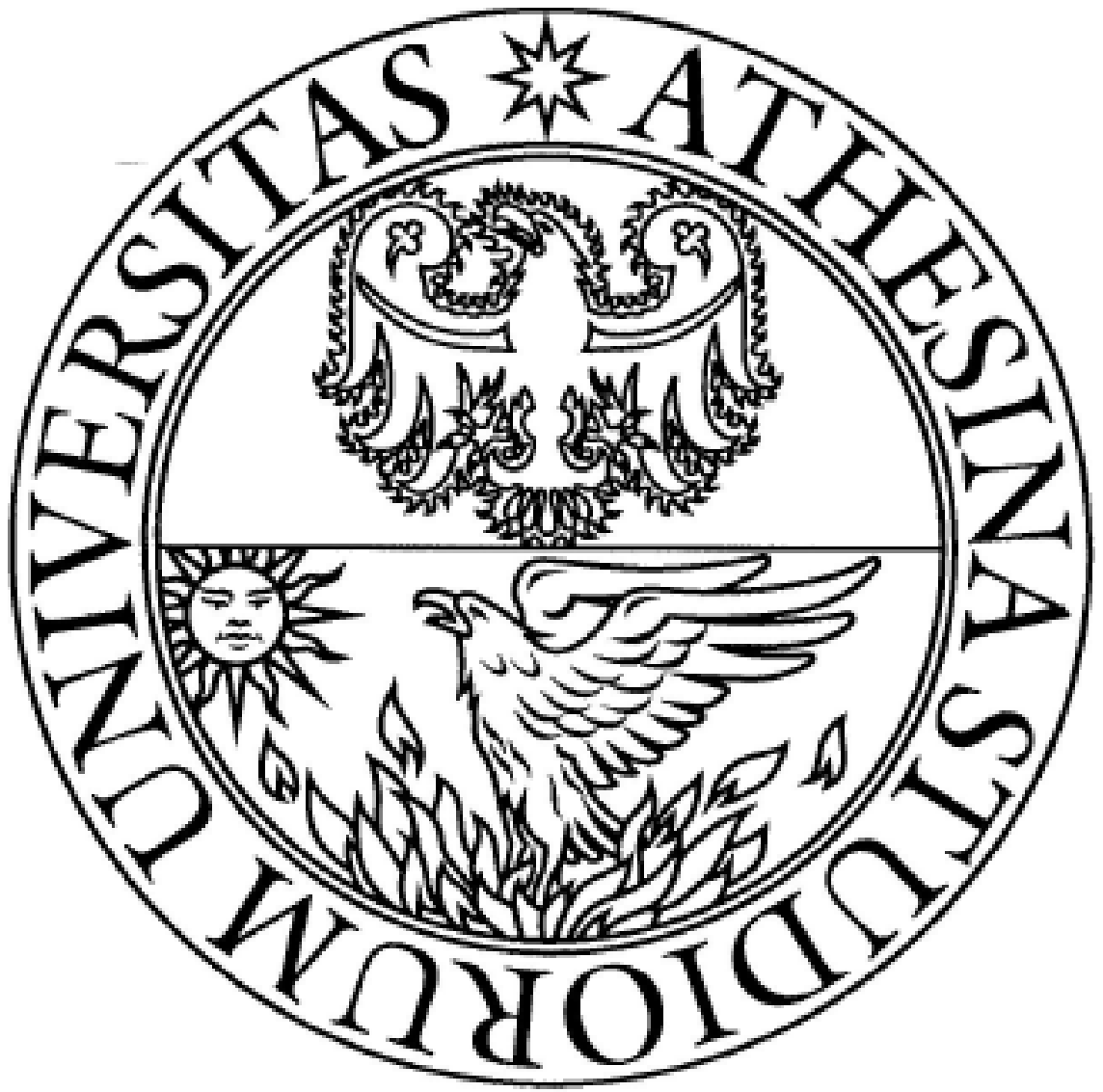}
\end{minipage}
\begin{minipage}{0.79\textwidth}
{\sc Universit\`a degli Studi di Trento\\Facolt\`a di Scienze Matematiche\\Fisiche e Naturali}
\end{minipage}
%  \vfill\vfill\vfill
   \vfill\vfill
   {%\bfseries
  \begin{huge}
   Tesi di Dottorato di Ricerca in Fisica\\
   \end{huge}
   \begin{LARGE}
   Ph.D. Thesis in Physics\\\hrulefill
   \end{LARGE}
   }
   \vfill\vfill\vfill\vfill
   \begin{minipage}{\textwidth}
    \center
    \Huge
%   \emph{\sc The Auxiliary Field Diffusion Monte Carlo method for nuclear physics and nuclear astrophysics}
    {\bfseries The Auxiliary Field Diffusion Monte Carlo Method for Nuclear Physics and Nuclear Astrophysics}
  \end{minipage}
{
	\vspace{\stretch{7}}
	
	{
		\large
		\begin{tabularx}{\textwidth}{X X}
			Candidate: &%
			{\hfill}%
			Supervisor:\\
			\textbf{Stefano Gandolfi} &%
			{\hfill}%
			\textbf{Dr. Francesco Pederiva}\\
		\end{tabularx}
	}
	\vspace{\stretch{3}}

	{
		\large
		November 2007
	}

}
\end{center}
%\cleardoublepage
%\thispagestyle{empty}
%\begin{flushright}
%To my parents\\[1ex]
%Ai miei genitori
%\end{flushright}
%\cleardoublepage
\setcounter{tocdepth}{2}
\tableofcontents
\cleardoublepage
\markboth{\sc \nomname}{}
%\printglossary[0.8in]
\mainmatter
\cleardoublepage
\chapter{Introduction}
Since the discovery of the neutron, immediately followed by their prediction, neutron stars are one of the more fascinating 
and exotic bodies in the universe. Matter inside reaches densities very far to be realized in
present experiments, and theoretical prediction can be verified only indirectly with astronomical 
observations.

Neutron stars\cite{heiselberg00,lattimer01,miller02,lattimer04} belong to a particular class of
stars whose mass is generally of the order of $\sim$1.5 solar masses (M$_{\bigodot}$) 
embodied in a radius of $\sim$12 Km, and have a central density ranging from 7 to 10 times the nuclear saturation 
density $\rho_0$=0.16 fm$^{-3}$ which is found in the center of heavy nuclei. The Fermi energy of Fermions at such 
densities is in excess of tens of MeV and then thermal effects are a very small perturbation 
to the structure of neutron stars. Therefore they exhibit the properties of cold matter at extremely 
high densities and have proved to be fantastic test bodies for theories from microscopic nuclear 
physics to general relativity\cite{raffelt96}.

\begin{figure}[h]
\begin{center}
\includegraphics[width=11cm]{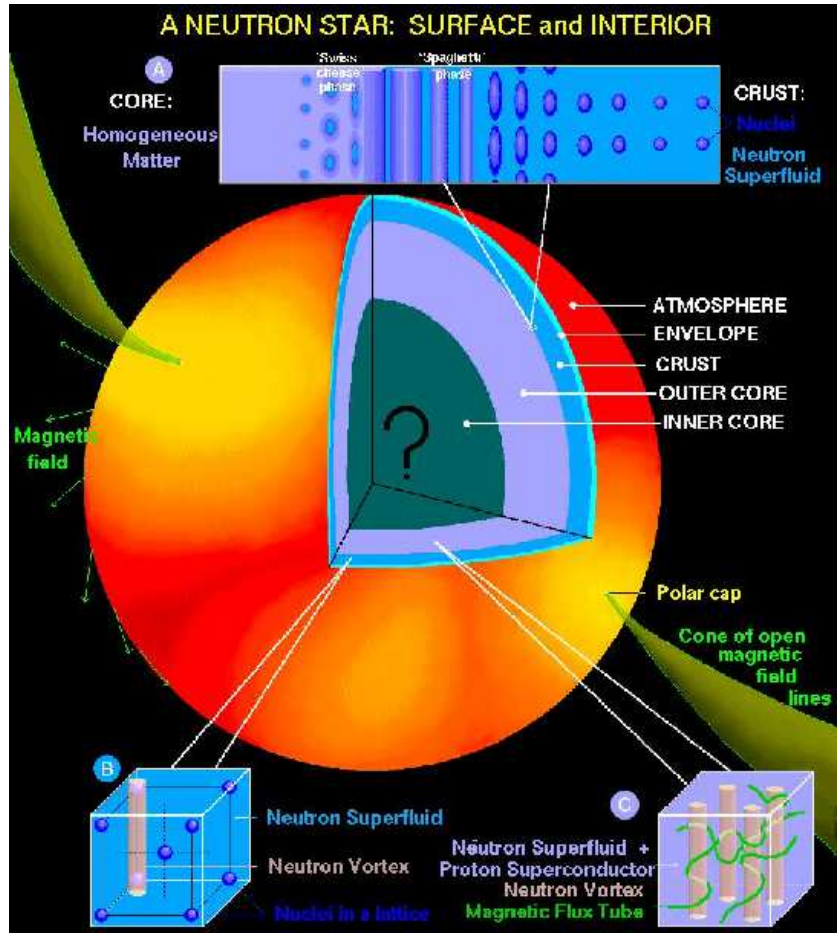}
\caption{
The possible composition inside a neutron star. Figure taken from Ref. \cite{lattimer04}.}
\label{fig:nstar}
\end{center}
\end{figure}

The crust of neutron stars (the outer 1-2 Km) primarily contains nuclei in the region where the density 
increases toward the crust-core interface at $\rho\simeq\rho_0/3$, and is made 
in large part of nuclei in which the proton fraction is $x\sim$ 0.1 to 0.2; such extremely 
neutron-rich nuclei cannot be directly observed, but can be created in laboratories in heavy-ion beam 
experiments\cite{danielewicz02}. At higher densities neutron rich nuclei undergo a number of transitions through phases
known as nuclear pasta\cite{lattimer04,lorenz93}, until the homogeneous neutron matter phase is reached.

When the mass of neutron-rich nuclei exceeds the so called drip-line, neutrons form a uniform 
gas. At the crust-core interface the mixing of nuclei and neutron matter gives rise to a phase in which 
nuclei are probably organized on a lattice.

The neutron matter in the inner crust may be in a $^1S_0$ superfluid phase\cite{baym75} that alters the specific heat 
and the neutrino emission of the star playing then a crucial role in the star cooling. 
The superfluid phase is also investigated in order to possibly explain sudden spin jumps 
observed in neutron stars, called glitches, probably related to the angular momentum stored in 
the rotating superfluid neutrons in the inner crust\cite{heiselberg00}.

The outer core of the neutron star consists of matter composed by nucleons, electrons and muons.
Nucleons are in the larger part neutrons with a small fraction of protons, and the degenerate 
electron gas contrasts the gravitational field by avoiding that the star becomes a black hole.

In the inner core higher hadrons could appear and play a crucial role in the matter (called then 
hadronic matter). Hadrons could be charged mesons, such pions or kaons that eventually could 
form a Bose condensate\cite{baym79,migdal78}, and other hyperons such as the $\Sigma^-$ or $\Lambda$. 
The nucleon-hyperon\cite{maessen89,rijken99} and hyperon-hyperon\cite{stoks99} interactions are not as well 
known\cite{vidana04} as those between 
nucleons. As a consequence the transition from nucleonic to hadronic matter is hard to be calculated.
Some attempt to explicitly include hyperons in the nuclear matter equation of state was 
recently proposed\cite{vidana00,schulze06}.
At higher densities there should also be a transition to quark matter in which quarks are deconfined
\cite{heiselberg93}.

The simpler model to predict observable interesting properties of neutron star starts with the 
study of pure neutron matter bulk, and after with the addition of a small fraction of protons.
The modelization and the knowledge of the nucleon-nucleon interaction from scattering data
is then the starting point for a realistic simulation of neutron stars. At the same time the research 
of an accurate technique to exploit such effective interactions is fundamental to attack ab-initio
problems of interest for nuclear astrophysics.

This thesis is concerned about of the interesting properties of neutron matter briefly described 
above.
It will be shown how the Auxiliary Field Diffusion Monte Carlo\cite{schmidt99,fantoni00,fantoni01,fantoni} (AFDMC) 
method can be used to accurately study a 
wide range of phenomena that govern the physics of a neutron star. The only needed starting point is
the choice of the degrees of freedom and consequently the Hamiltonian of the system.

In nuclear physics a calculation typically starts with the assumption that nucleons are nonrelativistic particles 
interacting with some effective potentials. This approach is a great simplification over reality.
Nucleons are composite systems with an internal structure due to quarks interacting by gluon exchange,
but presumably the quark degrees of freedom become important only at very short distances with high 
energies and momenta exchanged, so a nucleon-nucleon potential description should be adequate 
for the realm of low-energy nuclear physics.

At the present there are several forms of realistic Hamiltonians, able to accurately reproduce 
scattering processes and light nuclei properties. Unfortunately the development of modern
potentials and of the few body-methods used to study very light nuclei, was not followed 
by the development of an accurate method to test and apply the nuclear Hamiltonians to the study 
of heavier nuclei, neutron and nuclear matter, and the study of properties of four-body nuclei is 
the standard field of investigation\cite{viviani05,gazit06}.

Given a nuclear Hamiltonian, accurate calculations are limited to very few nucleons. The exact Faddeev-Yakubovsky 
equation approach is limited to systems with A=4\cite{glockle93}; other few-body variational 
calculations\cite{kamada01} solve the nuclear Schr\"odinger equation in good agreement with the exact method but are 
extended only up to A=6 nuclei\cite{bacca02,bacca04}. Techniques based on shell models calculations like the ab initio no-core
shell model were extended up to A=13\cite{navratil07} and very recently up to A=40\cite{roth07}
(but with some commented uncontrolled approximation\cite{dean07}). 
Quantum Monte Carlo methods, like the Variational Monte Carlo\cite{wiringa91,wiringa92} (VMC) or other techniques
based on recasting the Schr\"odinger equation into a diffusion
equation, like the Diffusion Monte Carlo (DMC) or Green's Function Monte Carlo\cite{carlson87,pudliner95} (GFMC)
allowed for performing calculations of nuclei up to A=12\cite{pieper05} and neutron matter 
with A=14 neutrons in a periodic box\cite{carlson03,carlson03b,chang04}.

Other many-body theories that contain uncontrolled approximations are used for heavier nuclei. The first 
class of them essentially modify the Hamiltonian to an easy-to-solve form like the Brueckner-Goldstone\cite{day67}
or the Hartree-Fock\cite{vautherin72} methods 
and all the theories constructing effective interactions from microscopic ones. The second class
include methods that work on the trial wave function typically used in variational calculations like 
the Correlated Basis Function (CBF) theory\cite{bisconti07,bisconti06,guardiola88}, Variational Monte 
Carlo\cite{pieper90,pieper92} or Coupled Cluster Expansion\cite{heisenberg99}.

\begin{figure}[h]
\begin{center}
\includegraphics[width=11cm]{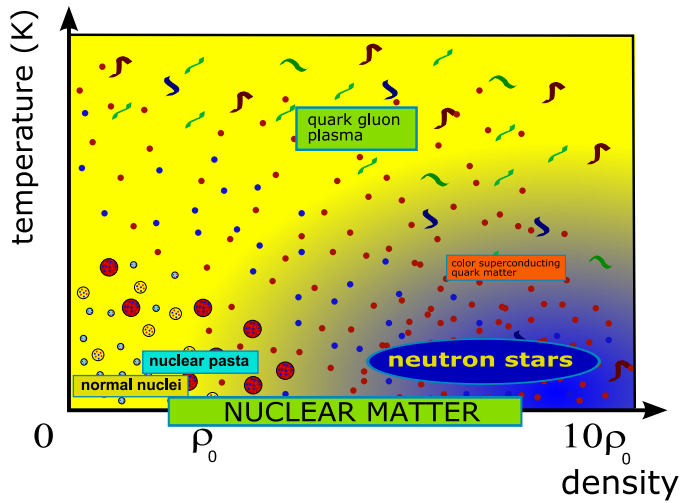}
\caption{
Schematic representation of the phase diagram temperature-density of matter. $\rho_0$ 
is the saturation density at T=0, extrapolated from the density of the core of massive nuclei.
The diagram shows the region in which the AFDMC calculations were performed, for 
densities that range from the crust and part of the interior of neutron stars to
ordinary nuclei at zero-temperature.}
\label{fig:diagramma}
\end{center}
\end{figure}

It will be shown that AFDMC can be used to simulate both nuclei\cite{gandolfi07b} and nuclear
matter\cite{gandolfi07} with high accuracy.
After the choice of the Hamiltonian for each different system considered,
AFDMC is applied to study properties of light nuclei and confined neutron drops to test the agreement of 
this algorithm with other exact methods, and then extended to heavier nuclei never studied with the same 
accuracy. This step is necessary to demonstrate that an accurate many-body calculation 
is now possible for realistic modern Hamiltonians. 

It will be then illustrated the application of AFDMC to study the symmetric nuclear matter 
equation of state, and to verify the limitations of other methods to deal with the more simple nucleon-nucleon 
interaction that contains the tensor-$\tau$ force generated by the one pion exchange process. 

The AFDMC method was then applied to the study of the equation of state of neutron matter, that is the simpler 
starting point to study the structure of a neutron star.
The global structure of a neutron star, given in terms of the mass-radius relation, is determined by the 
equations of hydrostatic equilibrium. For a non-rotating spherical object in general relativity, these 
equations are called Tolman-Oppenheimer-Volkov\cite{oppenheimer39} (TOV) equations.
The numerical solution of the TOV equations gives the M-R relation between the total mass versus the radius, 
or gives the total mass as a function of the central density $\rho_c$ of the neutron star.

By means of the AFDMC we studied the neutron matter in the lower density regime where neutrons form a $^1S_0$ superfluid 
phase\cite{fabrocini05}. It will be shown how the BCS phase of the 
system lower the energy with respect to the normal neutron gas, and therefore the superfluid energy gap
is evaluated.

In the last part of the thesis the AFDMC calculation of neutron-rich isotopes in a very simple model, will be presented.
In this case the external neutrons interact with 
a realistic interaction but all the excitation effects of the core interacting with valence nucleons 
are neglected\cite{gandolfi06,gandolfi06b}. The obtained results are in a very good agreement with experimental 
data, qualitatively better than other methods.

\chapter{Hamiltonian}
\label{ch:hamiltonian}
The goal of nuclear physics, and consequently of nuclear astrophysics, is to understand how nuclear binding, 
stability and structure arise from the interactions between individual nucleons. To achieve this goal 
the starting point consists in the determination of the Hamiltonian to be used. 

In principle the nuclear Hamiltonian should be directly derived from quantum chromodynamics (QCD) but the 
direct calculation starting from QCD is at present too complex to be attacked. 
Thus the structure of a nuclear Hamiltonian is usually determined phenomenologically and then fitted
to exactly reproduce the properties of few-nucleon systems\cite{carlson98}. However, in some case, its form is 
determined by the many-body methods used to solve the Schr\"odinger equation\cite{viviani07}.

In this chapter the modern nuclear Hamiltonians will be illustrated, while the many-body techniques 
used to solve the nuclear ground-state will be the subject of the following chapters.

Properties of a generic nuclear system can be studied starting from the non-relativistic Hamiltonian
\begin{equation}
\label{eq:hamiltonian}
H=-\frac{\hbar^2}{2m}\sum_i\nabla_i^2+\sum_{i<j}v_{ij}+\sum_{i<j<k}V_{ijk}+...
\end{equation}
which includes the kinetic energy operator, a two-nucleon (NN) interaction $v_{ij}$, a three-nucleon (TNI)
interaction $V_{ijk}$, and in principles could include higher many-body interactions that however 
are expected to be less important. It has been shown that in light nuclei the three-body force 
contributes only a few percent of the total potential energy, while for heavier nuclei the contribution 
of TNI could be of about 15-20\%. One expects that in the physics of nuclei a possible 
four-body interaction should be either negligible or just a very small perturbation to the system. 
This is not necessarily true in neutron and nuclear 
matter in the typical density regime of neutron star where the contribution 
of many-body interactions in the Hamiltonian could be more important.

The NN interactions are usually dependent on the relative spin and isospin state of the nucleons 
and therefore written as a sum of several operators. 
The coefficients and radial functions that multiply each operators are adjusted by fitting 
experimental scattering data, and the type and the number of these operators depend on the interaction.

A large amount of empirical information about the nucleon-nucleon (NN) scattering problem has been 
accumulated over time till 
1993, when the Nijmegen group analyzed all NN scattering data below 350 MeV published in a regular 
physics journal between 1955 and 1992\cite{stoks93b}. 
NN interaction models that fit the Nijmegen database with a $\chi^2/N_{data}\sim$1 are called
"modern" and the more diffuse of them include the Nijmegen models\cite{stoks94} (Nijm93, Nijm I, 
Nijm II and Reid-93), the Argonne models\cite{wiringa95,wiringa02} and the CD-Bonn\cite{machleidt96}.
In the last year also modern NN interaction derived directly from chiral effective field theory were
proposed\cite{epelbaum05,entem03}.
However all these NN interactions had been proved to underestimate the triton binding energy, by suggesting 
that the contribution of a TNI force is essential to reproduce the physics of nuclei.

The TNI contribution is mainly attributed to the possible $\Delta$ intermediate states that an 
excited nucleon could assume after and before exchanging a pion with other nucleons.
This process can be written as an effective three-nucleon interaction, and its parameters are 
fitted on light nuclei\cite{carlson81,pieper01} and eventually on properties of nuclear matter, such the empirical 
equilibrium density and the energy at saturation\cite{carlson83}. The TNI must accomplish the NN interaction
and the total Hamiltonian has then to be studied after the choice of NN\cite{wiringa83}, then different 
form of TNI also in the momentum space were proposed\cite{coon81}.

The possibility to explicitly include the $\Delta$ degrees of freedom to avoid the addition of the TNI
in the Hamiltonian were explored by Wiringa et al. with the Argonne AV28 potential\cite{wiringa84}, 
but its application to calculate properties of triton revealed to be very difficult (results took  
6 papers!\cite{picklesimer92,picklesimer92a,picklesimer92b,
picklesimer92c,picklesimer92d,picklesimer92e}), and also in nucleonic matter this model gave bad results
\cite{baldo91,baldo92}.

\section{Argonne NN interaction}
\label{sec:NNint}
A typical NN potential contains a strong short-range repulsion, an intermediate-range attraction 
and a long-range part.

The short-range part is usually treated phenomenologically, because for short distances and high momenta transferred
the quarks inside nucleons start to play an important role which in principle can not 
be well included in an effective potential without using explicit subnuclear degrees of freedom.

The intermediate-range attraction was initially phenomenologically described as a sum of Yukawa 
functions in the Reid\cite{reid68} potential, or as the exchange of a scalar meson $\sigma$ as in 
the one boson exchange (OBE) potentials\cite{smith76}.
The long-range part is due to the one pion exchange (OPE) between nucleons.

Other non-local terms like the linear and quadratic spin-orbit terms, as well as relative angular momentum 
or other operators coming from two pion exchange or from nucleon $\Delta$ excitations, are usually included in modern 
interactions.

The main contribution of NN potential at low energy can be derived by considering that nucleons 
exchange a meson. The long-range part of $v_{ij}$ is know to be mediated by the pion $\pi$, that is the 
lightest of all mesons. This is the common starting point of all recent and past NN interaction. 
The OPE potential is given by
\begin{equation}
v_{ij}^\pi=\frac{f_{\pi NN}^2}{4\pi}\frac{m_\pi}{3}X_{ij}\vec\tau_i\cdot\vec\tau_j \,,
\end{equation}
with 
\begin{equation}
\label{eq:Xop}
X_{ij}=Y(m_\pi r_{ij})\vec\sigma_i\cdot\vec\sigma_j+T(m_\pi r_{ij})S_{ij} \,,
\end{equation}
where 
\begin{equation}
\frac{f_{\pi NN}^2}{4\pi}=0.075\pm0.002
\end{equation}
is the pion-nucleon coupling constant\cite{stoks93}, 
$m_\pi$ is the averaged pion mass, $\vec\sigma$ and $\vec\tau$ are Pauli matrices acting on the 
spin or isospin of nucleons, $S_{ij}$ is the tensor operator
\begin{equation}
S_{ij}=3(\vec\sigma_i\cdot\hat r_{ij})(\vec\sigma_j\cdot\hat r_{ij})-\vec\sigma_i\cdot\vec\sigma_j \,,
\end{equation}
and the radial functions associated with the spin-spin and tensor parts are
\begin{eqnarray}
\label{eq:defYT}
Y(x)&=&\frac{e^{-x}}{x}\xi_Y(x) \,,
\nonumber \\
T(x)&=&\left(1+\frac{3}{x}+\frac{3}{x^2}\right)Y(x)\xi_T(x) \,.
\end{eqnarray}
The $\xi(x)$ are short-range cutoff functions defined by
\begin{equation}
\xi_Y(x)=\xi_T(x)=1-e^{-cx^2} \,.
\end{equation}

It is important to note that since $T(m_\pi r)\gg Y(m_\pi r)$ in the important region where $r\lesssim2$ fm, the OPE 
is dominated by the tensor part. 

The complete NN potential $v_{ij}$ is given by $v_{ij}^\pi+v_{ij}^R$, where $v_{ij}^R$ contains all the other 
intermediate-range and short-range parts. All the parameters in the $\xi$ short-range cutoff functions as well 
as other phenomenological $v_{ij}^R$ parts are fitted on NN Nijmegen scattering data.

The most recent interaction proposed by the Argonne group is called AV18\cite{wiringa95} NN potential because it
is written as a sum of 18 operators. This is a modern interaction fitting very well the scattering 
data of Nijmegen database. 
There are other simpler Argonne interactions called AVn'\cite{wiringa02} that are simpler versions of AV18; they contain 
$n<18$ operators, and the prime symbol indicates that such potentials are not just a simple truncation of AV18,
but a reprojection, which preserves the isoscalar part of AV18 in all $S$ and $P$ partial waves as well as in 
the $^3D_1$ wave and its coupling to $^3S_1$.

The Argonne NN potential between two nucleons $i$ and $j$ is written in the coordinate space as a sum of operators
\begin{equation}
\label{eq:vop}
v_{ij}=\sum_{p=1,n} v_p(r_{ij})O_{ij}^p \,,
\end{equation}
where $n$ is the number of operators depending of the potential, $v_p(r)$ are radial functions, and $r_{ij}$ 
is the inter-particle distance. 

The first eighth operators give the higher contribution to the NN interaction. The first six of them come 
from the long-range part of $v_{ij}^\pi$ OPE term. The last two terms depend on the velocity of nucleons and give the 
spin-orbit contribution.
These eight operators are
\begin{equation}
\label{eq:voperators}
O_{ij}^{p=1,8}=(1,\vec\sigma_i\cdot\vec\sigma_j,S_{ij},\vec L_{ij}\cdot\vec S_{ij})
\times(1,\vec\tau_i\cdot\vec\tau_j) \,,
\end{equation}
where $\vec L_{ij}$ is the relative angular momentum of couple $ij$
\begin{equation}
\vec L_{ij}=\frac{1}{2i}(\vec r_i-\vec r_j)\times(\vec\nabla_i-\vec\nabla_j) \,,
\end{equation}
and $\vec S_{ij}$ is the total spin of the pair
\begin{equation}
\vec S_{ij}=\frac{1}{2}(\vec\sigma_i+\vec\sigma_j) \,.
\end{equation}
In the modern interactions these eight operators are the standard ones required to fit $S$ and $P$ wave data 
in both triplet and singlet isospin states.

Following operators used in the Argonne AV14\cite{wiringa84} and AV18 are
\begin{equation}
O_{ij}^{p=9,14}=(L_{ij}^2,L_{ij}^2(\vec\sigma_i\cdot\vec\sigma_j),(\vec L_{ij}\cdot\vec S_{ij})^2)
\times(1,\vec\tau_i\cdot\vec\tau_j) \,.
\end{equation}
These additional operators were introduced to better describe the phase shift in both $P$ and $D$ state\cite{lagaris81}.
The four $L^2$ operators provide for differences between $S$ and $D$ waves, and $P$ and $F$ waves. In addition to 
$\vec S$ and $\vec L\cdot\vec S$ operators the $(\vec L\cdot\vec S)^2$ provide an additional way of splitting states 
with different $J$ values (for example $^3D_{1,2,3}$ states). 
The effect of the quadratic angular momentum operators $L^2$ instead of $p^2$ and consequences were 
studied by Wiringa et al.\cite{wiringa03}. However, the contribution of these operators is small respect to the 
total potential energy.

The last four additional operators of the AV18 potential break charge independence and are given by
\begin{equation}
O_{ij}^{p=15,18}=T_{ij},(\vec\sigma_i\cdot\vec\sigma_j)T_{ij},S_{ij}T_{ij},\tau_i^z+\tau_j^z \,,
\end{equation}
where $T_{ij}$ is the isotensor operator defined as
\begin{equation}
T_{ij}=3\tau_i^z\tau_j^z-\vec\tau_i\cdot\vec\tau_j \,.
\end{equation}

The Argonne AV18 potential has been used in a wide range of works about the ground-state\cite{pudliner95,pieper05,wiringa02,pieper02,
wiringa00,pudliner97} and excited states\cite{pieper04} of nuclei
and nucleonic matter\cite{akmal98}.

\section{Urbana and Illinois three-body forces}
\label{sec:TNIint}
The three-nucleon interaction (TNI) $V_{ijk}$ depends on the choice of the NN potential, but the final result 
with the total Hamiltonian should be independent of the choice.
For this purpose several TNI form have been used, fitted in addition to the Argonne NN interactions, namely
the Urbana IX\cite{carlson83} and the modern Illinois\cite{pieper01} forms. 

The Urbana and Illinois structure of the TNI interaction between three nucleons is written as a sum of several operators
\begin{eqnarray}
\label{eq:v3gen}
V_{ijk} = A_{2\pi}^{PW} O^{2\pi,PW}_{ijk}
+ A_{2\pi}^{SW} O^{2\pi,SW}_{ijk} 
+ A_{3\pi}^{\Delta R} O^{3\pi,\Delta R}_{ijk}
+A_R O^R_{ijk} \,,
\end{eqnarray}
where each term represents the interaction given by two-pion exchange in the $P$ and $S$ wave, the three-pion 
exchange and a phenomenological spin-isospin independent term.
The Urbana force only contains the first and the last terms, while the more sophisticated Illinois forces 
contain all of these operators.

\begin{figure}[ht]
\vspace{0.5cm}
\begin{center}
\includegraphics[scale=0.25]{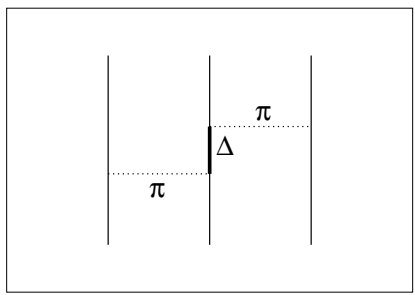}
\vspace{0.5cm}
\caption{Fujita-Miyazawa 2$\pi$-exchange in $P$ wave term}
\label{fig:v3a}
\end{center}
\end{figure}

The first term was introduced by Fujita and Miyazawa\cite{fujita57} to describe the process where two pions are exchanged 
between nucleons with an intermediate excited $\Delta$ resonance as shown in Fig. \ref{fig:v3a}.

This diagram with three nucleons can be schematically described as
\begin{equation}
\label{eq:vnucdel}
V_{ijk}^{2\pi,PW}=\sum_{cyc}\frac{v_{NN\rightarrow N\Delta}^\pi(ij) v_{\Delta N\rightarrow NN}^\pi(jk)}{(m_\Delta-m_N)^2} \,,
\end{equation}
where the sum is cycled to include all possible permutations, and 
$v_{NN\rightarrow N\Delta}^\pi(ij)$ describes the pion exchange transition potential that scatters the 
$\lvert NN\rangle$ to a $\lvert N\Delta\rangle$ state and viceversa.

The OPE interaction has to be the same form of NN potential, that for the AV18 is
\begin{equation}
v_{ij}^\pi = \frac{f_{\pi N N}^2}{4\pi}\frac{ m_\pi}{3} 
X_{ij}^{op} \vec \tau_i \cdot \vec \tau_j \,,
\end{equation}
with $X_{ij}^{op}$ defined as $X_{ij}$ in Eq. \ref{eq:Xop} and the functions $Y(x)$ and $T(x)$ defined in Eq. \ref{eq:defYT}.
The short-range cutoff functions are
\begin{equation}
\xi_Y(r) = \xi_T(r) = 1-e^{-cr^2} \,.
\end{equation}

The $X_{ij}$ is the same interaction of pion-exchange transition potential entering in Eq. \ref{eq:vnucdel} that,
with the inclusion of the appropriate spin-isospin operators, becomes
\begin{equation}
v_{NN\rightarrow N\Delta}^\pi(ij)=v_{tr}(ij)X_{ij}\vec\sigma_i\cdot \vec S_j^\dag\vec\tau_i\cdot\vec T_j^\dag \,,
\end{equation}
and
\begin{equation}
v_{\Delta N\rightarrow NN}^\pi(jk)=v_{tr}(jk)X_{jk}\vec S_j\cdot\vec\sigma_k \vec T_j\cdot\vec\tau_k \,,
\end{equation}
where $\vec\sigma$ and $\vec\tau$ are the usual spin and isospin operators acting on nucleon, $\vec S^\dag$ and $\vec T^\dag$ are 
the spin-isospin transition operators between the $N$ to the $\Delta$ space, and $\vec S$ and $\vec T$ for the reverse 
transition.

Using the Pauli identity
\begin{equation}
\label{eq:pau1}
\vec T^\dag\cdot\vec AT^\dag\cdot B=\frac{2}{3}\vec A\cdot\vec B-i\frac{1}{3}\vec\tau\cdot\vec A\times\vec B \,,
\end{equation}
the Eq. \ref{eq:vnucdel} can be rewritten in the usual form
\begin{eqnarray}
\label{eq:v2ppw}
&&V_{ijk}^{2\pi,PW}=A_{2\pi}^{PW}O^{2\pi,PW}_{ijk}= 
\nonumber \\
&&A_{2\pi}^{PW}\sum_{cyc}\left[\left\{X_{ij},X_{jk}\right\}\left\{\tau_i\cdot\tau_j,\tau_j\cdot\tau_k\right\}+
\frac{1}{4}\left[X_{ij},X_{jk}\right]\left[\tau_i\cdot\tau_j,\tau_j\cdot\tau_k\right]\right] \,,
\nonumber \\
\end{eqnarray}
where the constant $A_{2\pi}^{PW}$ is fitted to reproduce the ground state of light nuclei and properties of nuclear matter, and
is different for each model of three-body potential. In the Urbana forces the constant $A_{2\pi}^{PW}$ is typically denoted 
by $A_{2\pi}$.

\begin{figure}[ht]
\vspace{0.5cm}
\begin{center}
\includegraphics[scale=0.25]{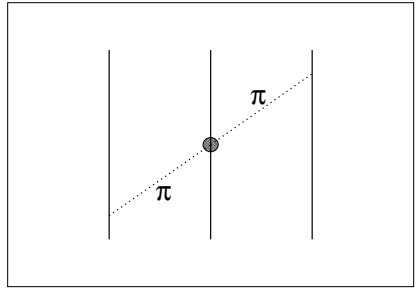}
\vspace{0.5cm}
\caption{2$\pi$-exchange in $S$ wave term}
\label{fig:v3b}
\end{center}
\end{figure}

The second term in Eq. \ref{eq:v3gen} is due to the $\pi N$ scattering in the $S$ wave as shown in 
Fig. \ref{fig:v3b}.

The form of this operator is
\begin{equation}
\label{eq:v2psw}
V_{ijk}^{2\pi,SW}=A_{2\pi}^{SW}O^{2\pi,SW}_{ijk}=A_{2\pi}^{SW}
\sum_{cyc}Z(m_\pi r_{ij})Z(m_\pi r_{jk}) \vec\sigma_i\cdot\hat r_{ij}\sigma_k\cdot\hat r_{kj}\tau_i\cdot\tau_k \,,
\end{equation}
with
\begin{equation}
Z(x)=\frac{x}{3}\left[Y(x)-T(x)\right] \,.
\end{equation}
This term is essentially a simplified form of the original Tucson model\cite{coon79}. The structure is the same of anticommutator 
term in Eq. \ref{eq:v2ppw} with a redefinition of radial functions and operators and the spin-isospin dependence reduces to 
a quadratic form. This term is not included in the Urbana forces.

\begin{figure}[ht]
\vspace{0.5cm}
\begin{center}
\includegraphics[scale=0.25]{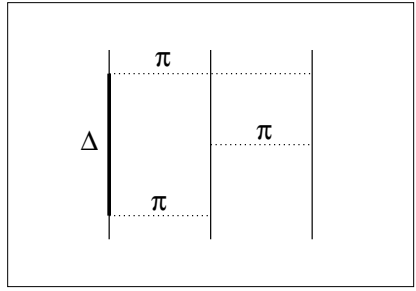}
\vspace{0.5cm}
\caption{3$\pi$-exchange with a $\Delta$-intermediate state}
\label{fig:v3c}
\end{center}
\end{figure}

\begin{figure}[ht]
\vspace{0.5cm}
\begin{center}
\includegraphics[scale=0.25]{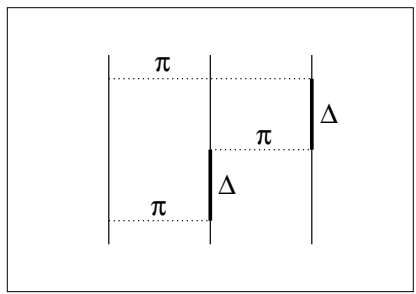}
\vspace{0.5cm}
\caption{3$\pi$-exchange with two $\Delta$-intermediate states}
\label{fig:v3d}
\end{center}
\end{figure}

The three-pion exchange term consists in two diagrams where pions excite one or two nucleons into intermediate $\Delta$-states 
as shown in Fig. \ref{fig:v3c} and \ref{fig:v3d}.

The scattering potentials of the process of Fig. \ref{fig:v3c} where only one $\Delta$ is created can be written in the same way as 
described for the Fujita and Miyazawa term by considering the possible excitation because of the pion exchange:
\begin{equation}
V_{1,ijk}^{3\pi,\Delta R}=\sum_{cyc}\frac{1}{(m_\Delta-m_N)^2}\left[ v_{NN\rightarrow \Delta N}^\pi(ij)
v^\pi(jk)v_{\Delta N\rightarrow NN}^\pi(ik)+j\rightleftharpoons k\right] \,,
\end{equation}
where the symbol $j\rightleftharpoons k$ means the exchange of nucleons $j$ and $k$ to consider all the possible 
permutations between nucleons. In the same way it is possible to describe the interaction of diagram in Fig. \ref{fig:v3d}
that is
\begin{equation}
V_{2,ijk}^{3\pi,\Delta R}=\sum_{cyc}\frac{1}{(m_\Delta-m_N)^2}\left[ v_{NN\rightarrow N\Delta}^\pi(ij)
v_{\Delta N\rightarrow N\Delta}^\pi(jk)v_{N\Delta\rightarrow NN}^\pi(ik)
+j\rightleftharpoons k\right] \,,
\end{equation}
with the Pauli identity given in Eq. \ref{eq:pau1} and the following
\begin{eqnarray}
\vec T^\dag\cdot\vec T&=&2 \nonumber \\
\vec T^\dag\times\vec T&=&-i\frac{2}{3}\vec\tau \nonumber \\
\vec\tau\cdot\vec A\vec\tau\cdot\vec B&=&\vec A\cdot\vec B
+i\vec\sigma\cdot\vec A\times\vec B \,,
\end{eqnarray}
the two diagrams can be rewritten in terms of other operators as reported in Ref. \cite{pieper01} in a very complex way. 

The last phenomenological part of TNI is due to all neglected physical effects. This $V^R$ part has the role to compensate 
the overbinding and the large equilibrium density of nuclear matter given by other TNI operators.

The $V^R$ is a spin-isospin independent operator given by
\begin{equation}
V_{ijk}^R=A_R O^R_{ijk}=A_R \sum_{cyc}T^2(m_\pi r_{ij})T^2(m_\pi r_{jk}) \,,
\end{equation}
where $T(x)$ is the same function defined in Eq. \ref{eq:defYT}.

The $A_{2\pi}$ term of Urbana UIX was originally fitted, in addition to the AV18, to reproduce the triton and alpha particle 
binding energy, while the $U_0$ strength was adjusted to obtain the empirical equilibrium 
density of nuclear matter\cite{pudliner95}. It has to be noted that the ground state of light nuclei can be exactly solved 
with few-body techniques, but the determination of the equation of state of symmetric nuclear matter can be 
evaluated only using many-body techniques that contain uncontrolled approximations.
Although the AV18+UIX Hamiltonian gives very good results for the triton and alpha particle ground-state,
a calculation of $p$-shell nuclei using GFMC revealed that this Hamiltonian cannot reproduce a binding energies 
in agreement with experimental data\cite{pudliner97}.

The more modern Illinois forces were fitted using GFMC to correctly reproduce all the 17 bound stated of 3$\leq$A$\leq$8 
nuclei\cite{pieper01}, and then they have been proved to well describe the ground state of nuclei up to $^{12}$C\cite{pieper05}.
However, the Illinois TNI presents some problems if used to evaluate the equation of state of pure neutron matter, as 
it will be discussed in the following chapters, possibly because they were fitted without any information coming from
nuclear matter calculations.

\chapter{Method}
The central problem in all many-body theories is the solution of the Schr\"odinger equation to evaluate 
the ground-state properties of a generic system. An exact analytic solution is impossible to be found for
most many-body Hamiltonians, and also the numerical brute-force solution is impossible in 
most interesting cases.

The Diffusion Monte Carlo (DMC) method, or the Green's Function Monte Carlo (GFMC) method, that is very similar, are ways
of exactly solving the many-body Schr\"odinger equation by means of a 
stochastic procedure. Although some attempts to develop such Quantum Monte Carlo (QMC) methods were just explored,
the first application of a GFMC code is due to Kalos\cite{kalos62}, and in 1962 the ground state of 
three- and four-body nuclei using some simple central interactions was solved.
The first DMC algorithm was implemented by Anderson in 1975\cite{anderson75} to study some molecular ground-state, 
and the first simulation of uniform matter was carried out by Ceperley and Alder\cite{ceperley80}.
The application of the GFMC algorithm to nuclear Hamiltonians containing spin and isospin operators was 
proposed in 1987 by Carlson\cite{carlson87}.

The extension of the DMC to deal with nuclear Hamiltonians using a large number of nucleons was invented 
by Schmidt and Fantoni\cite{schmidt99} with the Auxiliary Field Diffusion Monte Carlo (AFDMC)
method, used to solve the ground state for a generic system given a nuclear Hamiltonian, where 
the interaction contains spin and isospin operators as described in chapter \ref{ch:hamiltonian}.

In this chapter it will be review the DMC method to calculate the ground-state energy 
of a quantum system by starting from its Hamiltonian containing a kinetic term and a scalar interaction,
then it will be explained the extension to the AFDMC.

\section{Diffusion Monte Carlo}
The DMC method\cite{guardiola98,mitas99}, projects out the ground state from a trial wave function,
provided that it is not orthogonal to the true ground state. 

The Schr\"odinger equation in imaginary time $\tau=it$ is given by
\begin{equation}
\label{eq:schr}
-\frac{\partial}{\partial\tau}\psi(R,\tau)=H\psi(R,\tau) \,,
\end{equation}
with $R=\{\vec r_1...\vec r_N\}$ is a configuration of the system with all its degrees of freedom.

A formal solution written in integral form is
\begin{equation}
\label{eq:intschr}
\psi(R,\tau)=\int G(R,R',\tau)\psi(R',0)dR' \,,
\end{equation}
where the kernel $G(R,R',\tau)$ is the Green's function of the operator $H+\frac{\partial}{\partial\tau}$, 
and can be expressed as the matrix element
\begin{equation}
\label{eq:defgf}
G(R,R',\tau)=\langle R\lvert e^{-(H-E_0)\tau}\rvert R'\rangle=
\sum_n e^{-(E_n-E_T)\tau}\phi_n^*(R)\phi_n(R') \,,
\end{equation}
where $\{\phi_n\}$ is a complete set of eigenvectors of $H$. By considering a generic trial wave function 
$\psi_T$ as a linear combination of $\{\phi_n\}$, it is possible to consider the evolution in imaginary 
time $\tau$, given by
\begin{equation}
\label{eq:a}
\psi(R,\tau)=e^{-(H-E_0)\tau}\psi(R,0) \,,
\end{equation}
where $E_0$ is the exact ground state energy. In the limit $\tau\rightarrow\infty$, $\psi(R,\tau)$ 
approaches to the lowest eigenstate $\psi_0$ with the same symmetry of $\psi$. 
The evolution can be done by solving then the integral of Eq. \ref{eq:intschr} in the limit of infinite
imaginary time.

Let is considered the non interacting Hamiltonian of $N$ particles with mass $m$:
\begin{equation}
H_0=-\sum_i{\hbar^2\over2m}\nabla_i^2 \,.
\end{equation}

The Schr\"odinger equation \ref{eq:schr} 
becomes a $3N$-dimensional diffusion equation, and by writing the Green's function of Eq. \ref{eq:defgf} 
in the momentum space by means of the Fourier transform, it is possible to show that $G_0$ is a Gaussian 
with variance proportional to $\tau$:
\begin{equation}
G_0(R,R',\tau)=\left({m\over2\pi\hbar^2\tau}\right)^{3A\over2}
e^{-{m(R-R')^2\over2\hbar^2\tau}} \,.
\end{equation}
This equation describes the Brownian diffusion of a set of particles with a 
dynamic governed by random collisions. This interpretation can be implemented by representing $\psi$ 
by a set of discrete sampling points that generally are called \emph{walkers}. Therefore it can be defined as:
\begin{equation}
\psi(R)=\sum_k \delta(R-R_k) \,.
\end{equation}
The evolution for an imaginary time step $\Delta\tau$ is then realized by
\begin{equation}
\psi(R,\tau+\Delta\tau)=\sum_k G(R,R_k,\Delta\tau) \,.
\end{equation}
The result is a set of Gaussians that in the infinite limit of imaginary time $\tau$ represent a distribution 
of walkers according to the lowest state of Hamiltonian and can be used to calculate the ground state 
properties of the system.

Now, let be considered a more realistic case with a generic Hamiltonian
\begin{equation}
H=-\sum_i{\hbar^2\over2m}\nabla_i^2+V(R) \,.
\end{equation}
By using the Trotter formula
\begin{equation}
e^{(A+B)\Delta\tau}=e^{-A\frac{\Delta\tau}{2}}e^{-B\Delta\tau}e^{-A\frac{\Delta\tau}{2}}+O(\Delta\tau^3) \,,
\end{equation}
it is possible to break up the Green's function in two manageable parts, which can be written as:
\begin{equation}
G(R,R',\Delta\tau)\approx e^{-{V(R)+V(R')\over2}\Delta\tau}G_0(R,R',\Delta\tau) \,.
\end{equation}
This approximate expression is valid only in the limit $\Delta\tau\rightarrow0$.
Eq. \ref{eq:intschr} becomes:
\begin{equation}
\label{eq:evolution}
\psi(R,\tau+\Delta\tau)=\left({m\over2\pi\hbar^2\Delta\tau}\right)^{3A\over2}
\int e^{-{m(R-R')^2\over2\hbar^2\Delta\tau}}e^{-{V(R)+V(R')\over2}\Delta\tau}\psi(R',\tau)dR' \,,
\end{equation}
where the factor due to the interaction with the trial eigenvalue $E_T$ (that is a normalization factor) 
is the weight of the Green's function:
\begin{equation}
\label{eq:weight}
w=e^{-\left({V(R)+V(R')\over2}-E_T\right)\Delta\tau} \,.
\end{equation}
The integral \ref{eq:evolution} can be solved in a Monte Carlo way by propagating the particle coordinates 
by sampling a path according the diffusion term in the integral (Gaussians). The sampling of the weight term
$w$ is often realized with the \emph{branching} technique in which $w$ gives the probability of a configuration 
to multiply at the next step according to the normalization. 
Computationally, this is implemented by weighting estimators according to $w$, and generating from each single 
walker a number of walkers
\begin{equation}
n=\left [w+\xi \right ] \,,
\end{equation}
where $\xi\in[0;1]$ is a random number and $[x]$ means integral part of $x$. 

This is the most used technique to sample the weight term, but other techniques to sample $w$ are 
possible but not interesting in this work.

\section{Importance Sampling}
\label{sec:is}
The technique described in the above section is rather inefficient because the weight term of Eq. \ref{eq:weight} 
could suffer of very large fluctuations, for example when a particle moves close to another one.

A big improvement to overcome this problem is possible by means of \emph{importance sampling} techniques. 
The Green's function used for propagation can be modified as following:
\begin{equation}
\label{eq:is}
\tilde G(R,R',\Delta\tau)={\psi_I(R')\over\psi_I(R)}G(R,R',\Delta\tau) \,.
\end{equation}

The so called \emph{importance function} $\psi_I$ in the above equation is often (but not necessarily)
the same of that used for the projection.
In this case the walker's distribution is sampled according the function
\begin{equation}
f(R,\tau)=\psi_I(R)\psi(R,\tau) \,.
\end{equation}

By inserting Eq. \ref{eq:is} in \ref{eq:evolution} and expanding near $R$, the integration gives an additional 
term in $G_0$ and in the weight to keep the normalization of $G$.
The diffusion is now guided by an additional drift term
\begin{equation}
\label{eq:gis}
G_0(R,R',\Delta\tau)\rightarrow\tilde G_0(R,R',\Delta\tau)= 
\left({m\over2\pi\hbar^2\Delta\tau}\right)^{3A\over2}
e^{-{m\left(R-R'+2{\nabla\psi_I(R)\over\psi_I(R)}\right)^2\over2\hbar^2\Delta\tau}} \,,
\end{equation}
and the weight \ref{eq:weight} becomes
\begin{equation}
\label{eq:wis}
w\rightarrow \tilde w=e^{-\left({E_L(R)+E_L(R')\over2}-E_T\right)\Delta\tau} \,,
\end{equation}
where
\begin{equation}
E_L(R)=-{\hbar^2\over2m}{\nabla^2\psi_I(R)\over\psi_I(R)}+V(R) \,,
\end{equation}
is the local energy of the system.

It is possible to show that if one multiply the Eq. \ref{eq:schr} by the importance function $\psi_I$, the 
Schr\"odinger equation tunes into a Fokker-Plank equation:
\begin{equation}
-\frac{\partial}{\partial\tau}f(R,\tau)=-\frac{\hbar^2}{2m}\nabla^2 f(R,\tau)
+\frac{\hbar^2}{2m}\nabla\Bigg[2\frac{\nabla\psi_I}{\psi_I}f(R,\tau)\Bigg]+E_L(R)f(R,\tau) \,,
\end{equation}
and it can be verified that the corresponding Green's function is given by the two terms of Eq. \ref{eq:gis} 
and \ref{eq:wis}.

\section{Auxiliary Field Diffusion Monte Carlo}
In the case of nuclear Hamiltonians the potential $V$ contains quadratic spin and isospin and tensorial 
operators, so the many body wave function cannot be written as a product of single particle spin-isospin states. 

For example, let consider the generic quadratic spin operator $\vec\sigma_i\cdot\vec\sigma_j$
where the $\vec\sigma$ are Pauli's matrices operating on particles.
It is possible to write
\begin{equation}
\vec\sigma_i\cdot\vec\sigma_j=2\Big[(\vec\sigma_i+\vec\sigma_j)^2-\sigma_i^2-\sigma_j^2\Big]=2P_{ij}^\sigma-1 \,,
\end{equation}
where $P_{ij}^\sigma$ interchanges two spins, and this means that the wave function of each spin-pair must contain both components 
in the triplet and singlet spin-state\cite{pieper98,carlson99b}. 
By considering all possible nucleon pairs in the systems, the number of possible spin-states grows exponentially 
with the number of nucleons. Another way to understand this problem is simply connected to the fact that such 
operators cannot be diagonalized in a single particle spin-base.
These problems are obviously related also to quadratic isospin-operators.

In order to perform a DMC calculation with standard nuclear Hamiltonians, it is then necessary to sum over all 
possible single particle spin-isospin states of the system to build the trial wave function used for propagation. 
This is the standard approach in GFMC calculations for nuclear systems.

The idea of AFDMC is to rewrite the Green's function in order to reduce the quadratic dependence by 
spin and isospin operators to linear by using the Hubbard-Stratonovich transformation. 

Let us consider the first six operators of the NN potential of the form of Eq. \ref{eq:voperators}. The
operators can be recast in a more convenient form
\begin{eqnarray}
\label{eq:v6pot}
&&V=\sum_{i<j}\sum_{p=1}^6 v_p(r_{ij})O^{(p)}(i,j)=V_{SI}+V_{SD}= 
\nonumber \\
&&V_{SI}+{1\over2}\sum_{i\alpha,j\beta}\sigma_{i\alpha}A_{i\alpha,j\beta}^{(\sigma)}\sigma_{j\beta} 
+{1\over2}\sum_{i\alpha,j\beta}\sigma_{i\alpha}A_{i\alpha,j\beta}^{(\sigma\tau)}\sigma_{j\beta}\vec\tau_i\cdot\vec\tau_j
+{1\over2}\sum_{i,j}A_{i,j}^{(\tau)}\vec\tau_i\cdot\vec\tau_j \,,
\nonumber \\
\end{eqnarray}
where Latin indices label nucleons, Greek indices stay for Cartesian components, and
\begin{equation}
V_{SI}=\sum_{i<j}v_1(r_{ij}) 
\end{equation}
is the spin-isospin independent part of the interaction. The 3A by 3A matrices $A^{(\sigma)}$ and $A^{(\sigma\tau)}$, 
and the A by A matrix $A^{(\tau)}$ contain the interaction between nucleons of other terms:
\begin{eqnarray}
&&A_{i\alpha,j\beta}^{(\sigma)}=v_3(r_{ij})\delta_{\alpha\beta}
+v_5(r_{ij})(3\hat r_{ij}^\alpha\hat r_{ij}^\beta-\delta_{\alpha\beta}) \,,
\nonumber \\ 
&&A_{i\alpha,j\beta}^{(\sigma\tau)}=v_4(r_{ij})\delta_{\alpha\beta}
+v_6(r_{ij})(3\hat r_{ij}^\alpha\hat r_{ij}^\beta-\delta_{\alpha\beta}) \,,
\nonumber \\
&&A_{i,j}^{(\tau)}=v_2(r_{ij}) \,.
\end{eqnarray}
The matrices $A$ are zero along the diagonal (when $i=j$), in order to avoid self interaction, and are real and symmetric,
having then real eigenvalues and eigenvectors given by:
\begin{eqnarray}
&&\sum_{j\beta}A_{i\alpha,j\beta}^{(\sigma)}\psi_{n,j\beta}^{(\sigma)}=
\lambda_n^{(\sigma)}\psi_{n,i\alpha}^{(\sigma)} \,,
\nonumber \\
&&\sum_{j\beta}A_{i\alpha,j\beta}^{(\sigma\tau)}\psi_{n,j\beta}^{(\sigma\tau)}=
\lambda_n^{(\sigma\tau)}\psi_{n,i\alpha}^{(\sigma\tau)} \,,
\nonumber \\
&&\sum_jA_{i,j}^{(\tau)}\psi_{n,j}^{(\tau)}=\lambda_n^{(\tau)}\psi_{n,i}^{(\tau)} \,.
\end{eqnarray}
The matrices $A^{(\sigma)}$ and $A^{(\sigma\tau)}$ have $n=1...3A$ eigenvalues and eigenvectors, while $A^{(\tau)}$ has $n=1...A$.
One can then define a new set of operators written in terms of eigenvectors of matrices $A$:
\begin{eqnarray}
\label{eq:oper}
&&O_n^{(\sigma)}=\sum_{j\beta}\sigma_{j\beta}\psi_{n,j\beta}^{(\sigma)} \,,
\nonumber \\
&&O_{n\alpha}^{(\sigma\tau)}=\sum_{j\beta}\tau_{j\alpha}\sigma_{j\beta}\psi_{n,j\beta}^{(\sigma\tau)} \,,
\nonumber \\
&&O_{n\alpha}^{(\tau)}=\sum_{j}\tau_{i\alpha}\psi_n^{(\tau)} \,.
\end{eqnarray}
The potential \ref{eq:v6pot} becomes
\begin{equation}
V_{SD}={1\over2}\sum_{n=1}^{3A} O_n^{(\sigma)2}\lambda_n^{(\sigma)}
+{1\over2}\sum_{\alpha=1}^3\sum_{n=1}^{3A} O_{n\alpha}^{(\sigma\tau)2}\lambda_n^{(\sigma\tau)}
+{1\over2}\sum_{\alpha=1}^3\sum_{n=1}^{A} O_{n\alpha}^{(\tau)2}\lambda_n^{(\tau)} \,.
\end{equation}

Now it is possible to reduce the spin-isospin dependence of the operators from quadratic to linear by means of 
the Hubbard-Stratonovich transformation. Given a generic operator $\hat O$ and a parameter $\lambda$:
\begin{equation}
\label{eq:hs}
e^{-\frac{1}{2}\lambda \hat O^2}=\frac{1}{\sqrt{2\pi}}\int dx e^{-\frac{x^2}{2}+\sqrt{-\lambda}x\hat O} \,.
\end{equation}
With the Hubbard-Stratonovich transformation it is possible to linearize the quadratic form of 
the new set of operators $O_n$ defined in \ref{eq:oper}. Then the Green's function becomes:
\begin{eqnarray}
\label{eq:gf}
G(R,R,\Delta\tau)=\Big({m\over2\pi\hbar^2\Delta\tau}\Big)^{3A\over2}e^{-{m(R-R')^2\over2\hbar^2\Delta\tau}}
e^{-V_{SI}(R)\Delta\tau}
\nonumber \\
\times\prod_{n=1}^{15A}{1\over\sqrt{2\pi}}\int dx_ne^{-{x_n^2\over2}}
e^{\sqrt{-\lambda_n\Delta\tau}x_nO_n} \,,
\end{eqnarray}
the $O_n$ contain all the $3A$ $O_n^{(\sigma)}$ operators,
the $9A$ $O_{n\alpha}^{(\sigma\tau)}$ operators, and the $3A$ $O_{n\alpha}^{(\tau)}$ operators.
The newly introduced variables $x_n$, named \emph{auxiliary fields}, have to be sampled in order to evaluate the integral 
in Eq. \ref{eq:gf}.
The linear form of this operators gives the possibility to write the trial wave function as a 
product of single particle spin-isospin functions. In fact the effect of the new operators $O_n$ 
in the Green's function consists in a rotation of the spin-isospin of nucleons without mixing the spin-isospin 
state of nucleon pairs. 

The sampling of auxiliary fields to perform the integral in Eq. \ref{eq:hs} eventually gives 
the same effect as the propagator with quadratic spin-isospin operators acting on a trial wave function
containing all the possible good spin-isospin states.
The effect of the Hubbard-Stratonovich is then to reduce the number of terms in the trial wave function from exponential 
to linear. The price to pay is an additional computational cost due to the diagonalization of $A$ matrices and the sampling 
of the integral over auxiliary fields.

Sampling of auxiliary fields can be achieved in several ways. The more intuitive one, in the spirit 
of Monte Carlo sampling is to consider the Gaussian in the integral \ref{eq:hs} as a probability distribution.
The sampled values are then used to determine the action of the operators on the spin-isospin part of the 
wave function.
This is done exactly as in the diffusion process by noting that a each field value doesn't depend 
by the old value. Other techniques to solve the integral are possible (i.e. with the three-point Gaussian quadrature\cite{koonin97}) 
but they will not be considered in this work.

The introduced auxiliary fields might also be given a physical interpretation. In fact, they might be seen as a sort 
of meson field that 
generate the interaction between nucleons. By considering the simple one-pion exchange picture between two nucleons,
the operators in the Green's function \ref{eq:gf} represent exactly the interaction between a pion field with a 
nucleon. In fact the pion couple to nucleon with a term of the form:
\begin{equation}
\vec\sigma\cdot\vec\nabla\pi_\alpha \tau_\alpha \,,
\end{equation}
where $\pi_\alpha$ are the $\pi^0$ and $\pi^\pm$ fields.

By considering fixed nucleon coordinates, the faster pion degrees of freedom are integrated out giving 
the NN interaction as usual. In this way the Hubbard-Stratonovich is a reverse way of re-introducing the 
pion fields responsible of NN interaction. The fact that in sampling of integral \ref{eq:hs} each value 
of auxiliary fields does not depend on the previous one means that a pion fields does not have any correlation 
during the diffusion of slower nucleons, and the meson's kinetic energy can be included in counter terms of the 
interaction. This is exactly what is commonly done by integrating out the meson field to obtain a meson independent 
NN interaction.

The importance sampling in the Hubbard-Stratonovich transformation that rotate nucleon spinors can also be included. 
For auxiliary fields importance sampling is achieved by "guiding" the rotation given by each $O_n$ operator.
More precisely one can consider the following identity:
\begin{equation}
-\frac{x_n^2}{2}+\sqrt{-\lambda_n\Delta\tau}x_nO_n=-\frac{x_n^2}{2}+\sqrt{-\lambda_n}x_n\langle O_n\rangle
+\sqrt{-\lambda_n\Delta\tau}\left(O_n-\langle O_n\rangle\right) \,,
\end{equation}
where the mixed expectation value of the operator is calculated in the old spin-isospin configuration:
\begin{equation}
\langle O_n\rangle=\frac{\langle\psi_I\lvert O_n\rvert R,S\rangle}{\langle\psi_I\vert R,S\rangle} \,.
\end{equation}
This can be implemented by shifting the Gaussian used to sample auxiliary fields, and considering the extra terms in 
the weight for branching
\begin{equation}
\label{eq:hsis}
e^{-\frac{x_n^2}{2}+\sqrt{-\lambda_n\Delta\tau}x_nO_n}=
e^{-\frac{(x_n-\bar x_n)^2}{2}}e^{\sqrt{-\lambda_n\Delta\tau}x_nO_n}e^{2\bar x_nx_n-\frac{\bar x_n^2}{2}} \,,
\end{equation}
where
\begin{equation}
\bar x=\sqrt{-\lambda_n\Delta\tau}\langle O_n\rangle \,.
\end{equation}
The additional weight term in Eq. \ref{eq:hsis} to be added to Eq. \ref{eq:gf} can also be included as 
a local potential, so it becomes
\begin{equation}
e^{-\frac{\langle\psi_I\lvert V\rvert R,S\rangle}{\langle\psi_I\vert R,S\rangle}\Delta\tau} \,.
\end{equation}
By combining the diffusion, the rotation and all the additional factors it is possible to obtain two equivalent algorithm.
The first explicitly contains the drift correction included in the Gaussian used for the diffusion process 
and the extra terms given by the guidance of rotations:
\begin{equation}
\frac{e^{\frac{-m(R'-R)^2}{2 \hbar^2\Delta t}-V_c \Delta t}}
{e^\frac{-m(R'-R - d)^2}{2 \hbar^2 \Delta t}}
\frac{\langle \psi_I|R'S'\rangle }{\langle \psi_I|RS\rangle}
\prod_n e^\frac{-2x_n\bar x_n-\bar x_n^2}{2} \,,
\end{equation}
where the drift term $d$ can be arbitrarily fixed (for example, one could choose to guide the diffusion 
according to the modulus or with the real part of the trial wave function), by only correcting the 
factor $\frac{\langle \psi_I|R'S'\rangle }{\langle \psi_I|RS\rangle}$.

The second is calculated in a simpler 'local energy' scheme, as described in section \ref{sec:is} and 
derived in Eq. \ref{eq:wis}:
\begin{equation}
\exp \left [ - \left ( -\frac{\hbar^2}{2m}
\frac{\nabla^2 |\langle \psi_I |R S\rangle|}{|\langle \psi_I|R S\rangle|}
+ \frac{\langle \psi_I|V|RS\rangle}{\langle\psi_I|RS \rangle }
\right ) \Delta t
 \right ]
\frac{\langle \psi_I|R'S\rangle |\langle \psi_I|R S\rangle|}
{\langle \psi_I|RS\rangle |\langle \psi_I|R' S\rangle|} \,,
\end{equation}
where in the last expression the choice of the drift term is taken to be
\begin{equation}
d=\frac{\nabla\vert\langle\psi_I\vert R,S\rangle\vert}{\vert\langle\psi_I\vert R,S\rangle\vert} \,,
\end{equation}
but the generalization to a general drift term is trivial.

If the drift term is the same, the two algorithms are equivalent and should sample the same Green's function.

\section{Spin-orbit propagator for neutrons}
As described in Sec. \ref{sec:NNint} the nuclear Hamiltonian contains also a spin-orbit term that has to be 
treated separately from other terms described in the previous section.

The spin-orbit potential is defined
\begin{equation}
v_{LS}(i,j)=v_{LS}(r_{ij})\vec L_{ij}\cdot\vec S_{ij}=v_{LS}(r_{ij})(\vec L\cdot\vec S)_{ij} \,.
\end{equation}
Because the spin-orbit is a non-local operator the corresponding Green's function is not trivial to be derived.

One way to evaluate a $\vec L\cdot\vec S$ propagator is to consider the first order of expansion in $\Delta\tau$ 
\begin{equation}
\label{eq:LS1}
e^{-\frac{1}{2}v_{LS}(r_{ij}) (\vec L\cdot\vec S)_{ij}\Delta\tau}
\simeq\left[1-\frac{1}{2}v_{LS}(r_{ij}) (\vec L\cdot\vec S)_{ij}\Delta\tau\right] \,,
\end{equation}
and acting with this on the free propagator $G_0$\cite{pieper98}. The derivative terms of the above expression give
\begin{equation}
(\vec\nabla_i-\vec\nabla_j)G_0(R,R')=-\frac{m}{\hbar^2\Delta\tau}(\Delta\vec r_i-\Delta\vec r_j)
G_0(R,R') \,,
\end{equation}
where $\Delta\vec r_i=\vec r_i-\vec r_i'$.

Then
\begin{equation}
\label{eq:LS2}
(\vec L\cdot\vec S)_{ij}G_0(R,R')=-\frac{1}{4i}\frac{m}{\hbar^2\Delta\tau}
(\vec r_i-\vec r_j)\times(\Delta\vec r_i-\Delta\vec r_j)\cdot(\vec\sigma_i+\vec\sigma_j)G_0(R,R') \,,
\end{equation}
and after the exponentiation and the multiplication of $v_{LS}(r)$, the following propagator is obtained:
\begin{equation}
P_{LS}=\exp\left[-\sum_{i\neq j}\frac{m}{\hbar^2\Delta\tau}v_{LS}(r_{ij})(\vec L\cdot\vec S)_{ij}\right] \,.
\end{equation}

This propagator must be corrected for counter terms arising at order $\Delta\tau$ from the approximation 
implied in Eq. \ref{eq:LS1}.
Note that in Eq. \ref{eq:LS2} there are terms depending on $\sqrt{\Delta\tau}$ in the integration 
for the evolution of wave function in Eq. \ref{eq:intschr}:
\begin{equation}
r\times r'=r\times(r+\Delta r)=r\times\Delta r\sim \sqrt{\Delta\tau} \,,
\end{equation}
and one expects that other spurious terms linear in $\Delta\tau$ are intrinsically contained in $P_{LS}$.

In order to see the effect of these additional terms the trial wave function is expanded to the first order: 
\begin{equation}
\Psi(R')\simeq\Psi(R)+\sum_i\vec\nabla_{i}\Psi(R)(\vec r_i'-\vec r_i)+...
\end{equation}
with $R=\vec r_1,..,\vec r_N$.
Then the $P_{LS}$ is also expanded, giving
\begin{eqnarray}
\lefteqn{\quad P_{LS}\simeq 1-\sum_{i\neq j}\frac{m}{4i}\frac{v_{LS}(r_{ij})}{\hbar^2} 
(\Sigma_{ij}\times\vec r_{ij})\cdot\Delta\vec r_i }
\nonumber \\
&&-\sum_{i\neq j}\sum_{k\neq l}\frac{m^2v_{LS}(r_{ij})v_{LS}(r_{kl})}{32\hbar^4} 
(\Sigma_{ij}\times \vec r_{ij})\cdot\Delta\vec r_i(\Sigma_{kl}\times\vec r_{kl})\cdot\Delta\vec r_k+...
\nonumber \\
\end{eqnarray}
where
\begin{equation}
\Sigma_{ij}=\vec\sigma_i+\vec\sigma_j \,,
\end{equation}
and it was used the relation $\vec a\cdot(\vec b\times\vec c)=\vec c\cdot(\vec a\times\vec b)$.

By inserting $P_{LS}$ with the $G_0$ in the Eq. \ref{eq:intschr} gives
\begin{eqnarray}
\label{eq:monster}
&&\Psi(R,\tau+\Delta\tau)=\int dR'G_0(R,R',\Delta\tau)
\Bigg[1-\sum_{i\neq j}\frac{m}{4i}\frac{v_{LS}(r_{ij})}{\hbar^2} 
(\Sigma_{ij}\times\vec r_{ij})\cdot\Delta\vec r_i
\nonumber \\
&&-\sum_{i\neq j}\sum_{k\neq l}\frac{m^2v_{LS}(r_{ij})v_{LS}(r_{kl})}{32\hbar^4} 
(\Sigma_{ij}\times\vec r_{ij})\cdot\Delta\vec r_i(\Sigma_{kl}\times\vec r_{kl})\cdot\Delta\vec r_k+...\Bigg]
\nonumber \\
&&\times\Bigg[\Psi(R,\tau)
+\sum_i\vec\nabla_{i}\Psi(R,\tau)(\vec r_i'-\vec r_i)+...\Bigg] \,.
\end{eqnarray}

The integration in the variables $R'$ shows the presence of extra terms. Note that terms linear in $\Delta\vec r$ 
and quadratic with different components (like $\Delta\vec r_i\Delta\vec r_j$ with $i\neq j$)
integrate to zero, while quadratic terms in the same components give $\hbar^2\Delta\tau/m$.
The integration of Eq. \ref{eq:monster} of the part with quadratic $\Delta\vec r$ coming from 
linear terms of both $P_{LS}$ and the wave function gives
\begin{eqnarray}
-\frac{\hbar^2\Delta\tau}{m}\sum_{i\neq j}\frac{m}{4i}\frac{v_{LS}(r_{ij})}{\hbar^2}
(\Sigma_{ij}\times\vec r_{ij})\cdot\vec\nabla_i\Psi(R)
\nonumber \\
=-\Delta\tau\sum_{i\neq j}v_{LS}(r_{ij})(\vec L\cdot\vec S)_{ij}\Psi(R) \,,
\end{eqnarray}
that is the spin-orbit contribution of the propagator.

The integration of the quadratic terms in $\Delta\vec r$ coming from $P_{LS}$ gives the additional 
spurious terms to the propagator. These extra corrections have the structure of a two- and three-body 
terms:
\begin{eqnarray}
&&\Delta\tau(V_2+V_3)=\Delta\tau\frac{m}{32\hbar^2}\sum_i\sum_{j\neq i}\sum_{k\neq i}
v_{LS}(r_{ij})v_{LS}(r_{ik})(\Sigma_{ij}\times\vec r_{ij})\cdot(\Sigma_{ik}\times\vec r_{ik})= 
\nonumber \\
&&\Delta\tau\frac{m}{32\hbar^2}\sum_i\sum_{j\neq i}\sum_{k\neq i}
v_{LS}(r_{ij})v_{LS}(r_{ik})[\vec r_{ij}\cdot\vec r_{ik}\Sigma_{ij}
\cdot\Sigma_{ik}-\Sigma_{ik}\cdot\vec r_{ij}\Sigma_{ij}\cdot\vec r_{ik}] \,.
\nonumber \\
\end{eqnarray}
The term with $j=k$ is an extra two-body contribution
\begin{equation}
V_2^{add}=\sum_{i<j}{mr_{ij}^2v_{LS}^2(r_{ij})\over8\hbar^2}
[2+\vec\sigma_i\cdot\vec\sigma_j-\vec\sigma_i\cdot\hat r_{ij}\vec\sigma_j\cdot\hat r_{ij}] \,,
\end{equation}
while that with $j\neq k$ gives an additional three-body correction
\begin{eqnarray}
\label{eq:LSv3}
V_3^{add}&=&-\sum_{i<j<k}\sum_{cyc}\frac{m\vec r_{ij}\cdot\vec r_{ik} v_{LS}(r_{ij})v_{LS}(r_{ik})}
{16\hbar^2}[\hat r_{ij}\cdot\hat r_{ik}(2+\vec\sigma_i\cdot\vec\sigma_j
+\vec\sigma_i\cdot\vec\sigma_k
\nonumber \\
&&+\vec\sigma_j\cdot\vec\sigma_k)
-\vec\sigma_i\cdot\hat r_{ij}\vec\sigma_j\cdot\hat r_{ik}
-\vec\sigma_k\cdot\hat r_{ij}\vec\sigma_i\cdot\hat r_{ik}
-\vec\sigma_k\cdot\hat r_{ij}\vec\sigma_j\cdot\hat r_{ik}] \,.
\end{eqnarray}

Note that both the two- and the three-body additional terms contain a quadratic form of 
spin operators, and in order to be sampled they need some additional Hubbard-Stratonovich variables.

An alternate way to include the additional counter terms given by this spin-orbit propagator is 
to consider a different propagator of the form
\begin{eqnarray}
&&\exp\left[\sum_{i\neq j}\frac{1}{4i}\frac{m}{\hbar^2\Delta\tau}v_{LS}(r_{ij})
[\vec r_{ij}\times\Delta\vec r_{ij}]\cdot\vec\sigma_i\right]
\nonumber \\
&&\times\exp\left[-\frac{1}{2}\left[\sum_{i\neq j}\frac{1}{4i}\frac{m}{\hbar^2}v_{LS}(r_{ij})
[\vec r_{ij}\times\Delta\vec r_{ij}]\cdot\vec\sigma_i\right]^2\right] \,.
\end{eqnarray}

This different propagator contains the quadratic form of spin operators and need 
additional Hubbard-Stratonovich fields to be applied. However, by expanding the two exponential to keep 
terms linear in $\Delta\tau$, the above expression gives
\begin{eqnarray}
&&\Bigg[1+\sum_{i\neq j}\frac{1}{4i}\frac{m}{\hbar^2}v_{LS}(r_{ij})
[\vec r_{ij}\times\Delta\vec r_{ij}]\cdot\vec\sigma_i
\nonumber \\
&&-\sum_{i\neq j}\sum_{k\neq l}\frac{1}{32}\frac{m^2}{\hbar^4}v_{LS}(r_{ij})v_{LS}(r_{kl})
(\vec r_{ij}\times\Delta\vec r_{ij})\cdot\vec\sigma_i(\vec r_{kl}\times\Delta\vec r_{kl})
\cdot\vec\sigma_k+...\Bigg]
\nonumber \\
&&\times\Bigg[1+\frac{1}{2}\sum_{i\neq j}\sum_{k\neq l}\frac{1}{16}\frac{m^2}{\hbar^4}
v_{LS}(r_{ij})v_{LS}(r_{kl})
(\vec r_{ij}\times\Delta\vec r_{ij})\cdot\vec\sigma_i(\vec r_{kl}\times\Delta\vec r_{kl})
\cdot\vec\sigma_k\Bigg] \,,
\nonumber \\
\end{eqnarray}
the quadratic forms of $\Delta\vec r$ giving the extra terms after the integration cancel each selves
and others are of higher order in $\Delta\tau$.
The two forms of spin-orbit propagator are equivalent to the first order in $\Delta\tau$.

An important remark concerns the spin-orbit propagator in presence of isospin-operators. For neutrons 
the $\vec\tau_i\cdot\vec\tau_j$ operators is a constant, but the inclusion of these operators to the 
$\vec L\cdot\vec S$ is hard. The additional counter terms without isospin-operators have a three-body structure,
but the spin-operators appearing there are quadratic, and the Hubbard-Stratonovich transformation can be used safely. 
In presence of isospin-operators, the counter terms contain cubic spin-isospin operators. This fact prevents 
the straightforward use of the Hubbard-Stratonovich transformation.

\section{Three-body propagator for neutrons}
\label{sec:TNIprop}
For pure neutron systems the three-body forces in the Urbana or Illinois form can be rewritten as a 
two-body term, and then safely included in the Hubbard-Stratonovich transformation as described in the 
above section. 

Urbana and Illinois three-nucleon potentials can be written in the form
\begin{equation}
V_{ijk} = A_{2\pi}^{PW} O^{2\pi,PW}_{ijk}
+ A_{2\pi}^{SW} O^{2\pi,SW}_{ijk} + A_{3\pi}^{\Delta R} O^{3\pi,\Delta R}_{ijk}
+A_R O^R_{ijk} \,.
\end{equation}
Let us examine the structure of each of these operators. As explained in Sec. \ref{sec:TNIint} the 
one pion exchange interaction at the base of the three-body force is the same of that used in the NN interaction,
implying that the OPE operator $X^{\rm op}_{ij}$ defined in Eq. \ref{eq:Xop} has 
the same algebric structure as the 
spin dependent part of the two--body potential $V_{SD}$ of Eq. \ref{eq:v6pot}. Therefore, it can
be expressed in a similar way, in terms of a $3N$ by $3N$ matrix
$X_{i\alpha ;j\beta}$ as
\begin{equation}
X^{\rm op}_{ij} = \sigma_{i\alpha} X_{i\alpha ; j\beta} \sigma_{j\beta} \,,
\end{equation}
with $X_{i\alpha;j\beta}=0$ if $i=j$.  Notice that this matrix
is symmetric under Cartesian component interchange 
$\alpha \leftrightarrow \beta$
and under particle label interchange $i \leftrightarrow j$
and is also fully symmetric $X_{j\beta;i\alpha} = X_{i\alpha; j\beta}$.

For neutrons the commutator term in $O^{2\pi,PW}_{ijk}$ vanish, and the anticommutator can be reduced 
in a quadratic form of the spin operators. Thus the operator of Eq. \ref{eq:v2ppw} reduces to
\begin{equation}
\sum_{i<j<k} O^{2\pi,PW}_{ijk}=
4\sum_{i<j} \sigma_{i\alpha}\sigma_{j\beta} X^2_{i\alpha;j\beta} \,,
\end{equation}
where 
\begin{equation}
X^2_{i\alpha ; j\beta } = 
\sum_{k} X_{i\alpha;k\gamma} X_{k\gamma;j\beta} \,.
\end{equation}

The Tucson $S$ wave component $O^{2\pi,SW}_{ijk}$ of the Eq. \ref{eq:v2psw} is quadratic in the spin-operators
and it is convenient to write it as:
\begin{equation}
\sum_{i<j<k} O^{2\pi,SW}_{ijk} = \sum_{i<j} \sigma_{i\alpha}\sigma_{j\beta}
\sum_k G^S_{\alpha ;ik} G^S_{\beta ;jk} \,,
\end{equation}
where
\begin{eqnarray}
Z(x) &=& \frac{x}{3} \left [Y(x)-T(x) \right ] \,
\nonumber\\
\vec G^S_{i j} &=& \vec r_{ij} Z(m_\pi r_{ij}) \,.
\end{eqnarray}

The 3-pion exchange $O^{3\pi,\Delta R}_{ijk}$  terms have a central part and a spin dependent part.
\begin{equation}
\sum_{i<j<k} O^{3\pi,\Delta R}_{ijk} = V_c + V_s \,.
\end{equation}
The central part is given by
\begin{equation}
V_c = \frac{400}{9}\sum_{i<j} X^2_{i\alpha;j\beta}X_{i\alpha;j\beta} \,,
\end{equation}
which is $200/9$ times the trace of $X^3$. The spin dependent part is
\begin{equation}
V_s = \frac{200}{27}
\sum_{i<j} \sigma_{i\alpha} \sigma_{j\beta} X^2_{i\gamma;j\kappa}
X_{i\delta;j\omega}
\epsilon_{\alpha\gamma\delta} \epsilon_{\beta \kappa \omega} \,,
\end{equation}
and one can just write out the 4 nonzero terms for each combination
of $\vec\sigma_i$ and $\vec\sigma_j$.

Finally the spin independent $O^{R}_{ijk}$ terms can be written 
using just pair sums as
\begin{equation}
\sum_{i<j<k} O^R_{ijk}= G^R_0+\frac{1}{2} \sum_i (G^R_i)^2 \,,
\end{equation}
with
\begin{eqnarray}
G^R_i &=&\sum_{k \neq i} T^2(m_\pi r_{ik}) \,,
\nonumber\\
G^R_0 &=&-\sum_{i<j} T^4(m_\pi r_{ij}) \,.
\end{eqnarray}

Then the spin dependent part of the three--body interaction can be easily
included in the matrix $A_{i\alpha;j\beta}$ by
\begin{eqnarray}
A_{i\alpha;j\beta}\rightarrow A_{i\alpha;j\beta}&+&2 A_{2\pi}^{PW}
X^2_{i\alpha;j\beta} 
+\frac{1}{2} A_{2\pi}^{SW} \sum_k G^S_{\alpha;ik} G^S_{\beta;jk}
\nonumber \\
&+&\frac{200}{54} A_{3\pi}^{\Delta R}X^2_{i\gamma;j\kappa} X_{i\delta;j\omega}
\epsilon_{\alpha\gamma\delta} \epsilon_{\beta \kappa \omega} \,,
\end{eqnarray} 
and the central term to be included in the propagator is given by
\begin{eqnarray}
V_{SI}(R)\rightarrow V_{SI}(R)
&+&A_R\Big[G^R_0+\frac{1}{2} \sum_i (G^R_i)^2\Big] \nonumber \\
&+&A_{3\pi}^{\Delta R}\frac{400}{9}\sum_{i<j} X^2_{i\alpha;j\beta}X_{i\alpha;j\beta} \,.
\end{eqnarray}

The coefficients of three-body forces Urbana and Illinois can be found in Ref. \cite{pieper01}.

\section{The constrained-path and the fixed-phase approximation}
As described in the above sections, the DMC project out the ground-state of a given Hamiltonian
as a distribution of points in the configuration space.
On the other hands for the diffusion interpretation to be valid, the trial wave function must always be 
positive definite, since it represents a population density\cite{reynolds82}. Thus the use of DMC is restricted 
to a class of problem where the trial wave function is always positive such as for a 
many-Boson system in the ground-state.

For Fermionic system the DMC algorithm can be used by artificially splitting the space in regions corresponding to 
positive and negative regions
of the trial wave function. It is possible to define a nodal surface where the trial wave function is zero
and during the diffusion process a walker acrossing the nodal surface is dropped;
this approximate algorithm is called fixed-node\cite{schmidt84,anderson76}. It can be proved that 
it always provides an upper bound to the true Fermionic ground-state.

Other difficult techniques to overcome the Fermion sign problem exist, like the 
transient estimators analysis\cite{ceperley80} or other alternative methods to control the 
sign-problem\cite{kalos99,kalos00,kalos96}.
However, at present, none of these methods is efficient enough to compute physical properties with a sufficient 
accuracy (with the noticeable exception of the electron gas).

In the case of nuclear Hamiltonians, or for each problem where the trial wave function must be complex, 
the constrained-path approximation\cite{zhang03,zhang95,zhang97} is usually apply to avoid the Fermion sign problem.
The constrained-path method was originally proposed by Zhang et al. as a generalization of the fixed-node 
approximation to complex wave functions.

If the overlap between walkers and the trial-wave function is complex (as in our case), the usual sign 
problem becomes a phase problem. It is possible to deal with it by constraining the path of 
walkers to regions where the real part of the overlap with the trial wave function is positive.
The first AFDMC calculations were performed by applying this constrain\cite{schmidt99}.

Let be considered a complex importance function. In order to keep real the space coordinates of the system,
the drift term of Eq. \ref{eq:gis} that gives the importance sampling has to be a real term.
In the case of the constrained-path approximation, a good choice for the drift is
\footnote{It should be possible to consider a different drift, but this form was found to give better 
results and was employed in the past AFDMC calculations}
\begin{equation}
d=\frac{\nabla Re[\psi_I(R)]}{Re[\psi_I(R)]} \,,
\end{equation}
and to eliminate the decay of the signal-to-noise ratio it is possible to impose the constrained-path 
approximation, that is realized by requiring that the real part of the overlap of each walker 
with the trial wave function must keep the same sign. 
Thus, one has to impose that
\begin{equation}
\frac{Re[\psi_I(R')]}{Re[\psi_I(R)]}>0 \,,
\end{equation}
where $R$ and $R'$ denote the coordinates of the system after and before the diffusion of a time-step.
If this condition is violated, the walker is dropped.

An alternate way to control the sign problem is the fixed-phase approximation, that was
proposed for systems whose Hamiltonian contains a magnetic field\cite{ortiz93} following earlier explorations
by Carlson for nuclear systems\cite{carlson87}.

Let us start with the same assumption that spacial coordinates of the system must be real, and let then consider
the following drift term:
\begin{equation}
d={\nabla\lvert\psi_I(R)\rvert\over\lvert\psi_I(R)\rvert} \,.
\end{equation}
With this choice the weight for branching becomes
\begin{equation}
\label{eq:weight2}
\exp\left[{-\left(-{\hbar^2\over2m}{\nabla^2\lvert\psi_I(R)\rvert\over\lvert\psi_I(R)\rvert}+{V\psi_I(R)\over\psi_I(R)}
\right)\Delta\tau}\right]
\times\frac{\vert\psi_I(R)\vert}{\vert\psi_I(R')\vert}\frac{\psi_I(R')}{\psi_I(R)} \,.
\end{equation}
Note that in the above expression there is the usual importance sampling term as in Eq. \ref{eq:is}, and an 
additional term that corrects the choice of the drift term.

A generic complex wave function can be written as
\begin{equation}
\psi(R)=\vert\psi(R)\vert e^{i\phi(R)} \,,
\end{equation}
where $\phi(R)$ is the phase of $\psi(R)$, then the term appearing in Eq. \ref{eq:weight2} can be rewritten as 
\begin{equation}
\frac{\vert\psi_I(R)\vert}{\vert\psi_I(R')\vert}\frac{\psi_I(R')}{\psi_I(R)}=e^{i[\phi(R',S)-\phi(R,S)]} \,.
\end{equation}

The fixed-phase approximation constrains the walkers to have the same phase as the importance function $\psi_I$. 
It can be applied by keeping the real part of the last term. 
In the same way as adopted for the importance sampling,
after the expansion and the integration of the Green's Function to 
keep its normalization fixed, one has an additional term in the Green's function due to the phase, 
that must be included in the weight
\begin{equation}
1-{\hbar^2\over2m}\left(\nabla\phi\right)^2\Delta\tau \,,
\end{equation}
which can efficiently be included in the weight by keeping the real part of the kinetic energy. In fact:
\begin{equation}
Re{\nabla^2\psi_I(R)\over\psi_I(R)}={\nabla^2\lvert\psi_I(R)\rvert\over\lvert\psi_I(R)\rvert}-
\left(\nabla\phi(R)\right)^2 \,.
\end{equation}
Then the real part of the kinetic energy includes the additional weight term given by the fixed-phase approximation.

Both the constrained-path and the fixed-phase are approximations to deal with the Fermion sign-problem and in principle 
should be equivalent if the importance function is close to the real ground-state of the system. 

It is important to note that Carlson et at. showed that with the constrained-path approximation the DMC algorithm 
does not necessarily give an upper bound in the calculation of energy\cite{carlson99}. This was also observed by 
Wiringa et al. in some nuclear simulations using the GFMC technique\cite{wiringa00}.

Unfortunately it cannot be proved that the fixed-phase approximation gives or not an upper bound to the real energy.
However, it was observed by Francesco Pederiva\cite{pederiva06} that in some particular case the fixed-phase gives 
systematically energies higher than the fixed-node energies (that always is an upper bound). The system studied was a quantum 
dot in an open-shell configuration; in that case the trial wave function is generally complex but can be also written 
as a real function with properly linear combination of the single-particle orbitals (for more information on a generic 
trial wave function, see the following section).

\section{Trial wave function}
\label{sec:psitrial}
The trial wave function used as the importance and projection function for the AFDMC algorithm has
the following form:
\begin{eqnarray} 
\psi_T(R,S) =  \Phi_S(R) \Phi_A(R,S) \,,
\end{eqnarray}
where $ R\equiv (\vec r_1,\dots,\vec r_N) $ are the Cartesian coordinates
and $S\equiv (s_1,\dots ,s_N)$ are the spin and isospin coordinates of the system.
The spin-isospin assignments $s_i$ consist in giving the four-spinor components, namely
\[ 
s_i \equiv \left(\begin{array}{c} 
a_i \\ b_i \\ c_i \\ d_i
\end{array}\right)=a_i|p\uparrow\rangle+b_i|p\downarrow\rangle+c_i|n\uparrow\rangle+d_i|n\downarrow\rangle \,,
\]
where $a_i$, $b_i$, $c_i$ and $d_i$ are complex numbers, and the $\{|p\uparrow\rangle,p\downarrow\rangle
,|n\uparrow\rangle,|n\downarrow\rangle\}$ is the proton-up, proton-down, neutron-up and neutron-down base.

The Jastrow correlation function is symmetric under the exchange of two particles. Its role is to include 
the inter-particle correlation to the trial wave function for short distances. The generic form for the 
Jastrow is
\begin{equation}
\Phi_S(R)=\prod_{i<j}f_J(r_{ij}) \,,
\end{equation}
where the function $f_J$ has been taken as the scalar component of the
Fermi Hyper Netted Chain in the Single Operator Chain approximation
(FHNC/SOC) correlation operator $\hat F_{ij}$ which minimizes the energy per
particle of nuclear or neutron matter\cite{pandharipande79,wiringa88} at the correct density depending to the system to be studied.

The Jastrow part of the trial wave function in the AFDMC case only has the role of reducing the overlap
of nucleons, therefore reducing the energy variance. Since it does not change the phase of the wave function, it
does not influence the computed energy value in projection methods. 

The antisymmetric part of the trial wave function depends on the system to be studied; this function is generally 
given by the ground-state of non-interacting Fermions, that is usually written as a Slater determinant
\begin{eqnarray} 
\Phi_A(R,S) =A \left[\prod_{i=1}^N \phi_\alpha(\vec r_i,s_i)\right]=Det\{\phi_\alpha(\vec r_i,s_i)\} \,,
\end{eqnarray}
where $\alpha$ is the set of quantum numbers of single-particle orbitals $\phi_\alpha$ depending on the system to be studied.

For both nuclei and neutron drops orbitals are labelled with the set of quantum numbers $\alpha=\{n,j,m_j\}$:
\begin{eqnarray}
\phi_\alpha(\vec r_i,s_i)=R_{n,j}(r_i)\left[Y_{l,m_l}(\Omega)\chi_{s,m_s}(s_i)\right]_{j,m_j} \,,
\end{eqnarray}
where $R_{n,j}$ is a radial function (whose determination will be described in each specific case), 
$Y_{l,m_l}$ are spherical harmonics and $\chi_{s,m_s}$ are spinors in the usually proton-neutron-up-down base. 
The angular functions are coupled to spinors using the Clebsh-Gordan to have orbitals in the $\{n,j,m_j\}$ base 
according to the usual shell-model classification of the nuclear single-particle spectrum.

In the case of nuclei or neutron drops, a summation of different determinants might be needed to 
build a trial wave function with the same symmetry of the ground-state of the nucleus considered. Because the AFDMC 
projects out the lower energy state not orthogonal to the starting trial wave function, it is possible to 
study a state with given symmetry imposing to the trial wave function the total angular momentum $J$ 
experimentally observed. This can be achieved by taking a summation over a different set of determinants, thus
\begin{equation}
D(R,S)=\left[\sum_{\beta} c_\beta Det_\beta\right]_{J,M_J} \,,
\end{equation}
where the $c_\beta$ coefficients are determined in order to have the eigenstate of total angular 
momentum $J=j_1+...+j_N$.

For a nuclear or neutron matter calculation the antisymmetric part is the ground state of the Fermi gas, built 
over a set of plane waves. The infinite uniform system is simulated with $N$ nucleons in a cubic periodic 
box of volume $L^3$. The momentum vectors in this box are
\begin{equation}
\vec k_\alpha=\frac{2\pi}{L}(n_{\alpha x},n_{\alpha y},n_{\alpha z}) \,,
\end{equation}
where $\alpha$ labels the quantum state and $n_x$, $n_y$ and $n_z$ are integer numbers describing the state.
The single-particle orbitals are given by
\begin{equation}
\phi_\alpha(\vec r_i,s_i)=e^{i\vec k_\alpha\cdot\vec r_i}\chi_{s,m_s}(s_i) \,.
\end{equation}

The system has a shell structure that must be closed, in order to meet the requirement of homogeneity and 
isotropy, then the total number of Fermions in a particular spin-isospin 
configuration must be 1, 7, 19, 27, 33,...
It is also possible to modify the trial wave function to deal with an arbitrary number of Fermions and 
keeping zero the total momentum of the system\cite{lin01} but this was not applied 
to the calculations discussed in this work.

A particular phase of the system can also be simulated, for example to study the gap of a superfluid system.
This is possible by including in the Slater determinant some pairing functions as done in several works about superfluid Fermi 
gases\cite{carlson03c,chang04b,astrakharchik04}, or by pairing Fermions with a Pfaffian\cite{fabrocini05,bajdich06}.

In neutron matter the paired orbitals used in the Pfaffian are taken to be
\begin{equation}
\phi_{12}(\vec r_i,\vec r_j,s_i,s_j)=
\sum_\alpha\frac{v_{\vec k_\alpha}}{u_{\vec k_\alpha}} e^{i\vec k_\alpha\cdot\vec r_{ij}}\xi(s_i,s_j)
\end{equation}
where the $\xi(s_i,s_j)$ is a spin function of two neutrons that correctly describe the spin-coupling, that in  
this case is in the $^1S_0$-channel, and $v_{\vec k_\alpha}$ and $u_{\vec k_\alpha}$ are BCS coefficients to be 
determined variationally.

\chapter{Results: nuclei and neutron drops}
In this and in the following chapters I will give a description of the results obtained by applying the AFDMC 
method to the study of nuclear and neutron matter and nuclei.

This chapter is devoted to the analysis of finite systems, nuclei and neutron drops. The main concern is the 
comparison of AFDMC results with those computed by other methods. This first step is necessary 
to provide a benchmark test to assure the high accuracy of AFDMC calculations in dealing with Hamiltonians
containing tensor forces.
Some results on heavier nuclei, which can be tackled only by AFDMC, will also be given.

The accuracy of AFDMC in calculating properties of neutron drops with a more sophisticated Hamiltonian 
with a spin-orbit and a TNI interaction in addition to the tensor force, and a comparison with the GFMC  
will be shown. The role of the TNI interaction for heavier neutron drop with an increasing of the density 
will be discussed.

\section{Open and closed shell nuclei}
\label{sec:nuclei}
As shown in chapter \ref{ch:hamiltonian} the nuclear Hamiltonians are more complicated than those commonly used 
in condensed-matter problems. The increasing sophistication of NN and TNI interactions was not followed 
by a parallel development of a computational technique providing enough accuracy to predict 
properties of medium heavy nuclear systems, and progresses in accurate ground-state calculations has been quite slow.

GFMC calculations was originally applied to the ground-state calculation of the alpha particle in 1987.
The most recent progress was achieved in 2005 with the application to $^{12}$C. 
In table \ref{tab:Avstime} a chronological sequence of GFMC computations for heavier and heavier nuclei is 
displayed.

\begin{table}[ht]
\begin{center}
\vspace{0.2cm}
\begin{tabular}{||l|cc||}
\hline
nucleus           & year & Ref. \\
\hline
\hline
$^4$He            & 1987 & \cite{carlson87,carlson88} \\
$^5$He            & 1993 & \cite{pieper98} \\
$^6$Li and $^6$He & 1995 & \cite{pudliner95} \\
$^7$Li and $^7$He & 1997 & \cite{pudliner97} \\
A=8 nuclei        & 2000 & \cite{wiringa00} \\
A=9,10 nuclei     & 2002 & \cite{pieper02} \\
$^{12}$C          & 2005 & \cite{pieper05} \\
\hline
\end{tabular}
\vspace{0.2cm}
\caption{The ground-state calculations performed with the GFMC technique of nuclei up to the present. All the 
reported calculations were performed with some state-dependent nuclear Hamiltonian, and in the second column
the year of publication of corresponding results is reported.}
\label{tab:Avstime}
\end{center}
\end{table}

Unfortunately the GFMC, that at present is probably the most accurate method applicable up to A=12 nuclei, 
will not be useful for nuclei heavier than A$>$12 at least for the future years because the huge power 
computing resources needed\cite{pieper06b,carlson06}.
At present, the major goal in nuclear physics is the development of a many-body technique as accurate as 
the GFMC but applicable to medium nuclei.

In this section it will be shown the capability of the AFDMC technique to deal with heavier nuclei problems 
by keeping the same accuracy of few-body techniques.

\subsection{Comparison between CP-AFDMC, FP-AFDMC and other techniques}
The possibility of applying the AFDMC with the constrained-path approximation (CP-AFDMC) to nuclei was originally explored by 
Schmidt et al.
Early calculations yielded good results for nuclear Hamiltonians with a $v_4$-like NN interaction\cite{fantoni01b}, but in the 
presence of tensorial forces there was a very poor agreement with estimates obtained with other methods,
as can be seen in table \ref{tab:CP-AFDMCnuclei} where the CP-AFDMC results are taken from \cite{schmidt03}.

\begin{table}[ht]
\begin{center}
%\vspace{0.2cm}
\begin{tabular}{||c|c|c|c||}
\hline
nucleus & potential & $E_{AFDMC}$ & $E_{other}$\\
\hline
\hline
$^4$He & UV14       & -20.7(3)    & \\
$^4$He & AV8'       & -17.9(1)    & -22.8(2) \\
\hline
$^{16}$O & UV14     & -84(2)      & -82.40 \\
$^{16}$O & AV14     & -57(3)      & -84.0  \\
\hline
\end{tabular}
\vspace{0.2cm}
\caption{The CP-AFDMC ground-state energy in MeV for the indicated potentials truncated to the $v_6$ level compared 
with other results\cite{schmidt03}. The $^4$He energy was calculated with the GFMC using the Argonne AV8'\cite{carlson02},
while the binding of the oxygen were performed with the CVMC\cite{pieper92} and 
with the CBF\cite{fabrocini98,fabrocini00} using the Urbana $v_{14}$\cite{lagaris81}
and the Argonne AV14\cite{wiringa84} interactions.}
\label{tab:CP-AFDMCnuclei}
\end{center}
\end{table}

Both the alpha particle and the oxygen ground-state were calculated using a different NN interactions,
namely the Urbana UV14\cite{lagaris81} and the Argonne AV14\cite{wiringa84} truncated to include only the first six
operators. The alpha particle ground-state was solved with the GFMC method\cite{carlson02}, that cannot be used for
oxygen. For the latter the only microscopic calculations available are the Cluster Variational Monte Carlo (CVMC)\cite{pieper92} and
the Correlated Basis Functions (CBF)\cite{fabrocini98,fabrocini00}.

The effect of the constrained-path approximation is apparently very different for the two interactions: 
for the AV14 interaction the CP-AFDMC does not reproduce well the binding energy of the alpha particle 
with a difference of about 1.2 MeV per nucleon higher than the GFMC estimate. 
In the case of oxygen with the UV14 the CP-AFDMC energy is roughly on par with the CBF calculation, while 
the result with the AV14 is very far from that of other variational calculations. It has to be noted that in the AV14 interaction 
the strength of the isoscalar-tensor and the isovector-tensor operators is higher than in UV14.
Hence, the CP-AFDMC seems to well project the ground-state of a $v_4$-like interaction but not the isovector-tensor component
of the potential.
This fact suggests that such component induce a strong change in the nodal/phase structure of the many-nucleons wave function.

These calculations were then repeated using the fixed-phase rather then the constrained-path approximation. 
The trial wave functions were modified with respect to the calculations of Ref. \cite{schmidt03}. 
Radial orbitals were numerically computed in a self-consistent Hartree-Fock potential well generated using Skyrme 
forces\cite{vautherin72,shen96} with parameters fitted for light nuclei\cite{xinhua97}. 
The fixed-phase (FP) approximation turns out to be quite more effective than the constrained-path. 

The FP-AFDMC energy for alpha particle compares very well with the estimates from other methods, within 1\% as 
can be seen in table \ref{tab:FP-AFDMCalpha}. 

\begin{table}[ht]
\begin{center}
\vspace{0.2cm}
\begin{tabular}{||l|c||}
\hline
method   & $E$ \\
\hline
\hline
FP-AFDMC & -27.13(10)  \\
GFMC     & -26.93(1)   \\
EIHH     & -26.85(2)   \\
\hline
\end{tabular}
\vspace{0.2cm}
\caption{The FP-AFDMC ground-state energy of the alpha particle in MeV for the AV6' interaction compared 
with other results\cite{gandolfi07b}. The GFMC result is that reported in Ref. \cite{wiringa02} after the subtraction
of the Coulomb contribution of 0.7 MeV\cite{pieper06}, while the EIHH result is for the pure AV6' interaction
\cite{orlandini06}.}
\label{tab:FP-AFDMCalpha}
\end{center}
\end{table}

The AV6' interaction employed\cite{argonnev18} is a simplification of the full AV18 as explained in section \ref{sec:NNint}.
FP-AFDMC energy is compared with the GFMC result of Ref. \cite{wiringa02} after the subtraction of the Coulomb 
term that is of about 0.7MeV\cite{pieper06}. The other result were obtained with the Effective Interaction 
Hyperspherical Harmonic (EIHH)\cite{barnea00,barnea01} that does not contain the Coulomb interaction\cite{orlandini06}.
Both the GFMC and EIHH methods were proved to be in an excellent agreement with other few-body 
techniques\cite{kamada01}, and the fact that FP-AFDMC is in par with this methods is the first important test to 
show its accuracy for light nuclei calculation\cite{gandolfi07b}.

In order to have another comparison with GFMC, FP-AFDMC was used to calculate the ground-state 
energy of $^8$He\cite{gandolfi07b} (note that the A=8 nucleus is just beyond the capability of few-body techniques).
The AV6' interaction was employed. The $^8$He is an open-shell nucleus and in order to have a trial 
wave function which is an eigenstate of total angular momentum $J=0$ a linear combination of different Slater 
determinants is needed. Results are reported in table \ref{tab:AFDMChe8}.

\begin{table}[ht]
\begin{center}
\begin{tabular}{||l|c||}
\hline
method & $E$ \\
\hline
\hline
FP-AFDMC & -23.6(5)    \\
GFMC     & -23.6(1)    \\
\hline
\end{tabular}
\vspace{0.2cm}
\caption{The calculated FP-AFDMC ground-state energy in MeV for $^8$He nucleus\cite{gandolfi07b} with AV6' interaction
compared with the GFMC result, 
whose result was taken from Ref. \cite{wiringa02} and the Coulomb term of 0.7 MeV\cite{pieper06} were 
subtracted.}
\label{tab:AFDMChe8}
\end{center}
\end{table}

It should be noted that $^8$He is an open shell nucleus. This fact does not seem to affect the agreement between
FP-AFDMC and the GFMC, which is excellent. 
One more the calculation was performed using the AV6' interaction, after the subtraction of the Coulomb term 
from the GFMC calculation.
An important further step has been the computation of the ground-state energy of $^{16}$O. This calculation is 
way beyond the limits that presently affect GFMC, and comparison is possible only with variational 
calculations (either Monte Carlo or CBF based). FP-AFDMC overcomes the problems observed in previous CP-AFDMC 
calculations.

The interaction used is the AV14 cut to the $v_6$-level. Results are reported in 
table \ref{tab:AFDMCox}. As it can be seen, the FP-AFDMC estimates is -90.8(1)MeV against -57(3)MeV of the CP-AFDMC.
The effect of the constrained-path in this case is apparently catastrophic.
The FP-AFDMC result is about 7\% lower than other variational estimates, consistent with 
the average difference between VMC and GFMC in light nuclei\cite{pieper04}.

\begin{table}[ht]
\vspace{0.2cm}
\begin{center}
\begin{tabular}{||l|c||}
\hline
method & $E$ \\
\hline
\hline
FP-AFDMC & -90.8(1)    \\
VCMC     & -83.2       \\
CBF      & -84.0       \\  
\hline
\end{tabular}
\vspace{0.2cm}
\caption{The calculated FP-AFDMC ground-state energy in MeV for $^{16}$O nucleus\cite{gandolfi07b} with AV14 interaction
cut to the $v_6$-level. The VCMC\cite{pieper92} and CBF\cite{fabrocini00} results of the corresponding 
interaction are also reported.}
\label{tab:AFDMCox}
\end{center}
\end{table}

The results on $^{16}$O show that AFDMC within the fixed-phase approximation can definitely deal with realistic nuclear 
Hamiltonians also for medium nuclei.

\subsection{Application of FP-AFDMC up to $^{40}$Ca}
\label{nucleiAV6}
The capability of FP-AFDMC to treat medium-heavy nuclei can be exploited to understand some features of the 
interactions commonly used in literature. One important open question is the portability of potentials fitted 
to reproduce light nuclei for heavier systems. How important the detailed operatorial structure does become?
Can reprojections of sophisticated AV18 potential like AV6' be accurate enough to reproduce the main physical 
features of medium heavy nuclei? To this end 
it could be interesting to analyze the contribution to the total energy of nuclei with respect to the experimental 
data.

The ground-state energy estimates of $^{16}$O and $^{40}$Ca, in addition to those of alpha particle and 
$^8$He nucleus previously reported, are shown in table \ref{tab:AFDMCnuclei}, together with the estimate of the energy of 
symmetric nuclear matter (SNM) at the empirical equilibrium density $\rho_0=$0.16$fm^{-3}$, calculated 
with the FP-AFDMC by simulating 28 nucleons in a periodic box\cite{gandolfi07} (see the following sections for more details).

\begin{table}[ht]
\begin{center}
\begin{tabular}{||c|c|c|c||}
\hline
nucleus   & $E_{AFDMC}$ & $E_{AFDMC}/A$ & $E_{exp}/A$ \\
\hline
\hline
$^4$He    & -27.13(10)  & -6.78         & -7.07 \\
$^8$He    & -23.6(5)    & -2.95         & -3.93 \\
$^{16}$O  & -100.7(4)   & -6.29         & -7.98 \\
$^{40}$Ca & -272(2)     & -6.80         & -8.55 \\
$\infty$  &             & -12.8(1)      & -16   \\
\hline
\end{tabular}
\vspace{0.2cm}
\caption{The FP-AFDMC ground-state energy in MeV of $^4$He, $^8$He, $^{16}$O and $^{40}$Ca for the Argonne AV6' 
interaction\cite{gandolfi07b}. 
Experimental energies are also reported\cite{exp00}. 
The estimate of the energy of SNM at equilibrium density, simulated with 28 nucleons in a 
periodic box\cite{gandolfi07} is also included.}
\label{tab:AFDMCnuclei}
\end{center}
\end{table}

The AV6' interaction is not sufficient to build the whole experimental binding energy\cite{exp00} of the nuclei considered.
This NN interaction gives about 96\% of total binding energy for alpha particle, 75\% for $^8$He, 79\% for $^{16}$O 
and 79\% for $^{40}$Ca. By neglecting the missing NN terms, the lack of the TNI seems to reach a saturation around the 
value of $\sim$20\%.
However the $^{16}$O is unstable to break up into 4 alpha particles, and the $^{40}$Ca has the same
energy of 10 alpha particles. This behavior is consistent with the simple pair counting argument 
of Ref. \cite{wiringa06}.

The estimate of SNM energy as well as the ground-state energy of $^{40}$Ca can be used to fit the surface 
energy coefficient in the Weizsacker formula\cite{bethe99} that is given 
as a function of the mass number $A$ and atomic number $Z$:
\begin{equation}
\label{eq:massformula}
\frac{BE(A,Z)}{A}=a_v-a_s\frac{1}{A^{1/3}}-a_c\frac{Z(Z-1)}{A^{4/3}}-a_{sym}\frac{(A-2Z)^2}{A^2}+\frac{\delta}{A} \,,
\end{equation}
where $a_v$ gives the volume energy, $a_s$ is the coefficient of the surface energy,
$a_c$ is the term of the Coulomb energy, $a_{sym}$ gives the asymmetry energy and 
$\delta$ describes the pairing energy.
The AFDMC calculations do not contain the Coulomb term. From the comparison of the binding 
energies per nucleon of symmetric nuclear matter and $^{40}$Ca, the fitted surface coefficient 
is 20.5 MeV, not too far from the experimental value of 17.23 MeV\cite{povh94}.

\section{Neutron drops}
\label{sec:neutdrop}
A neutron drop is a simple system, useful to benchmark theoretical predictions, and significant both for nuclear 
and astronuclear physics.
Several works on neutron drops were presented to extract some information for mean field 
calculations\cite{pudliner96,smerzi97} of neutron matter.

In order to test the ability to reproduce the ground-state of neutron drop the CP-AFDMC was applied to calculate the 
energy of a $^8n$ drop as well as the $^7n$ drop in both the $J$=3/2 and $J$=1/2 states\cite{pederiva04}, and results were compared
to the GFMC energy. The results indicated that the ground-state of $^8n$ drop was in agreement with GFMC within 2\%.
Despite the agreement on the total energies,
the yielded spin-orbit splitting (SOS) derived as the difference between the two $^7n$ states with different $J$, was too 
small compared to the GFMC estimate. 
This fact was interpreted as a problem in dealing with the spin-orbit contribution of the AFDMC algorithm, as just suggested 
by earlier neutron matter calculations (that however will be discussed in the following sections).

In addition, as just discussed in chapter \ref{ch:hamiltonian} the NN interaction is not sufficient to well describe properties of 
light nuclei and different forms of TNI were proposed to build a nonrelativistic Hamiltonian that well reproduces 
experimental results\cite{pieper01} such ground-state energy, density profile, square mean radius and others.
The Urbana IX TNI (UIX) was implemented to well describe properties of nuclei with A$\leq$4\cite{gazit06}, but it was not
satisfactory to reproduce heavier nuclei. An alternate form is the TNI Illinois-1 to 5 (IL1 to IL5),
developed to well describe nuclei up to A$\leq$8\cite{pieper01}, and later employed to compute the ground state
energy of nuclei with A$\leq$10\cite{pieper02} and subsequently up to A=12\cite{pieper05}.
However it was recently shown that these accurate TNI
interactions have some problems in reproducing the properties of neutron-rich nuclei like $^{10}$He.

CP-AFDMC calculation on neutron matter revealed that Illinois-like (ILx) TNI give very unexpected results. In particular, 
ILx gives a large attractive contribution to the energy when the density of neutron  
matter increases\cite{sarsa03}. With the same NN interaction, at equilibrium density $\rho_0$=0.16 fm$^{-3}$ all the ILx TNI 
interactions give similar results (within 10\%), but when density increases their contribution vary 
in a sensible way. 

Both points are hence discussed. 1) It will be shown that FP-AFDMC better reproduces the SOS, which is now comparable to 
that given by GFMC. 2) In order to better clarify the role of the nuclear TNI in pure neutron systems, 
I will discuss some ground-state energy of neutron drops made of 8 and 20 neutrons at several densities. 

The ground-state energy of a neutron drop can be calculated by starting from a nonrelativistic Hamiltonian of the form 
of Eq. \ref{eq:hamiltonian} with the addition of an external field:
\begin{equation}
H=-\sum_i{\hbar^2\over2m}\nabla_i^2+\sum_iV_{ext}(r_i)+\sum_{i<j}v_{ij}+\sum_{i<j<k}V_{ijk} \,,
\end{equation}
where the $v_{ij}$ NN interaction is the Argonne AV8' (see section \ref{sec:NNint}), the TNI considered 
were the Urbana IX and Illinois IL1-4 forms (see section \ref{sec:TNIint}), 
and the external Wood-Saxon well is needed to have a bound state for pure neutron system, otherwise unbound. The 
external field has the following form:
\begin{equation}
V_{ext}(r)=-\frac{V_0}{1+e^{(r-R)/a}} \,,
\end{equation}
where the parameter $a=$0.65 fm is fixed, while the $V_0$ and $R$ were varied to modify the density of the drop.

The trial wave function has the usual form with the omission of the isospin degrees of freedom, and radial orbitals 
were obtained by solving the Hartree-Fock problem with the Skyrme SKM force\cite{pethick95}.
The wave functions employed for drops with 8 and 20 neutrons are built as for closed shells.
For 8 neutrons we only need $1S_{1/2}$, $1P_{3/2}$ and $1P_{1/2}$ orbitals, and with the bigger droplet with 20 
neutrons we also add $1D_{5/2}$, $2S_{1/2}$ and $1D_{3/2}$ orbitals. In the case of 7 neutrons, the two states 
with $J=3/2$ and $J=1/2$ are obtained by dropping the orbitals $\phi_{1,3/2,3/2}$ and $\phi_{1,1/2,1/2}$ respectively.

The results of $^7n$ drop with the corresponding SOS contributions for various Hamiltonian are reported 
in table \ref{tab:ndropsos}.
The fixed-phase approximation noticeably improves the AFDMC results for SOS. In fact, the energy of 
$^7n(1/2^+)$ is lower respect that obtained with the constrained-path\cite{pederiva04} and in a better agreement 
with GFMC\cite{pieper01}. The two-body interaction employed is not the same, in fact we use the Argonne AV8' while in GFMC the 
more accurate AV18 is employed. However, the SOS given by GFMC with AV8' and UIX is 1.6 MeV, and the overall 
agreement is good.
Small differences are present with the IL3 where the SOS is a bit different but however well within error bars.

\begin{table}[!ht]
%\vspace{0.2cm}
\begin{center}
\begin{tabular}{||c|cccc||}
\hline
TNI & $^7n(1/2^+)$ & $^7n(3/2^+)$ & AFDMC-SOS & GFMC-SOS \\
\hline
\hline
UIX & -33.06(3)    & -31.51(2)    & 1.55(5) & 1.5(1) \\
IL1 & -35.28(3)    & -32.58(2)    & 2.70(5) & 2.8(3) \\
IL2 & -35.36(4)    & -32.43(1)    & 2.93(5) & 2.8(3) \\
IL3 & -36.06(4)    & -32.67(2)    & 3.39(6) & 3.6(4) \\
IL4 & -35.00(3)    & -32.53(3)    & 2.47(6) & 2.4(4) \\
\hline
\end{tabular}
\vspace{0.2cm}
\caption{The FP-AFDMC ground-state energy of 7 neutrons in a Wood-Saxon well with $V_0$=--20 MeV, R=3.0 fm and a=0.65 fm. 
Calculations were performed both with $J$=3/2 and $J$=1/2 with Argonne AV8' two nucleon interaction and with 
different TNI forces. In the last column the GFMC SOS from Ref. \cite{pieper01} were also reported. 
With the UIX the CP-AFDMC gives a SOS=0.7(2) MeV\cite{pederiva04}. All the energies are express in MeV.} 
\label{tab:ndropsos}
\end{center}
\end{table} 

The first main conclusion is that AFDMC with the fixed-phase can correctly build the spin-orbit 
correlations that seemed to be not satisfactory with the constrained-path approximation.

Let then be examined the ground-state energy as a function of the external well parameters of $^8n$ and $^{20}n$ 
drops respectively reported in table \ref{tab:ndrop8} and \ref{tab:ndrop20}. 
The $a$ parameter is fixed to the value of a=0.65 fm, while the $V_0$ and R are varied to change the density of the drop.
Reported results were obtained using the Argonne AV8' NN interaction, and with the addition of various TNI considered 
as indicated.
The TNI contribution, obtained by subtracting for each drop the corresponding AV8' result, are reported in table
\ref{tab:ndropTNI8} and \ref{tab:ndropTNI20}.

\begin{table}[!h]
%\vspace{0.2cm}
\begin{center}
\begin{footnotesize}
\begin{tabular}{||cc|cccccc||}
\hline
$V_0$ &  R  &  AV8'    & AV8'/UIX & AV8'/IL1 & AV8'/IL2 & AV8'/IL3 & AV8'/IL4   \\
\hline                                                                                   
\hline
20 &  2.5   & -18.66(4)& -17.98(4)& -19.69(4)& -19.72(4)& -20.02(4)& -19.61(4)  \\
20 &  3.0   & -38.51(6)& -37.55(2)& -39.76(6)& -39.72(3)& -40.16(7)& -39.59(4)  \\
20 &  3.5   & -57.18(5)& -56.30(6)& -58.35(5)& -58.25(5)& -58.66(5)& -58.19(4)  \\
30 &  2.5   & -56.94(5)& -54.82(5)& -60.08(4)& -59.8(1) & -61.2(2) & -59.60(6)  \\
30 &  3.0   & -87.96(1)& -86.11(6)& -91.2(2) & -90.98(5)& -92.15(9)& -90.69(4)  \\
30 &  3.5   &-115.38(5)&-113.80(5)&-117.69(4)&-117.5(1) &-118.24(8)&-117.51(4)  \\
40 &  2.5   &-101.01(2)& -97.92(3)&-107.01(8)&-106.58(9)&-108.82(9)&-106.6(1)   \\
40 &  3.0   &-142.81(3)&-140.11(3)&-148.07(8)&-147.7(1) &-149.5(1) &-147.52(8)  \\
40 &  3.5   &-177.82(5)&-175.65(4)&-181.30(5)&-181.04(4)&-182.32(7)&-180.84(6)  \\
50 &  2.5   &-149.12(7)&-144.89(7)&-158.0(3) &-157.4(1) &-160.8(2) &-157.05(9)  \\
50 &  3.0   &-200.46(5)&-196.77(4)&-206.95(6)&-206.81(4)&-209.3(1) &-206.42(6)  \\
50 &  3.5   &-242.05(4)&-239.36(5)&-246.62(7)&-246.4(1) &-248.17(8)&-246.26(7)  \\
60 &  2.5   &-199.66(7)&-194.28(7)&-211.1(2) &-210.5(2) &-214.9(3) &-210.2(2)   \\
60 &  3.0   &-260.14(4)&-255.59(3)&-268.9(1) &-268.5(1) &-271.7(2) &-267.92(8)  \\
60 &  3.5   &-308.01(6)&-304.84(4)&-313.8(1) &-313.5(1) &-315.7(1) &-313.25(6)  \\
\hline
\end{tabular}
\end{footnotesize}
\vspace{0.2cm}
\caption{AFDMC ground-state energy of 8 neutrons in a Wood-Saxon well as a function of parameters 
$V_0$ (in MeV) and R (in fm) of $V_{ext}$, as reported in the first two columns. Only the a=0.65 fm parameter was fixed. 
In the third column AV8' is the result of energy with only Argonne AV8' NN interaction, and from the fourth to eighth are 
those of NN plus various TNI forces. All the energies are express in MeV.} 
\label{tab:ndrop8}
\end{center}
\end{table} 

\begin{table}[!h]
\begin{center}
\begin{footnotesize}
\begin{tabular}{||cc|cccccc||}
\hline
$V_0$ &  R  &  AV8'    & AV8'/UIX & AV8'/IL1 & AV8'/IL2 & AV8'/IL3 & AV8'/IL4   \\
\hline                                                                                   
\hline
20 &  3.0   &  -24.0(3) & -19.8(2) &  -27.8(2)& -27.1(1) & -28.9(1) & -27.4(1)  \\
20 &  3.5   &  -65.3(2) & -60.5(1) & -73.2(1) & -72.0(1) & -74.63(8)& -72.3(1)  \\
20 &  4.0   & -109.30(9)&-104.4(1) &-117.3(1) &-116.21(3)&-118.4(2) &-116.30(8) \\
30 &  3.0   & -111.10(5)&-101.60(9)&-130.5(2) &-129.1(3) &-135.7(2) &-129.9(2)  \\
30 &  3.5   & -184.72(5)&-173.9(2) &-203.8(2) &-202.41(5)&-208.0(3) &-202.5(4)  \\
30 &  4.0   & -250.99(5)&-241.48(7)&-266.0(2) &-264.7(1) &-269.0(2) &-265.0(2)  \\
40 &  3.0   & -221.00(8)&-203.81(6)&-258.3(3) &-255.9(2) &-265.9(3) &-256.8(1)  \\
40 &  3.5   & -319.77(7)&-303.65(9)&-351.5(3) &-348.7(1) &-357.7(1) &-349.9(3)  \\
40 &  4.0   & -404.51(5)&-390.96(9)&-427.7(3) &-426.0(1) &-431.9(2) &-426.0(2)  \\
50 &  3.0   & -341.73(8)&-317.45(6)&-398.6(4) &-396.9(1) &-408.5(4) &-397.8(3)  \\
50 &  3.5   & -464.0(1) &-442.31(3)&-506.9(3) &-505.4(4) &-516.5(3) &-505.7(3)  \\
50 &  4.0   & -564.93(8)&-547.7(1) &-595.4(3) &-594.1(2) &-601.9(1) &-594.2(1)  \\
\hline
\end{tabular}
\end{footnotesize}
\vspace{0.2cm}
\caption{AFDMC ground-state energy of 20 neutrons as explained in table \ref{tab:ndrop8}.
All the energies are express in MeV.} 
\label{tab:ndrop20}
\end{center}
\end{table} 

\begin{table}[!h]
\vspace{0.2cm}
\begin{center}
\begin{tabular}{||cc|ccccc||}
\hline
$V_0$ &  R  & UIX   & IL1    & IL2   & IL3   & IL4 \\
\hline                                                      
\hline
20 &  2.5   & 0.68  & -1.03 & -1.06 & -1.36 & -0.95 \\
20 &  3.0   & 0.96  & -1.25 & -1.21 & -1.65 & -1.08 \\
20 &  3.5   & 0.88  & -1.17 & -1.07 & -1.48 & -1.01 \\
30 &  2.5   & 2.12  & -3.14 & -2.86 & -4.26 & -2.66 \\
30 &  3.0   & 1.85  & -3.24 & -3.02 & -4.19 & -2.73 \\
30 &  3.5   & 1.58  & -2.31 & -2.12 & -2.86 & -2.13 \\
40 &  2.5   & 3.09  & -6.00 & -5.57 & -7.81 & -5.59 \\
40 &  3.0   & 2.70  & -5.26 & -4.89 & -6.69 & -4.71 \\
40 &  3.5   & 2.17  & -3.48 & -3.22 & -4.50 & -3.02 \\
50 &  2.5   & 4.23  & -8.88 & -8.28 &-11.68 & -7.93 \\
50 &  3.0   & 3.69  & -6.49 & -6.35 & -8.84 & -5.96 \\
50 &  3.5   & 2.69  & -4.57 & -4.35 & -6.12 & -4.21 \\
60 &  2.5   & 5.38  &-11.44 &-10.84 &-15.24 &-10.54 \\
60 &  3.0   & 4.55  & -8.76 & -8.36 &-11.56 & -7.78 \\
60 &  3.5   & 3.17  & -5.79 & -5.49 & -7.69 & -5.24 \\
\hline
\end{tabular}
\vspace{0.2cm}
\caption{The TNI contributions to the $^8n$ drop as a function of parameters 
$V_0$ (in MeV) and R (in fm) of $V_{ext}$, as reported in the first two columns. Only the a=0.65 fm parameter was fixed. 
The TNI contributes to the total energy were obtained by subtracting from the full Hamiltonian with each TNI the AV8' result.
All the energies are express in MeV.} 
\label{tab:ndropTNI8}
\end{center}
\end{table} 

\begin{table}[!h]
\begin{center}
\begin{tabular}{||cc|ccccc||}
\hline
$V_0$ &  R  & UIX   & IL1   & IL2   & IL3   & IL4  \\
\hline                                                      
\hline
20 &  3.0   &  4.2  & -3.8  & -3.1  & -4.9  & -3.4 \\
20 &  3.5   &  4.8  & -7.9  & -6.7  & -9.33 & -7.0 \\
20 &  4.0   &  4.9  & -8.0  & -6.91 & -9.1  & -7.0 \\
30 &  3.0   &  9.5  &-19.4  &-18.0  &-24.6  &-18.8 \\
30 &  3.5   & 10.82 &-19.08 &-17.69 &-23.28 &-17.78\\
30 &  4.0   &  9.51 &-15.01 &-13.71 &-18.01 &-14.01\\
40 &  3.0   & 17.19 &-37.3  &-34.9  &-44.9  &-35.8 \\
40 &  3.5   & 16.12 &-31.73 &-28.93 &-37.93 &-30.13\\
40 &  4.0   & 13.55 &-23.19 &-21.49 &-27.39 &-21.49\\
50 &  3.0   & 24.28 &-56.87 &-55.17 &-66.77 &-56.07\\
50 &  3.5   & 21.69 &-42.9  &-41.4  &-52.5  &-41.7 \\
50 &  4.0   & 17.23 &-30.47 &-29.17 &-36.97 &-29.27\\
\hline
\end{tabular}
\vspace{0.2cm}
\caption{The TNI contributions to the $^{20}n$ drop as explained in table \ref{tab:ndropTNI8}.
All the energies are express in MeV.} 
\label{tab:ndropTNI20}
\end{center}
\end{table} 

The energy of the drop strongly depends on the parameters of external well. The first observation is that all Illinois 
forces give always an attractive contribution to the drop while the UIX term is repulsive. 
This is due to the three pions exchange term that only appears in the Illinois forces; 
this term should be always attractive in neutron drops and gives a large absolute 
contribution to the TNI with respect of two pions terms as reported by Pieper et al. in \cite{pieper01} for $^7n$ 
and $^8n$ neutron drops. Hence it is clear that for pure neutron systems the main physics of TNI comes from 3$\pi$ 
exchange and not from 2$\pi$ exchange terms.

A global trend of the Illinois potentials is observed by changing parameters of the external well (therefore varying the density of the system). 
For $^8n$ drop with $V_0$=20 MeV and 30 MeV the larger contribution of TNI is given with R=3.0 fm, and in the case of $V_0$=30 MeV 
the TNI with R=2.5 fm is similar to that of R=3.0 fm. When $V_0$ increases to 40 MeV, 50 MeV and 60 MeV the contribution of TNI drastically 
changes and the more bound is given if R is smaller (then when the density increase). 
For $^{20}n$ drop with $V_0$=20 MeV the larger TNI contribution is for the higher value of R instead of lower for $^8n$ drop. 
However when $V_0$ increases to 30 MeV, 40 MeV and 50 MeV the larger attractive contribution from TNI is given when R is smaller.
In the case of UIX a similar behavior is observed. In this case the TNI contribution is always repulsive, but the absolute 
value of TNI essentially has the same trend of Illinois forces. However, it has to be pointed out that the UIX contribution 
seems to change more slowly than that of Illinois when the density of the system increases.
These observations suggest that probably the attractive contribution given by 3$\pi$ exchange dependence by density is
stronger compared to the 2$\pi$ and central terms. Hence, for smaller densities the repulsion grows as the attraction, but when density 
increases the negative 3$\pi$ exchange term becomes more important than repulsion. This observation is in agreement of observations in 
pure neutron matter calculations\cite{sarsa03}.

The IL2 and IL4 seem to give substantially the same contribution in all the systems we simulated, but a small different trend is 
observed by using a different number of neutrons. In fact in the $^8n$ drop the IL2 is as always slightly more attractive than IL4 but 
in the case of $^{20}n$ the IL4 gives more binding, although the absolute values are very close. The IL2 and IL4 have a similar 
coefficient for 3$\pi$ exchange term, 0.0026 and 0.0021 respectively, and the parameter of the $O_{2\pi}^{PW}$ term are -0.037 and -0.028.
This suggest that probably the contribution of $O_{2\pi}^{PW}$ is always very small as reported in ref. \cite{pieper01}.

The IL1 is always more bound than IL2, and the difference between them never exceeds the 20\% of total TNI. The relative 
difference between IL1 and IL2 with respect to the total TNI seems to be higher when the $V_0$ of external well is smaller. The 
strength of $O_{2\pi}^{PW}$ is a bit different but not important as suggested in observing the behavior of IL2 and IL4.
However IL2 differs from IL1 because it has the term of $O_{2\pi}^{SW}$ that is zero in IL1 and this suggest that this term 
gives always repulsion.

The IL3 always gives more binding with respect to IL1, IL2 and IL4 and this probably is a consequence of the different strength of 
the $A_{3\pi}^{\Delta R}$ parameter that for IL3 is 0.0065 respect to 0.0026 of IL1 and IL2 and to 0.0021 of IL4. 

The radial density of a given drop can be calculated as a mixed operator\cite{pieper98}:
\begin{equation}
O_{mix}=2\frac{\langle\psi_T\lvert O \rvert \phi_0\rangle}{\langle\psi_T\vert \phi_0\rangle}
-\frac{\langle\psi_T\lvert O \rvert \psi_T\rangle}{\langle\psi_T\vert\psi_T\rangle} \,,
\end{equation}
where $\phi_0$ is the walker distribution in the limit of infinity imaginary time that represent the ground-state 
of the system.
Because the trial wave function $\psi_T$ does not contain any spin correlations, this mixed operator should be 
not well accurate.
The mixed-densities for each drop was calculated anyway to show the effect of using different Hamiltonians and 
for different parameters of external well.

\begin{figure}[!h]
\vspace{1.0cm}
\begin{center}
\includegraphics[width=10cm]{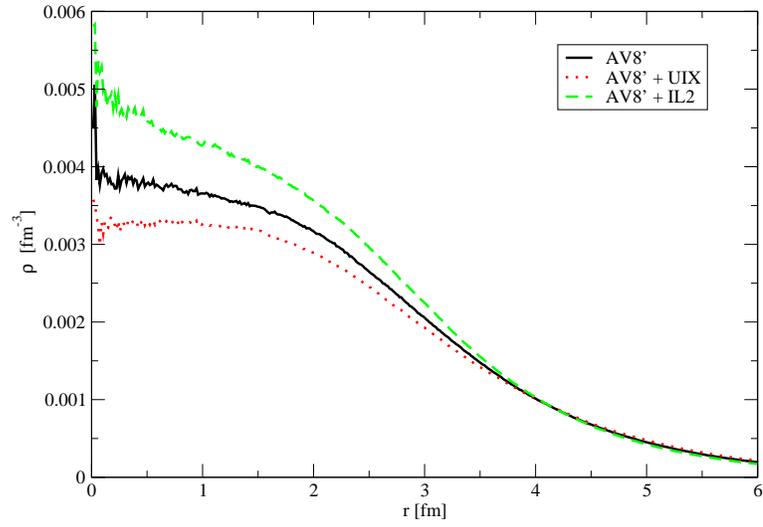}
%\vspace{0.5cm}
\caption{
Radial densities of $^{20}n$ drop interacting with the AV8' and AV8'+UIX and IL2 three-body force
in an external well with $V_0$=20 MeV, R=3.0 fm and a=0.65 fm.}
\label{figure:v20v3}
\end{center}
\end{figure}

\begin{figure}[!h]
\vspace{1.0cm}
\begin{center}
\includegraphics[width=10cm]{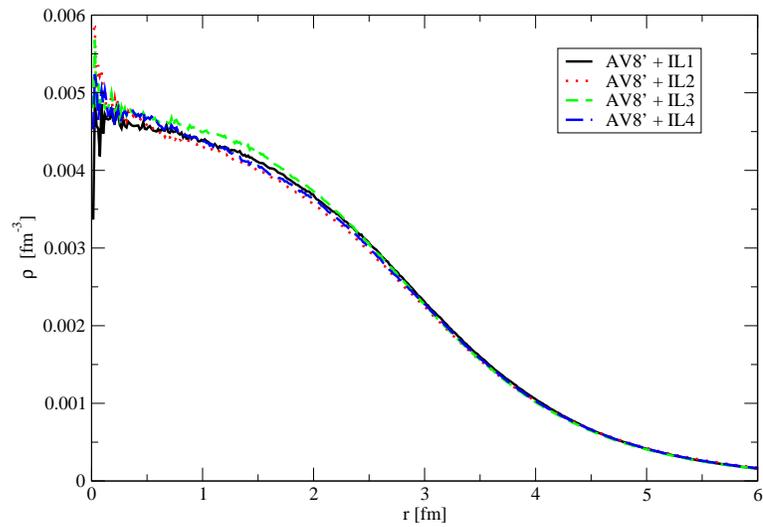}
%\vspace{0.5cm}
\caption{The same of Fig.\ref{figure:v20v3} with AV8' and different Illinois TNI forces.}
\label{figure:v20ilx}
\end{center}
\end{figure}

In Fig.\ref{figure:v20v3} we plot the radial densities of the $^{20}n$ drop in an external well with 
$V_0$=20 MeV, R=3.0 fm and a=0.65 fm for different Hamiltonians. It is clear that UIX and IL2 give completely 
different contributions to the density profile compared
to the pure two-body interaction. In particular the UIX lowers the density at the center of 
the drop and the IL2 increases it. This reflects the fact that UIX gives a repulsive contribution to the drop energy,
while IL2 is overall attractive.
All the Illinois TNI have this peculiarity as seen in Fig.\ref{figure:v20ilx} showing that all the densities computed 
with different Illinois forces are similar.

\begin{figure}[!ht]
\vspace{0.8cm}
\begin{center}
\includegraphics[width=10cm]{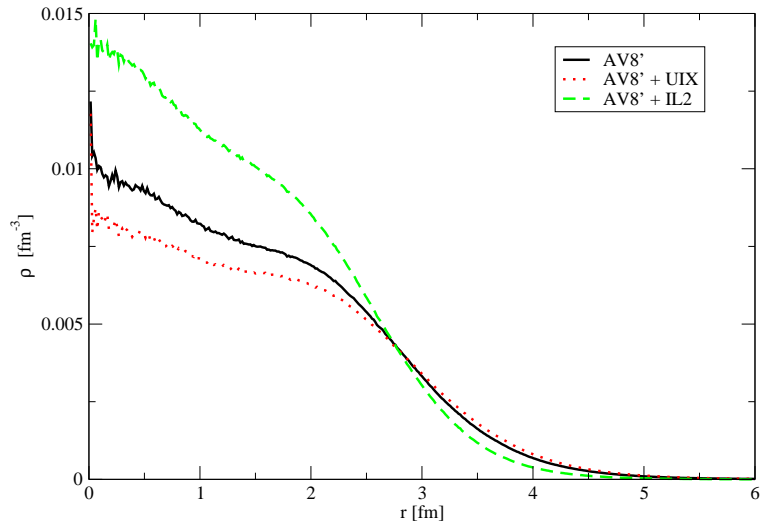}
%\vspace{0.5cm}
\caption{The same of Fig.\ref{figure:v20v3} with $V_0$=50 MeV.}
\label{figure:v50v3}
\end{center}
\end{figure}

The effect of using different TNI with the two-body interaction becomes bigger also when the density increases. 
The $V_0$ parameter of the external well was changed to 50 MeV to increase the density at the center of the drop. 
As it can be seen in Fig.\ref{figure:v50v3} the presence of TNI drastically increases the center density value.
The density computed with UIX is a bit lower with respect the pure AV8' NN interaction. However, the difference with the IL2 
case is evident.

\begin{figure}[!h]
\vspace{1.0cm}
\begin{center}
\includegraphics[width=10cm]{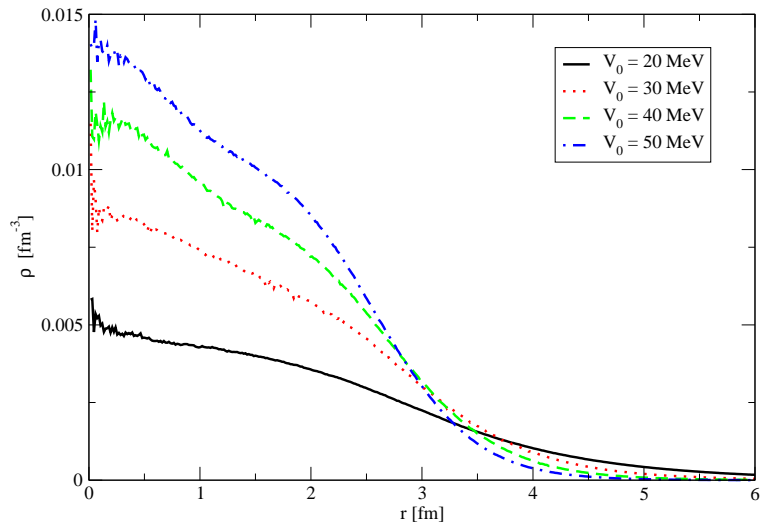}
%\vspace{0.5cm}
\caption{
Radial densities of neutron drop with 20 neutrons interacting with AV8'+IL2 for different 
values of the parameter $V_0$ of the external well with R=3.0 fm and a=0.65 fm.}
\label{figure:r30il2}
\end{center}
\end{figure}

\begin{figure}[!h]
\vspace{1.0cm}
\begin{center}
\includegraphics[width=10cm]{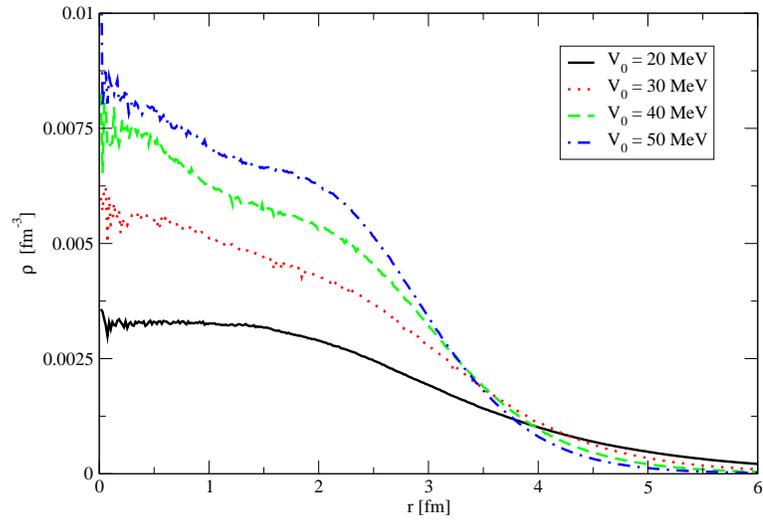}
%\vspace{0.5cm}
\caption{The same of Fig.\ref{figure:r30il2} with UIX three-body force instead of IL2.}
\label{figure:r30uix}
\end{center}
\end{figure}

In Fig.\ref{figure:r30il2} and \ref{figure:r30uix} were reported the densities of AV8' plus IL2 
and UIX and by varying the external well depth $V_0$=20 MeV, 30 MeV, 40 MeV and 50 MeV.

\begin{figure}[!h]
\vspace{1.0cm}
\begin{center}
\includegraphics[width=10cm]{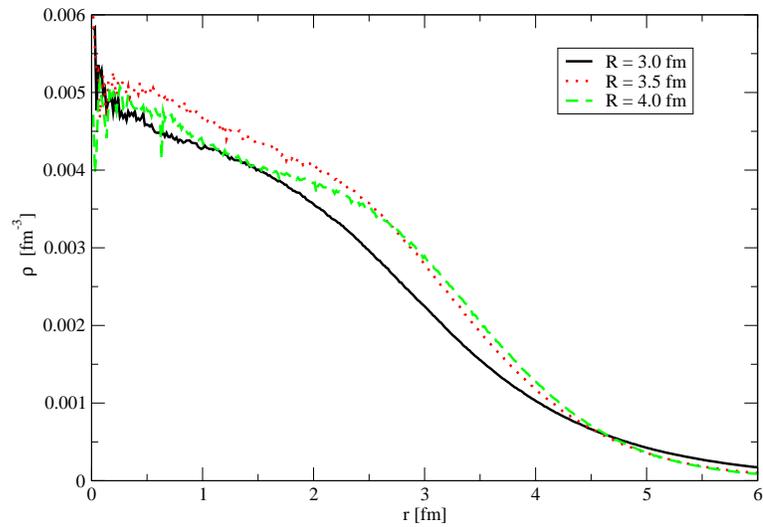}
%\vspace{0.5cm}
\caption{
Radial densities of neutron drop with 20 neutrons interacting with AV8'+IL2 interaction for different 
values of the parameter R of the external well with $V_0$=20 MeV and a=0.65 fm.}
\label{figure:v20il2}
\end{center}
\end{figure}

Finally we studied the density varying the R parameter of external well by keeping 
fixed $V_0$=20 MeV using the AV8'+IL2 interaction. As it can be seen in Fig.\ref{figure:v20il2} 
there are some non significant changes to the surface density, as expected.

\begin{figure}[!h]
\vspace{1.0cm}
\begin{center}
\includegraphics[width=10cm]{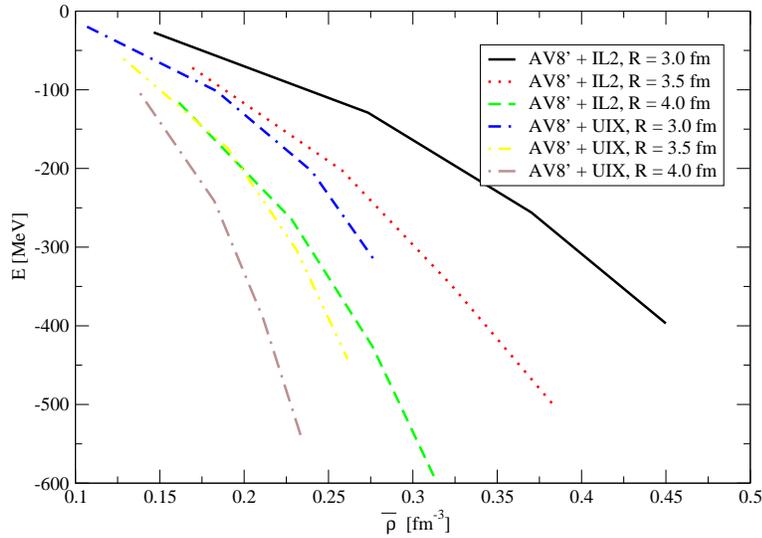}
%\vspace{0.5cm}
\caption{
Energy as a function of average density $\bar\rho$ for Hamiltonians with AV8' plus IL2 and UIX. Each curve correspond 
to different values of R parameter of the external well.}
\label{figure:n0}
\end{center}
\end{figure}

In order to extract some information about the energy dependence on density
we define an average density of the drop $\bar\rho$ as:
\begin{equation}
\bar\rho={N\over V} \,,
\end{equation}
where N is the number of neutrons in the drop, and V=$\frac{4}{3}\pi R_0^3$ is the volume of the drop, 
with $R_0$ calculated as
\begin{equation} 
<R_0^2>=\int r^2 \rho(r) dr \,.
\end{equation}
The energy as a function of the $\bar\rho$ for the $^{20}n$ is reported in Fig.\ref{figure:n0}.

A strong dependence of the energy on the parameter R appears. The effect of R is related 
only to the surface of the drop and not to the center, as noted above.
The UIX and IL2 give completely different contributions for the same R, indicating that the effect of different 
TNI is probably not only due to the different density in the center of the drop, but also to the surface effects.

\chapter{Results: nuclear and neutron matter}
The main challenging and fundamental problem in nuclear physics and nuclear astrophysics is the 
determination of the equation of state (EOS) of nuclear matter. The stability of neutron-rich 
nuclei\cite{todd03}, the dynamics of heavy-ion collisions\cite{danielewicz02,li02}, the structure 
of neutron stars\cite{lattimer01}, and the simulation of core-collapse supernova\cite{buras03}, all
depend on the nuclear matter EOS\cite{piekarewicz04}.

The calculation of the EOS of nuclear matter models with realistic NN interactions
is considered of primarily importance in astronuclear physics. Starting from
the famous "Bethe Homework problem" invented by Hans Bethe to test the existing many--body theories
in dealing with dense and cold hadronic matter\cite{owen76,howes78}, considerable advances have 
been made\cite{morales02,akmal97}.
However, yet, it is not possible to firmly ascertain the degree of accuracy of the approximations one has to 
introduce in any of such theories, and substantial discrepancies still exist among the different
theoretical estimates of the EOS, the response functions and Green's Functions of nuclear matter.

At present the theoretical uncertainties on the equation of state, coming from both
the approximations one has to introduce in the many-body methods, and the lack of 
knowledge of the nuclear interaction in the high density regime,
do not allow for definite conclusions in the comparison with the 
data from astronomical observations\cite{baldo07}.

In this chapter it will be shown the application of the AFDMC algorithm, within the fixed-phase approximation, 
to calculate the EOS of 
symmetric nuclear matter (SNM) using a simplified NN interaction which still contains full tensor correlations. 
This first work reveals and focuses
the problematic differences in principal many-body techniques used at present for the EOS 
calculation and also employed to fit the TNI.

The determination of the pure neutron matter (PNM) EOS obtained with a modern and realistic Hamiltonian 
will be presented, and some conclusion on the structure of neutron stars will be given and discussed.

Finally it will be discussed the application of AFDMC to study the low-density regime of PNM where 
a superfluid phase is present.

\section{Nuclear matter}
\label{sec:SNM}
As shown in Sec. \ref{sec:nuclei} the AFDMC combined with the fixed-phase approximation works very well with 
nuclear Hamiltonian that contains a tensor force, and permits to calculate the ground-state energy of nuclei in
good agreement with other accurate techniques.

In order to better understand the status of modern many-body theories and their accuracy\cite{bombaci05}
the AFDMC algorithm can be applied to perform a simulation of homogeneous nuclear matter\cite{gandolfi07}, 
simply by modifying the trial wave function used for the simulation. 

\subsection{Symmetric nuclear matter, comparison with other techniques}
In order to simulate the nucleonic matter in the normal state, a trial wave function using the 
Fermi-gas antisymmetric part as described in Sec. \ref{sec:psitrial} is employed.
As just explained, the use of plane waves as single-particle orbitals constrains the number of nucleons that
can be used in the calculation. The simulation is carried out with nucleons in a periodic box the volume of which
depends on the density and nucleon's number.

Finite-size effects are then present for two reasons: the first is the dependence of the kinetic energy of 
Fermi-gas to the number of particles, and the second is due to the interaction. 
A Monte Carlo simulation of an infinite system is performed in a finite periodic box with size $L$, and all 
inter-particle distances are truncated to $L/2$. Usually, tail corrections due to this truncation are 
estimated with an integration of the interaction from $L/2$ to infinity. However, this is possible 
only for spin-independent terms.
In order to include some tail correction to the potential it is possible to include the contributions
given by neighboring cells to the simulation box\cite{sarsa03}. Then, each two-body contribution to the potential is given by
\begin{equation}
V_p(x,y,z)=\sum_{mno} v_p\left(|(x+mL)\hat x+(y+nL)\hat y+(z+oL)\hat z|\right)
\end{equation}
where $L$ is the box size, and $m$, $n$ and $o$ are $\pm$1, $\pm$2, $\pm$3... depending on the number of shells of the 
boxes considered. For all the simulations performed, 
the inclusion of 26 additional neighbor (corresponding to $m$, $n$ and $o$ to be -1, 0 and 1) have convergent results.
In this way the finite-size corrections due to the interaction should be correctly included, but the presence of finite size 
effects in the kinetic energy should be verified.

The SNM-EOS is calculated by using 28 nucleons (7 for each spin-isospin state) interacting with the $v_6'$ 
NN interaction for several densities. The $v_6'$ interaction is a truncation of the Argonne AV8' to the $v_6$ level.
This interaction was chosen because in a recent work different many-body theories using this interaction 
predicted very different results to the SNM-EOS\cite{bombaci05}.

Before presenting the AFDMC EOS, let us examine the efficacy of the procedure to correct finite-size effects. 
In order to assess the magnitude of finite-size effects different simulations were performed using 
different numbers of nucleons for the highest and lowest densities considered. Results are reported in 
table \ref{tab:nucmatsize}. As it can be seen for both $\rho=$0.08 fm$^{-3}$ and $\rho=$0.48 fm$^{-3}$ the 
results given by 76 and 108 coincide with that given by 28 nucleons within 3\%. Then the effect due to the 
finite-size of the system is relatively small.

\begin{table}[ht]
\vspace{0.4cm}
\begin{center}
\begin{tabular}{||c|ccc||}
\hline
$\rho/\rho_0$ & E/A(28)  & E/A(76)  & E/A(108) \\
\hline
\hline
0.5           & -7.64(3) & -7.7(1)  & -7.45(2) \\
3.0           & -10.6(1) & -10.7(6) & -10.8(1) \\
\hline
\end{tabular}
\end{center}
\caption{AFDMC energies per particle in MeV of 28, 76 and 108 nucleons in a periodic box at various densities.}
\label{tab:nucmatsize}
\end{table}

Then the simulation of 28 nucleons interacting with the $v_6'$ interaction was repeated for different densities 
in the range with 0.5$\leq$($\rho/\rho_0$)$\leq$3, being $\rho_0=$0.16 fm$^{-3}$ the empirical equilibrium density.
The energy per nucleon for considered densities obtained with the AFDMC in the fixed-phase approximation are 
reported in table \ref{tab:eneSNM}. 
The first observation regard the empirical equilibrium density that is far to be reproduced. However this 
was expected because a pure NN interaction is not sufficient and a TNI is necessary to have the correct 
saturation density in SNM\cite{carlson83}.

\begin{table}[ht]
\begin{center}
\begin{tabular}{||cc|c||}
\hline
$\rho$ [fm$^{-3}$] & $\rho/\rho_0$ & E/A \\
\hline
\hline
0.08   & 0.5  &  -7.64(3) \\
0.12   & 0.75 &  -9.81(4) \\
0.16   & 1.0  & -11.5(1)  \\
0.20   & 1.25 & -13.0(1)  \\
0.24   & 1.5  & -13.73(7) \\
0.28   & 1.75 & -14.1(2)  \\
0.32   & 2.0  & -14.0(3)  \\
0.36   & 2.25 & -13.5(3)  \\
0.40   & 2.5  & -12.7(2)  \\
0.44   & 2.75 & -11.7(2)  \\
0.48   & 3.0  & -10.6(1)  \\
\hline
\end{tabular}
\end{center}
\caption{AFDMC energies per particle in MeV of 28 nucleons in a periodic box at various densities.}
\label{tab:eneSNM}
\end{table}
 
AFDMC results are then compared with two of mainly used many-body theories\cite{gandolfi07}, 
namely the Brueckner-Hartree-Fock (BHF) and the Fermi Hyper Netted Chain in the Single Operator Chain (FHNC/SOC).

The BHF method essentially recasts the Hamiltonian to a single-particle operator and  
the residual part is evaluated by solving the Brueckner-Bethe-Goldstone (BBG) equation. The accuracy of the results 
obtained with BBG depends on the expansion level adopted; the perturbative expansion of BBG is not convergent, but 
the cluster diagrams can be grouped with the number of independent hole lines\cite{pandharipande79}.
Several attempt to study the convergence of hole-line expansion were recently explored\cite{song98,gad05}.

The Correlated Basis Functions (CBF) theory is particularly well suited to deal with strongly interacting systems, that 
are described with a many-body wave function written with some correlation operators acting on the 
non-interacting state. Unfortunately the energy evaluation of the CBF state is not easy to do. The Variational 
Monte Carlo (VMC) could be used to numerically solve the integration of CBF state, but the complexity of 
wave function make impractical the applicability of VMC when the number of nucleons increases.
Another possibility to calculate the CBF energy is given by the FHNC/SOC theory\cite{pandharipande79}. 
Unfortunately the solution of 
FHNC/SOC equations provides the summation of 
an infinite class of cluster diagrams like that called "elementary diagrams".
Hence the CBF is a variational theory that can be solved with the FHNC/SOC that however is not exact because 
the summation is not complete and some elementary diagrams are not fully considered.

\begin{figure}[ht]
\vspace{1.5cm}
\begin{center}
\includegraphics[width=11cm]{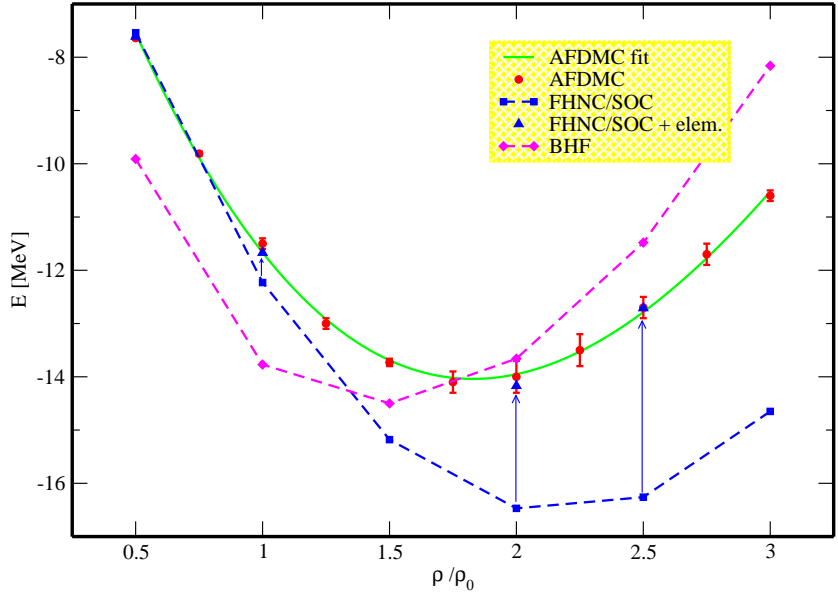}
\vspace{0.5cm}
\caption{Equation of state of symmetric nuclear matter calculated with different methods. Red circles 
represent AFDMC results with statistical error bars and the green line is the fitted functional.
Dashed lines correspond to calculations performed with other methods\cite{bombaci05}
(blue line with squares: FHNC/SOC; magenta with diamonds: BHF). Blue triangles represent 
the FHNC/SOC energies corrected by including the low order of elementary diagrams.
Blue arrows show the corresponding energy shift, which increases at higher densities.
Figure taken from Ref. \cite{gandolfi07}.}
\label{fig:EOS-SNM}
\end{center}
\end{figure}

The AFDMC EOS is shown in Fig. \ref{fig:EOS-SNM} where different EOS obtained with FHNC/SOC and BHF are also 
reported\cite{bombaci05}. 

By assuming that FP-AFDMC for SNM calculation is as accurate as shown for nuclei, and that finite size-effects are 
small as suggested by table \ref{tab:nucmatsize}, the comparison of the various EOS suggests 
the following comments: FHNC/SOC leads to an overbinding at high density.
A similar indication was found by Moroni et al.\cite{moroni95} after a DMC calculation 
of the EOS of normal liquid $^3$He
at zero temperature, with a guiding function including triplet and backflow correlations.
The comparison with the equivalent FHNC/SOC calculations of Refs. \cite{manousakis83,viviani88} 
have shown similar discrepancies.

On the other hand
there are two main intrinsic approximations in variational FHNC/SOC calculations, which violate
the variational principle. The first one consists in neglecting a whole class of 
cluster diagrams, the elementary diagrams, which are not summed up in the
FHNC integral equations. The calculation of the lowest order diagram of this class
shows a substantial effect from this diagram represented by the blue arrows in Fig. \ref{fig:EOS-SNM}, and
bring the FHNC/SOC estimates very close to AFDMC results.
An accurate study of corrections given by higher diagrams was performed by Wiringa\cite{wiringa80} showing that their effects
are not negligible.
The second approximation is related to the non-commutativity of the correlation operators
entering the variational wave function. The only class of cluster diagrams contributing
to such non-commuting terms, which can be realistically calculated, is that characterized by single
operator chains (SOC approximation). It is believed that such an approximation is reliable in
nuclear matter, but there is no clear proof of this.

BHF calculations of Ref. \cite{bombaci05} includes corrections up to the two hole-line expansion and 
predict an EOS with a shallower binding than the AFDMC one. It has been shown for symmetric nuclear matter, 
using the Argonne AV18 and AV14 potentials, that contributions from three hole--line 
diagrams add a repulsive contribution up to $\sim$ 3MeV 
at densities below $\rho_0$\cite{song98}, and decrease the energy at high 
densities\cite{baldo01}. Such corrections, if computed with the $v_6'$ based on AV8' 
potential, would probably preserve the same general behavior, and bring the BHF EOS 
closer to the AFDMC one. Therefore, the AFDMC
calculations show that the two hole--line approximation used in Ref. \cite{bombaci05}
is too poor, particularly at high density.

Although the EOS calculated with the AFDMC started from a simplified Hamiltonian, it is particularly important 
because it emphasizes some limitations of the FHNC/SOC calculation that was used to fit the TNI 
on nuclear matter. 
At the present the TNI forces are fitted using the GFMC to simulate light-nuclei, in the addition to 
the FHNC/SOC calculation to reproduce the equilibrium density of nuclear matter; differences 
shown between AFDMC and FHNC/SOC open some questions about the physical meaning of such potentials in the bulk 
matter calculations, and stress the fact that more work should be addressed to study the role of TNI, 
especially at higher densities.

The computed EOS with AFDMC is fitted with a functional form as following:
\begin{equation}
\label{eq:nucmatfit}
\frac{E}{A}=\frac{E_0}{A}+\alpha(\tilde\rho-\tilde\rho_0)^2+\beta(\tilde\rho-\tilde\rho_0)^3 \,,
\end{equation}
where $\tilde\rho$ is the nucleon density respect to the equilibrium density
\begin{equation}
\tilde\rho=\frac{\rho}{\rho_0} \,,
\end{equation}
and the $E_0$, $\alpha$, $\beta$ and $\tilde\rho_0$ coefficients are given by the fit:
\begin{eqnarray}
\frac{E_0}{A}&=&-14.04(4)MeV \,,\nonumber \\
\alpha&=&3.09(6)MeV \,,\nonumber \\
\beta&=&-0.44(8)MeV \,,\nonumber \\
\tilde\rho_0&=&1.83(1) \,.
\end{eqnarray}
The nuclear matter compressibility K is given by
\begin{equation}
K=9\tilde\rho_0^2\left[\frac{\partial(E/A)}{\partial\tilde\rho^2}\right]_{\tilde\rho_0} \,.
\end{equation}

The value obtained with the functional form of Eq. \ref{eq:nucmatfit} is K$\sim$190 MeV.
The fitted EOS obtained with the AFDMC could be used to calibrate other many-body theories.

The FP-AFDMC is an accurate method to calculate the EOS of SNM, but it could be easily generalized to microscopically 
simulate the asymmetric nuclear matter particular important to study neutron stars. At the present the 
asymmetric nuclear matter can be studied using mean field calculations\cite{frick05,vandalen07}.

\newpage
\subsection{Equation of state of symmetric nuclear matter with the Argonne AV6' interaction}
The AFDMC with the fixed-phase approximation was used to calculate the equation of state of nuclear matter 
using the realistic Argonne AV6' NN interaction.

The Argonne AV6' is obtained by dropping the spin-orbit operators from the AV8'. 
The spin-orbit does not contribute to the $S$ wave in the NN scattering as well as the $P$ wave in the spin-singlet 
$^1P_1$ channel, but it is important in differentiating between $^3P_{0,1,2}$ channels.
Then the AV6' was obtained to preserve the deuteron energy by including the energy difference from the AV8' in the 
central part in the $S,T$=1,0 channel.

\begin{table}[ht]
\vspace{0.4cm}
\begin{center}
\begin{tabular}{||cc|cc||}
\hline
$\rho$ [fm$^{-3}$] & $\rho/\rho_0$ & E/A($v_6'$) & E/A(AV6') \\
\hline
\hline
0.04   & 0.25 &           &  -5.01(4)  \\
0.08   & 0.5  &  -7.64(3) &  -8.14(4)  \\
0.12   & 0.75 &  -9.81(4) & -10.86(3)  \\
0.16   & 1.0  & -11.5(1)  & -12.77(7)  \\
0.20   & 1.25 & -13.0(1)  & -14.67(4)  \\
0.24   & 1.5  & -13.73(7) & -16.0(2)   \\
0.28   & 1.75 & -14.1(2)  & -16.4(1)   \\
0.30   & 1.875&           & -16.6(1)   \\
0.32   & 2.0  & -14.0(3)  &            \\
0.36   & 2.25 & -13.5(3)  & -16.9(2)   \\
0.40   & 2.5  & -12.7(2)  & -16.7(1)   \\
0.44   & 2.75 & -11.7(2)  & -15.9(1)   \\
0.48   & 3.0  & -10.6(1)  & -14.3(2)   \\
0.52   & 3.25 &           & -13.0(2)   \\
\hline
\end{tabular}
\end{center}
%\vspace{0.2cm}
\caption{AFDMC energies per particle in MeV of 28 nucleons in a periodic box at various densities.
The reported ground state energy were obtained by considering the Argonne AV6' interaction and the 
AV8' without the spin-orbit contribution (called $v_6'$).}
\label{tab:SNM_AV6}
\end{table}

\begin{figure}[ht]
\vspace{1.0cm}
\begin{center}
\includegraphics[width=11cm]{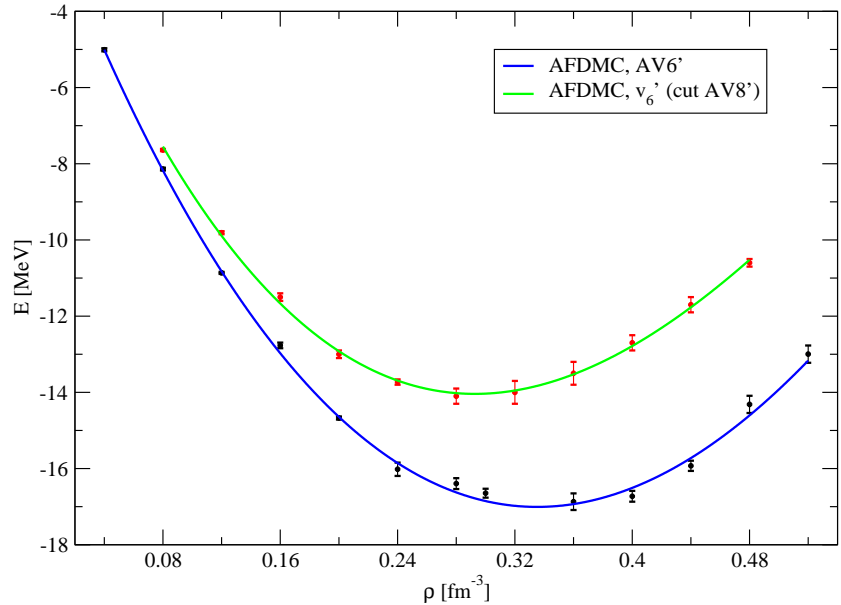}
\vspace{0.5cm}
\caption{The AFDMC equation of state of symmetric nuclear matter. The energies were evaluated by simulating
28 nucleons in a periodic box interacting with different Hamiltonians.
The red points (taken from Ref. \cite{gandolfi07}) were obtained with the Argonne AV8' NN
interaction without the spin-orbit contribution (called $v_6'$), while the black points 
were calculated with the Argonne AV6' interaction. The lines correspond to the fitted functionals.}
\label{fig:SNM_AV6}
\end{center}
\end{figure}

The AFDMC was employed to simulate 28 nucleons interacting with the AV6' as described in the previous section.
The tail corrections to the potential were also included. The energy of 28 nucleons in a periodic box are reported 
in table \ref{tab:SNM_AV6} were they are also included the energies of Ref. \cite{gandolfi07} obtained with the 
$v_6'$ interaction.

It is interesting to note how the reprojection of the AV8' to the AV6' by dropping the spin-orbit significantly 
changes the EOS of the symmetric nuclear matter as it can also be seen in Fig. \ref{fig:SNM_AV6}. 
The spin-orbit term of AV8' is very small in the alpha-particle, and its
contribution is -4.75 MeV with respect to the total potential energy of -128.25 MeV\cite{kamada01}.
The alpha-particle total energy is -25.14 MeV with the AV8' and -26.15 MeV with the AV6'\cite{wiringa02}, and the 
energy with the AV8' without the spin-orbit is -21.18 MeV\cite{kamada01}. Then the difference between the $v_6'$ and the AV6'
is of about 1.2 MeV per nucleon in the alpha-particle ground state.
In the nuclear matter at densities $\rho=$0.08 fm$^{-3}$, 0.12 fm$^{-3}$ and 0.16 fm$^{-3}$ the difference 
between the $v_6'$ and the AV6' energies is respectively of 0.5 MeV, 1.05 MeV and 1.27 MeV, not far from that 
in the alpha-particle. 
When the density increases the spin-orbit contribution becomes larger, and also the saturation density 
increases from 0.29 fm$^{-3}$ to 0.34 fm$^{-3}$. 
However this effect could also be due to the different strength of the central term in the $S,T$=10 channel 
between the two interactions.
For densities higher than $\sim$0.36 fm$^{-3}$ the behavior of the two EOS seems to be similar with a constant shift.

The calculated EOS with AV6' is fitted with the same functional form of Eq. \ref{eq:nucmatfit} as in the previous section:
and in this case the $E_0$, $\alpha$, $\beta$ and $\tilde\rho_0$ coefficients are:
\begin{eqnarray}
\frac{E_0}{A}&=&-17.0(1)MeV \,, \nonumber \\
\alpha&=&3.1(1)MeV \,, \nonumber \\
\beta&=&-0.22(7)MeV \,, \nonumber \\
\tilde\rho_0&=&2.09(2) \,.
\end{eqnarray}
The nuclear matter compressibility for the AV6' results to be K$\sim$244 MeV with the functional form of Eq. \ref{eq:nucmatfit}.

\subsection{Equation of state of isospin-asymmetric nuclear matter with the Argonne AV6' interaction}
In order to test the reliability of AFDMC to deal with the asymmetric nuclear matter calculation, several simulations
with $N\ne Z$ were performed. 

In according to the mass formula of Eq. \ref{eq:massformula}, the asymmetry parameter $\alpha$ can be 
defined as
\begin{equation}
\alpha=\frac{N-Z}{N+Z} \,,
\end{equation}
and, in order to have a closed-shell configuration in the trial wave function, only particularly $(N,Z)$ pairs are admitted.
The trial wave function has the same structure of that used for symmetric nuclear matter, and both 
$N$ and $Z$ must be numbers corresponding to a closed shell of plane waves.

The finite-size effects for the potential were included as described in the symmetric nuclear matter section,
and for density $\rho$=0.08 fm$^{-3}$, 0.16 fm$^{-3}$ and 0.24 fm$^{-3}$ the energy dependence to the $\alpha$ parameter
were studied.

\begin{figure}[!h]
\vspace{1.2cm}
\begin{center}
\includegraphics[width=11cm]{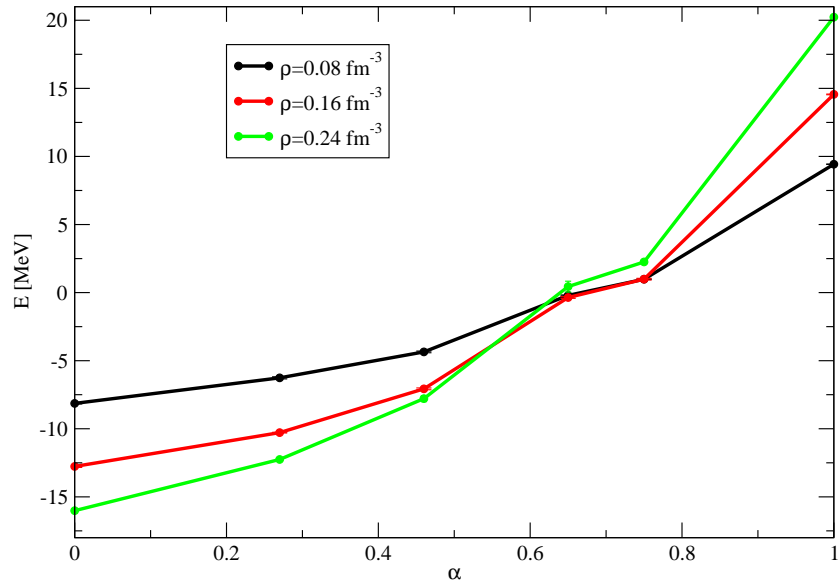}
\vspace{0.5cm}
\caption{The AFDMC energy of asymmetric nuclear matter as a function of the asymmetry parameter $\alpha$ at
different densities as indicated. The considered Hamiltonian is the AV6', and the estimate of the energies were obtained 
by simulating different set of $(N,Z)$ in a periodic box.}
\label{fig:asym}
\end{center}
\end{figure}

The AFDMC energies of asymmetric nuclear matter for the considered densities are reported in table \ref{tab:asym} 
and plotted in Fig. \ref{fig:asym}.
As it can be seen the energy dependence to the $\alpha$ parameter becomes stronger when the density increases. 
In particular for $\alpha>$ 0.5 there is a strange trend of the energy, probably due to the finite-size effect in the 
kinetic energy. The potential is corrected to avoid these effects, but the kinetic energy could still suffer of 
the fact that different shells are considered. 
This trend is less evident for small values of $\alpha$ probably because in the region of $\alpha<$ 0.5 the 
energy dependence to the asymmetry parameter is weaker.

\begin{table}[!ht]
\vspace{0.2cm}
\begin{center}
\begin{tabular}{||ccc|ccc||}
\hline
N  & P  & $\alpha$ & E/A($\rho$=0.08fm$^{-3}$) & E/A($\rho$=0.16fm$^{-3}$) & E/A($\rho$=0.24fm$^{-3}$) \\
\hline
\hline
14 & 14 & 0.0      & -8.14(4) & -12.77(7)  & -16.0(2)  \\
66 & 38 & 0.27     & -6.26(8) & -10.29(3)  & -12.3(1)  \\
38 & 14 & 0.46     & -4.35(6) &  -7.07(7)  &  -7.8(2)  \\
66 & 14 & 0.65     & -0.21(4) &  -0.36(6)  &   0.4(4)  \\
14 & 2  & 0.75     &  0.98(2) &   0.99(6)  &   2.25(7) \\
38 & 2  & 0.9      &          &   8.36(2)  &           \\
66 & 2  & 0.94     &          &  12.22(2)  &           \\
14 & 0  & 1.0      &  9.43(1) &  14.56(2)  &  20.23(3) \\
\hline
\end{tabular}
\end{center}
%\vspace{0.2cm}
\caption{AFDMC energies per particle in MeV of asymmetric nuclear matter in a periodic box at various densities.
The number of neutrons and protons ($N$ and $Z$) and the corresponding asymmetry parameter $\alpha$ is 
indicated. The Hamiltonian contains the Argonne AV6' NN interaction.}
\label{tab:asym}
\end{table}

\begin{figure}[!ht]
\vspace{1.5cm}
\begin{center}
\includegraphics[width=8cm]{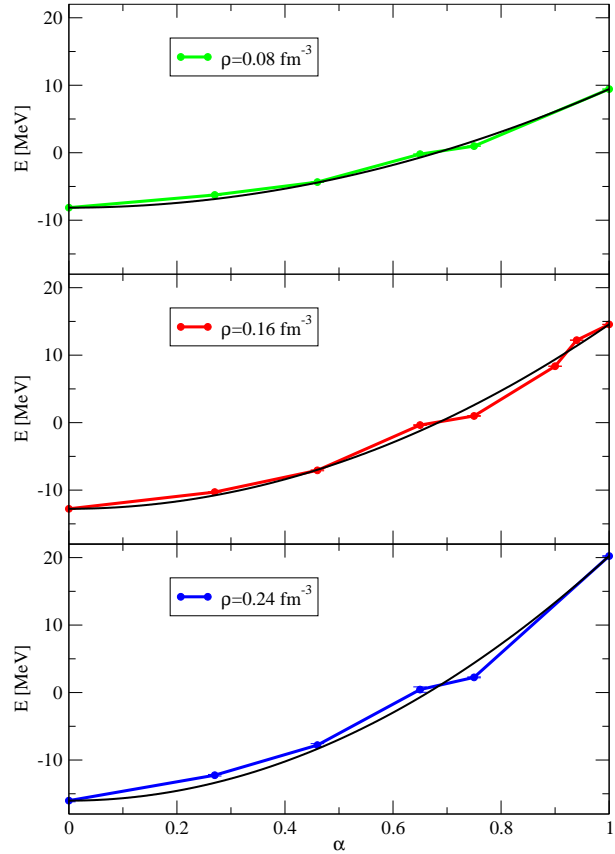}
\vspace{0.5cm}
\caption{The AFDMC energy of asymmetric nuclear matter at different densities as a function of the asymmetry parameter.}
\label{fig:asymall}
\end{center}
\end{figure}

The energy per nucleon as a function of $\alpha$ is plotted if Fig. \ref{fig:asymall}, where for each density 
is also displayed the quadratic functional of E($\alpha$) fitted on the SNM ($\alpha=0$) and PNM ($\alpha=1$) points.
The fit at the equilibrium density $\rho$=0.16 fm$^{-3}$ represent the coefficient of the asymmetry energy in the 
Weizsacker formula of Eq. \ref{eq:massformula}. 

The energy dependence to the asymmetry parameter in the infinite limit is
\begin{equation}
\frac{E(A,Z)}{A}=-a_v+a_{sym}\frac{(A-2Z)^2}{A^2}
\end{equation}
and by taking $-a_v$=E/N($\alpha$=0)=--12.77 MeV, the fit gives $a_{sym}$=27.33 MeV while the experimental value is
23.2875 MeV\cite{povh94}. 

The effect of the tensor term is essential in the asymmetry parameter calculation. In fact a previously AFDMC 
using a $v_4$-like interaction predicted $a_{sym}$=36.4 MeV\cite{fantoni01b} that is sensibly larger with 
respect to both the experimental fit and to the AFDMC calculation with the AV6' interaction.

However, an accurate calculation of the asymmetric nuclear matter equation of state requires a different treatment of the 
finite-size effects due to the kinetic energy.
One possible way to make shell effects milder is the use of the twisted averaged boundary conditions\cite{lin01}.

\section{Neutron matter in the normal phase}
\label{sec:PNM}
An approximation of matter as it is found in neutron stars is a system made of neutrons only, called pure neutron matter (PNM).
The AFDMC algorithm combined with the constrained-path approximation was already used by Sarsa et al. to study 
the equation of state of PNM at zero temperature\cite{sarsa03}. The Hamiltonian considered contains both a realistic 
AV8' NN and the Urbana IX TNI interactions; at present such Hamiltonian is the most commonly used to calculate properties 
of both SNM and PNM.

The CP-AFDMC proved to give a very satisfactory result for PNM calculations with a NN and TNI interaction, but some 
problem were encountered in the evaluation of the spin-orbit contribution as just discussed in the neutron drop 
section \ref{sec:neutdrop}. The inclusion of spin-backflow correlations reduced the discrepancies\cite{brualla03}, 
but a detailed study just considering a pure NN interaction emphasized this problem of 
AFDMC in dealing with the spin-orbit correlations\cite{baldo04}.
A similar behavior was found by comparing the AFDMC\cite{sarsa03} with the GFMC evaluation of the energy of 14 neutrons in 
a periodic box\cite{carlson03,carlson03b}. By considering the same Hamiltonian with the same box truncation used 
in GFMC calculation, the AFDMC overestimates the energy of 14 neutrons with a AV8' interaction. 

The FP-AFDMC overcomes the problem of the spin-orbit correlations as shown in 
table \ref{tab:PNMbench}. Without tail corrections the CP-AFDMC energy of 14 neutrons at $\rho=$0.16 fm$^{-3}$ 
is 20.32(6) MeV compared to 17.00(27) MeV given by UC-GFMC, while the FP-AFDMC energy is 17.586(6) MeV, in 
better agreement (within 3\%) with the UC-GFMC. For higher densities reported in the table it should be noted 
that the CP-GFMC significantly differs from UC-GFMC, probably because the Fermion sign problem in that case is more 
severe\cite{carlson03}, and the convergence of the energy can be very problematic, because the evolution
of the energy can be carried out only for a very short imaginary-time. These shortcomings could introduce some 
spurious effects limiting the accuracy of GFMC for the neutron matter calculation\cite{carlson06}.
Anyway this comparison in addition to the SOS calculation in neutron drops demonstrates the high accuracy of
FP-AFDMC to deal with neutron systems interacting with realistic NN Hamiltonians.

\begin{table}[!ht]
\vspace{0.2cm}
\begin{center}
\begin{tabular}{||c|cc|cc||}
\hline
$\rho$ [fm$^{-3}$] & FP-AFDMC  & CP-AFDMC & CP-GFMC   & UC-GFMC \\
\hline
\hline
0.04   & 6.69(2)   &          &  6.43(01) &  6.32(03) \\
0.08   & 10.050(8) &          & 10.02(02) & 9.591(06) \\
0.16   & 17.586(6) & 20.32(6) & 18.54(04) & 17.00(27) \\
0.24   & 26.650(9) &          & 30.04(04) & 28.35(50) \\
\hline
\end{tabular}
\end{center}
\caption{FP-AFDMC energies per particle of 14 neutrons interacting with the AV8' NN interaction 
in a periodic box at various densities. The CP-AFDMC of Ref. \cite{sarsa03}, the constrained path CP-GFMC 
and the unconstrained UC-GFMC of Ref. \cite{carlson03} are also reported for a comparison.
All the energies are expressed in MeV.}
\label{tab:PNMbench}
\end{table}

The FP-AFDMC was applied to study the PNM by simulating periodic boxes containing a different number of neutrons interacting with the 
Argonne AV8' NN potential, including the finite-size corrections to the potential energy as described in the 
above section of symmetric nuclear matter. All FP-AFDMC results, as well as those of Ref. \cite{sarsa03} with 
CP-AFDMC are reported in table \ref{tab:neutmatAV8}. Where available, the CP-AFDMC 
calculation using the same Hamiltonian and the same corrections are included for comparison.

\begin{table}[!ht]
\vspace{0.2cm}
\begin{center}
\begin{footnotesize}
\begin{tabular}{||c|ccc|c||}
\hline
$\rho$ [fm$^{-3}$] & FP-AFDMC(14)  & FP-AFDMC(38) & FP-AFDMC(66) & CP-AFDMC(14) \\
\hline
\hline
0.12   & 11.843(9) & 10.69(2) &  11.99(3) &  12.32(5) \\
0.16   & 14.566(7) & 13.23(3) &  15.02(7) &  14.98(7) \\
0.20   & 16.76(1)  & 15.13(3) &  17.11(4) &  17.65(7) \\
0.24   & 20.05(2)  & 17.66(3) &  19.87(4) &           \\
0.28   & 23.142(8) & 20.76(2) &  23.36(4) &           \\
0.32   & 26.038(8) & 23.88(3) &  26.74(4) &  27.3(1)  \\
0.36   & 30.28(2)  & 27.35(2) &  30.20(5) &           \\
0.40   & 33.344(7) & 30.87(6) &  34.32(4) &  35.3(1)  \\
0.48   &           &          &  43.14(5) &           \\
0.56   &           &          &  53.34(6) &           \\
0.64   &           &          &  64.90(6) &           \\
0.80   &           &          &  91.28(3) &           \\
\hline
\end{tabular}
\end{footnotesize}
\end{center}
\caption{FP-AFDMC energies per particle of 14, 38 and 66 neutrons interacting with the AV8' NN interaction 
in a periodic box at various densities. The finite-size effects due to the NN truncation are included. 
The CP-AFDMC energy of 14 neutrons with the same interaction and finite-size corrections\cite{sarsa03} 
are also reported for a comparison.
All the energies are expressed in MeV.}
\label{tab:neutmatAV8}
\end{table}

The first consideration concerns the inclusion of finite-size effects that sensibly reduce the gap between the FP-AFDMC 
and CP-AFDMC with 14 neutrons. In fact without this corrections for AV8' at $\rho=$0.16 fm$^{-3}$ the difference 
is less than 14\%, while in this case is less than 3\%. For higher densities the discrepancy never exceed the 5\% 
respect to the total energy.

Some finite-size effects are still present, as it can be seen from the energies 
obtained by varying the number of neutrons. The same behavior is consistently followed at each density, and typically 
$E(38)<E(14)<E(66)$, although small deviations are present, and in some case $E(14)>E(66)$. This trend
directly follows from behavior of the kinetic energy of the Fermi gas.
As discussed above, finite-size effects due to the kinetic energy could however be included using 
the twisted averaged boundary conditions\cite{lin01}.
It should also be noted that the finite-size contributions due to the spin-orbit potential are neglected, and the 
spin-orbit potential is obtained by a summation of pairs inside the simulation box. 
The effect of this approximation should be small, but remains to be studied.

For PNM the Urbana UIX TNI force reduces to a NN spin-independent interaction as explained in Sec. \ref{sec:TNIprop}.
Moreover it can be easily included into the propagator. The finite-size corrections due to the UIX 
are not included because they increase the computational time and rather cumbersome. 
The FP-AFDMC results compared to available CP-AFDMC of Ref. \cite{sarsa03} are reported in 
table \ref{tab:AFDMC-PNM}; results were obtained by simulating different systems with 14, 38 and 66 neutrons.
For 14 and 66 neutrons with the same Hamiltonian the CP-AFDMC using spin-backflow correlations are also reported.
The effect of using the fixed-phase approximation rather then the constrained-path to the spin-orbit evaluation 
are once more evident, also comparing the FP-AFDMC with the available CP-AFDMC using spin-backflow correlations.

\begin{table}[!t]
\vspace{0.4cm}
\begin{center}
\begin{small}
\begin{tabular}{||c|c|c|c||}
\hline
$\rho$ [fm$^{-3}$] &FP-AFDMC(14)&CP-AFDMC(14)&JSB-CP-AFDMC(14) \\
\hline
\hline
0.12   &14.12(1)    &14.80(9) & \\
0.16   &18.76(1)    &19.76(6) & \\
0.20   &23.57(2)    &25.23(8) & \\
0.32   &45.51(1)    &48.4(1)  & 46.8(1) \\ 
0.40   &65.86(2)    &70.3(2)  & \\
\hline
\hline
$\rho$ &FP-AFDMC(38)&CP-AFDMC(38)& \\
\hline
\hline
0.12   &13.09(1)    &13.96(5) & \\
0.16   &17.84(2)    &18.67(6) & \\
0.20   &22.76(4)    &24.7(1)  & \\
0.32   &46.84(3)    &46.8(2)  & \\
0.40   &70.07(5)    &76.3(2)  & \\
\hline
\hline
$\rho$ &FP-AFDMC(66)&CP-AFDMC(66)&JSB-CP-AFDMC(66) \\
\hline
\hline
 0.12  & 14.36(1)   &15.26(5) & \\  
 0.16  & 19.64(1)   &20.23(9) & \\
 0.20  & 24.71(3)   &27.1(1)  & \\
 0.32  & 49.94(3)   &54.4(6)  & 52.9(2)\\
 0.40  & 74.2(1)    &81.4(3)  & \\
\hline
\end{tabular}
\end{small}
\end{center}
\vspace{0.2cm}
\caption{FP-AFDMC energies per particle of 14, 38 and 66 neutrons interacting with the AV8'+UIX interaction 
in a periodic box at various densities compared with the available CP-AFDMC ones of Ref. \cite{sarsa03}.
The available CP-AFDMC using spin-backflow correlations taken from Ref. \cite{brualla03} are also 
reported.
All the energies are expressed in MeV.}
\label{tab:AFDMC-PNM}
\end{table}

\begin{figure}[!ht]
\vspace{0.8cm}
\begin{center}
\includegraphics[width=10cm]{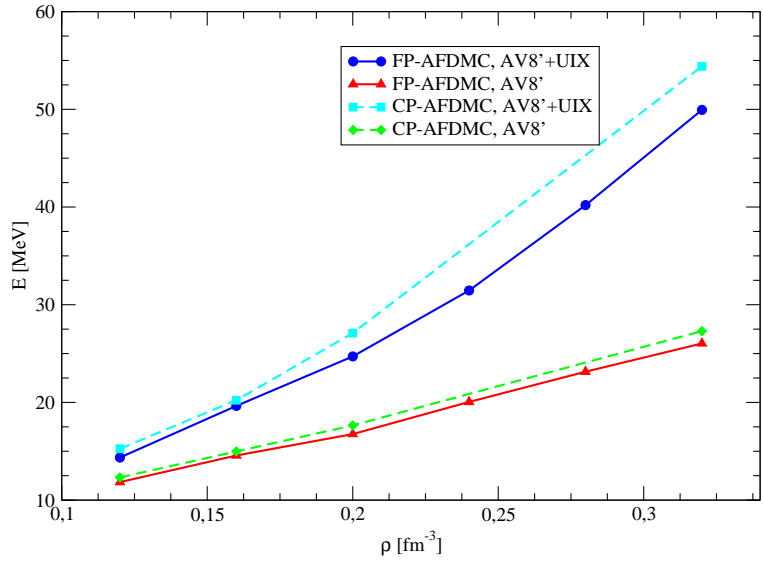}
%\vspace{0.5cm}
\caption{The FP-AFDMC and CP-AFDMC\cite{sarsa03} equation of state for AV8' and AV8'+UIX. See the legend for details.
The EOS of CP-AFDMC\cite{sarsa03} and FP-AFDMC with AV8' were calculated by simulating 14 neutrons in a periodic box, while 
in the case of AV8'+UIX 66 neutrons were used.}
\label{fig:PNMold}
\end{center}
\end{figure}

To better show the differences between the FP-AFDMC and CP-AFDMC some results are plotted in 
Fig. \ref{fig:PNMold}

All the FP-AFDMC results for 14, 38, 66 and 114 neutrons interacting with the AV8'+UIX Hamiltonian are 
summarized in table \ref{tab:allPNM}, where it were also reported the extrapolated CP-AFDMC energies to 
the infinite system. 

\begin{table}[!ht]
\begin{center}
\begin{footnotesize}
\begin{tabular}{||c|cccc|c||}
\hline
$\rho$[fm$^{-3}$] &FP-AFDMC &FP-AFDMC &FP-AFDMC &FP-AFDMC  &CP-AFDMC   \\
                  &     (14)&     (38)&     (66)&     (114)&($\infty$) \\
\hline
\hline
0.12   &14.12(1) &13.09(1) &14.36(1) &15.13(1)  &15.5 \\
0.16   &18.76(1) &17.84(2) &19.64(1) &20.73(2)  &20.6 \\
0.20   &23.57(2) &22.76(4) &24.71(3) &25.75(6)  &27.6 \\
0.24   &30.51(1) &29.24(3) &31.46(3) &32.85(5)  &     \\
0.28   &37.91(1) &37.46(1) &40.19(3) &41.83(4)  &     \\
0.32   &45.51(1) &46.84(3) &49.94(3) &51.83(1)  &55.6 \\
0.36   &56.336(9)&57.84(2) &63.5(1)  &63.5(2)   &     \\
0.40   &65.86(2) &70.07(5) &74.2(1)  &76.74(6)  &83.5 \\
0.48   &         &         &105.9(1) &          &     \\
0.56   &         &         &145.3(2) &          &     \\
0.64   &         &         &193.5(1) &          &     \\
0.80   &         &         &313.7(2) &          &     \\
\hline
\end{tabular}
\end{footnotesize}
\end{center}
\caption{FP-AFDMC energies per particle of 14, 38, 66 and 114 neutrons interacting with the AV8'+UIX interaction 
in a periodic box at various densities. 
The extrapolated CP-AFDMC energy for $\infty$-neutrons of Ref. \cite{sarsa03} are also reported.
All the energies are expressed in MeV.}
\label{tab:allPNM}
\end{table}

The obtained results suggest some consideration: finite-size effects are important, 
especially using 14 or 38 neutrons. However,
in the range of considered densities the energy of 66 neutrons is always very close to that of 114 within few 
percent, indicating a possible saturation near the infinite limit. This behavior was observed after the study 
of finite-size effects using the periodic box FHNC (PB-FHNC) technique\cite{fantoni01c}, and applied to 
extrapolate to the infinite limit the CP-AFDMC results reported in table \ref{tab:allPNM}.
The PB-FHNC might also be used to correct the truncation of inter-particle distances entering in the 
spin-orbit and TNI potential; anyway the deep analysis of Sarsa et al.\cite{sarsa03} suggests that 
the energy of the system in the infinite limit is somewhere between the energies of 66 and 114 neutrons.

\begin{figure}[!ht]
\vspace{1.0cm}
\begin{center}
\includegraphics[width=10cm]{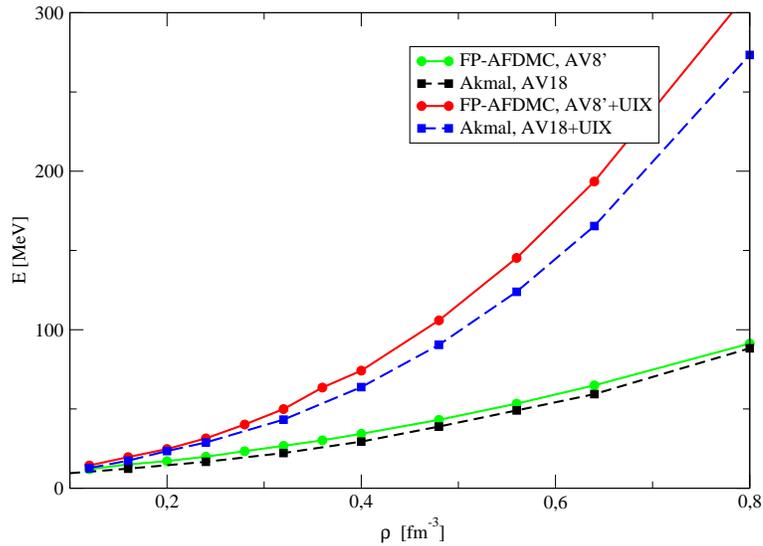}
%\vspace{0.5cm}
\caption{The FP-AFDMC equation of state evaluated by simulating 66 neutrons in a periodic box; the 
AV8' and AV8'+UIX Hamiltonians were considered. The EOS were compared with the variational 
calculations of Ref. \cite{akmal98} using the AV18 and AV18+UIX Hamiltonians.
See the legend for details.}
\label{fig:PNMeos}
\end{center}
\end{figure}

In Fig. \ref{fig:PNMeos} we plot
the FP-AFDMC equation of state obtained as the energy of 66 neutrons, and the variational 
calculation of Akmal et al. of Ref. \cite{akmal98},where the AV18 NN interaction was used 
in combination with the Urbana UIX TNI. 
As it can be seen both AV8' and AV18 essentially give an EOS with the same behavior, but the addition 
of the TNI adds some differences, in particular at higher densities. 

\subsection{Neutron star structure}
The PNM results reported in table \ref{tab:allPNM} were used to predict some property of a neutron 
star. The structure of a non-rotating neutron star can be fully determined by solving the 
Tolman-Oppenheimer-Volkoff (TOV) equations\cite{heiselberg00,lattimer01,lattimer04}:
\begin{equation}
\frac{dP}{dr}=-\frac{G[m(r)+4\pi r^3P(r)/c^2][\epsilon(r)+P(r)/c^2]}{r[r-2Gm(r)/c^2]} \,,
\end{equation}
\begin{equation}
\frac{dm(r)}{dr}=4\pi\epsilon(r)r^2 \,,
\end{equation}
where $P(r)$ and $\epsilon(r)$ are the pressure and the energy density of the matter, $m(r)$ is the gravitational 
mass enclosed within a radius $r$, and $G$ is the Gravitational constant. 
To solve the TOV, the EOS of baryonic matter is needed, whose main bulk is given by the neutron matter physics.

It is commonly accepted that in the lower density regime where $\rho\approx\rho_0$ the matter inside a neutron 
star is dominated by nucleons and leptons, but when density increases other hyperonic species may play an 
important and not negligible role in the physics of the neutron star.
The inclusion of hyperons in the EOS is far to be accessible because the of lack of knowledge about the nucleon-hyperon 
and hyperon-hyperon interactions; only few experimental data were extracted from nucleon-hyperon scattering 
and some potential were proposed\cite{maessen89}.

As a first approximation, the PNM equation of state calculated with 66 neutrons with the FP-AFDMC was employed to solve the TOV 
equations.
For this purpose the results of table \ref{tab:allPNM} were fitted to the functional form 
\begin{equation}
E(\rho)=\alpha\rho^a+\beta\rho^b \,,
\end{equation}
where $E(\rho)$ is the energy per neutron as a function of the density, and $\alpha$, $a$, $\beta$ and $b$ are 
parameters given by the fit.

Using the fitted energy density $\epsilon(\rho)$ defined as
\begin{equation}
\epsilon(\rho)=\rho E(\rho)
\end{equation}
and the pressure $P(\rho)$ at zero temperature given by
\begin{equation}
P(\rho)=\rho^2\frac{\partial E(\rho)}{\partial\rho}
\end{equation}
it is possible to solve the TOV equation using as the initial condition the $\rho_c$ parameter that represents
the density at the center of the star. The distance from the center at which $P=0$ is the radius $R$ of the star.
From the value of the radius it is possible to calculate the gravitational mass $M$ defined as
\begin{equation}
M(R)=\int_0^R dr 4\pi\epsilon(r)r^2 \,.
\end{equation}

\begin{figure}[!ht]
\vspace{1.5cm}
\begin{center}
\includegraphics[width=11cm]{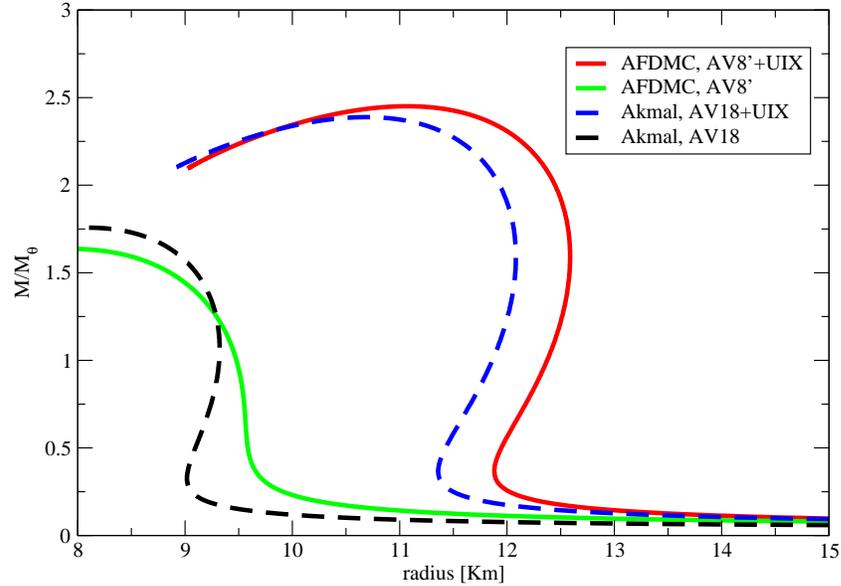}
\vspace{0.5cm}
\caption{Neutron star gravitational mass as a function of the radius given by the AFDMC and 
VCS taken from Ref. \cite{akmal98}. 
The considered Hamiltonians are the AV8' and AV8'+UIX for AFDMC, and AV18 and AV18+UIX for VCS.}
\label{fig:mass-r}
\end{center}
\end{figure}

\begin{figure}[!ht]
\vspace{1.5cm}
\begin{center}
\includegraphics[width=11cm]{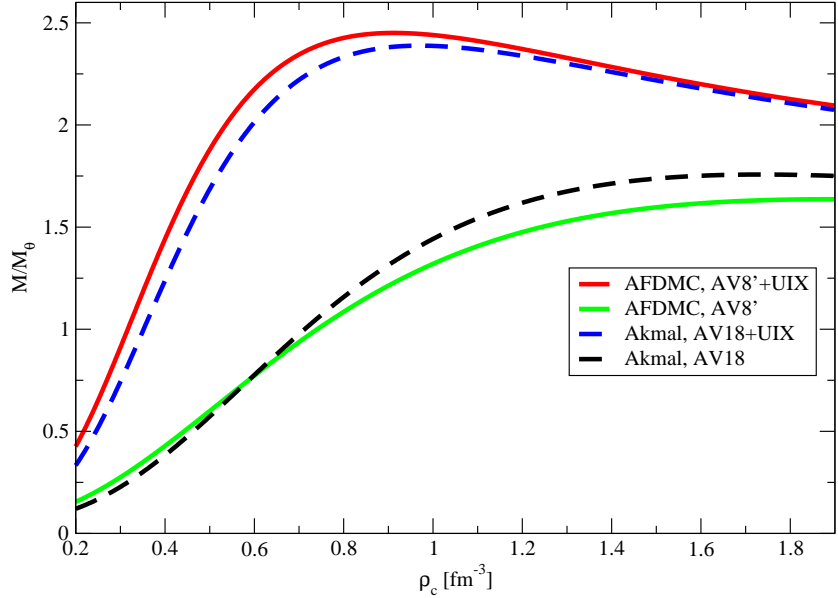}
\vspace{0.5cm}
\caption{Neutron star gravitational mass as a function of the central density $\rho_c$ calculated 
with the AFDMC and VCS taken from Ref. \cite{akmal98}.
The considered Hamiltonians are the same of Fig. \ref{fig:mass-r}.}
\label{fig:mass-rho}
\end{center}
\end{figure}

The gravitational mass in unit of solar masses M$_{\bigodot}$ as a function of $\rho_c$ or $R$ can be directly 
compared to the experimental observations of neutron stars\cite{lattimer04}. The TOV solutions are plotted in Fig. \ref{fig:mass-r} and 
\ref{fig:mass-rho}, together with the TOV solutions given by the variational chain summation (VCS) 
of Akmal et al. from Ref. \cite{akmal98}, where the full AV18 as NN interaction instead of the AV8' was considered. 

In the M-R diagram it is possible to understand the dependence of the gravitational mass by the 
radius of the star. The predicted maximum mass and radius are not too far from observations despite the 
great simplifications introduced, and the absence of important physics coming from the inclusion of hyperons and 
quark matter. The global behaviour is similar for both the AFDMC and VCS, especially in the case of the full 
Hamiltonian; in particular, the AFDMC predicts a maximum mass and radius slightly higher with respect to the VCS with 
the addition of TNI, while the maximum mass is lower for pure NN Hamiltonian.

In the M-$\rho_c$ diagram the dependence to the central density $\rho_c$ of the gravitational mass of the 
star is displayed. $\rho_c$ is a parameter used to solve the TOV equations but can be determined by
measurements of the maximum mass. Also in this case the AFDMC and VCS results are slightly different, especially 
in the case of pure NN interaction.

The predicted maximum gravitational masses, with the corresponding central density and radius obtained by 
the TOV solution are shown in table \ref{tab:starsummary}. The results of Akmal et al. taken from Ref. \cite{akmal98} are 
also included for sake of comparison.
The properties computed in this work are not so far with respect to VCS calculations. The different methods
considered give almost the same gravitational mass if the only NN interactions is used, and the difference 
does not change much after the addition of the UIX interaction.

\begin{table}[!ht]
\vspace{0.2cm}
\begin{center}
\begin{tabular}{||l|l|ccc||}
\hline
Hamiltonian & method & $M/M_{\bigodot}$ & $\rho_c$ [fm$^{-3}$] & R [Km] \\
\hline
\hline
AV8'        & AFDMC  & 1.64              & 1.89                 & 7.9 \\
AV8'+UIX    & AFDMC  & 2.45              & 0.91                 & 11.1 \\
AV18        & VCS    & 1.68 & & \\
AV18+UIX    & VCS    & 2.39 & & \\
AV18+$\delta$v+UIX$^*$ & VCS & 2.21      & 1.14 & \\
\hline
\end{tabular}
\end{center}
\caption{Maximum gravitational masses, in M$_{\bigodot}$, the corresponding central density $\rho_c$ and 
the radius. The AFDMC results are obtained from the EOS of 66 neutrons using the indicated Hamiltonians, 
while the VCS results are taken from Ref. \cite{akmal98}.}
\label{tab:starsummary}
\end{table}

It has again to be stressed the fact that a better knowledge and investigation of the TNI structure as well as 
an accurate method to solve the Hamiltonian is of primary importance to predict neutron star properties.

\newpage
\subsection{Pure neutron matter and Illinois TNI forces}
The simulation of PNM using the Illinois TNI forces\cite{pieper01} was performed. There are signs of
some problems\cite{sarsa03} similar to those discussed in the neutron drop section above.
The $3\pi$ exchange term seems to overbind the neutrons when density increases, and the phenomenological 
central term is probably not repulsive enough.

The Illinois forces were fitted using the GFMC for light nuclei; the IL2 was extended up to $^{12}$C 
and provided binding energies in very good agreement with experimental data\cite{pieper05}.
Then, in addition to the AV8' interaction, the IL2 was considered, and by means of the FP-AFDMC 
several simulations of 66 neutrons as described in the above section were performed by varying the 
strength of $V_{3\pi}$ and $V_R$ to see the effect of the behavior of the PNM-EOS. Results are reported 
in table \ref{tab:IL2v3p} and \ref{tab:IL2vr}.

\begin{table}[ht]
\vspace{0.3cm}
\begin{center}
\begin{small}
\begin{tabular}{||c|ccccc|c||}
\hline
$\rho$[fm$^{-3}$]&0.0      & 0.0005  & 0.001   & 0.0015  & 0.002   & 0.0026 (IL2) \\
\hline
\hline
0.12 & 15.55(1) & 14.76(2)& 13.88(4)& 12.87(4)& 11.68(3)&  9.96(6) \\
0.16 & 21.78(6) & 20.25(5)& 18.51(7)& 16.51(4)& 14.34(6)& 10.9(2)  \\
0.20 & 28.31(6) & 25.86(4)& 22.95(5)& 19.69(3)& 15.86(6)& 10.2(1)  \\
0.32 & 61.1(3)  & 53.3(1) & 44.6(2) & 34.5(1) & 22.7(1) &  5.56(9) \\
0.40 & 92.7(2)  & 79.8(2) & 64.9(1) & 47.5(2) & 27.2(2) & -2.2(3)  \\
\hline
\end{tabular}
\end{small}
\end{center}
\caption{FP-AFDMC energies per particle calculated with 66 neutrons and AV8'+IL2 interaction
in a periodic box at various densities. The strength of the $V_{3\pi}$ was varied 
from the original IL2 also reported.
All the values are expressed in MeV.}
\label{tab:IL2v3p}
\end{table}

\begin{figure}[!ht]
\vspace{1.0cm}
\begin{center}
\includegraphics[width=10cm]{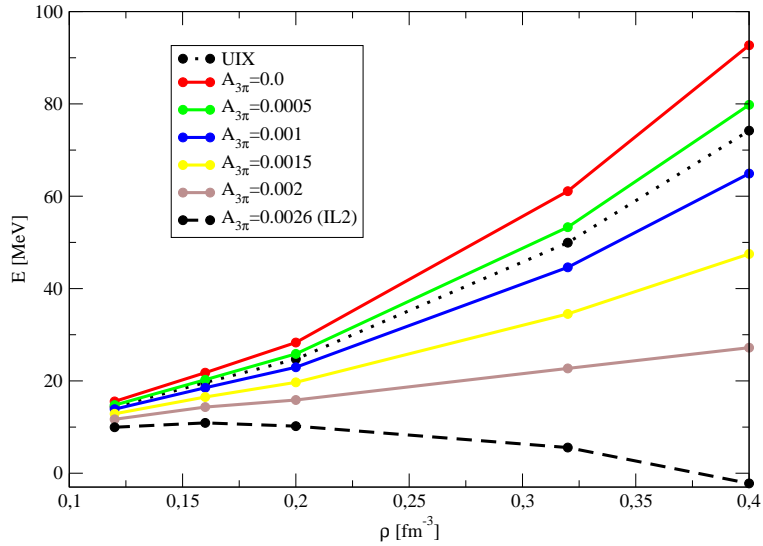}
%\vspace{0.5cm}
\caption{AFDMC equation of state calculated with 66 neutrons interacting with AV8'+IL2 
interaction. Different EOS were obtained by varying the strength of the $A_{3\pi}$ parameter of IL2.
The AV8'+UIX is also reported for a comparison. 
The $A_{3\pi}$ values are expressed in MeV.}
\label{fig:IL2v3p}
\end{center}
\end{figure}

As it can be seen in Fig. \ref{fig:IL2v3p}, the effect of the $3\pi$ exchange term of IL2 is dominant 
and by varying its strength the EOS drastically changes. Only if $A_{3\pi}\leq0.01$ MeV the EOS is comparable
to that obtained using the UIX TNI.

\begin{table}[!ht]
%\vspace{0.5cm}
\begin{center}
\begin{tabular}{||c|cccc||}
\hline
$\rho$[fm$^{-3}$]&0.0      & 0.0075  & 0.008   & 0.0085   \\
\hline
\hline
0.12 &    3.5(4)& 10.18(1)& 10.48(3)& 10.73(2)\\
0.16 &   -0.1(7)& 11.50(2)& 12.11(3)& 12.53(4)\\
0.20 &  -27(3)  & 11.01(4)& 11.89(4)& 12.80(3)\\
0.32 &  -85(4)  &  7.95(2)& 10.95(4)& 13.57(6)\\
0.40 &  -147(10)&  2.18(9)&  6.66(4)& 10.89(8)\\
\hline
\hline
$\rho$[fm$^{-3}$]& 0.009   & 0.0095  & 0.01    & 0.0105  \\
\hline
\hline
0.12 & 10.99(2)& 11.18(3)& 11.47(2)& 11.74(2)\\
0.16 & 13.01(2)& 13.45(4)& 13.97(4)& 14.42(2)\\
0.20 & 13.72(4)& 14.41(4)& 15.17(3)& 15.96(3)\\
0.32 & 15.99(7)& 18.3(1) & 20.83(4)& 23.24(6)\\
0.40 & 15.6(1) & 19.6(1) & 23.85(4)& 27.99(4)\\
\hline
\end{tabular}
\end{center}
\caption{FP-AFDMC energies per particle calculated with 66 neutrons and AV8'+IL2 interaction 
in a periodic box at various densities. The strength of the $V_R$ was varied 
from the original IL2 that is $A_R$=0.00705 MeV.
All the values are expressed in MeV.}
\label{tab:IL2vr}
\end{table}

\begin{figure}[ht]
\vspace{1.0cm}
\begin{center}
\includegraphics[width=10cm]{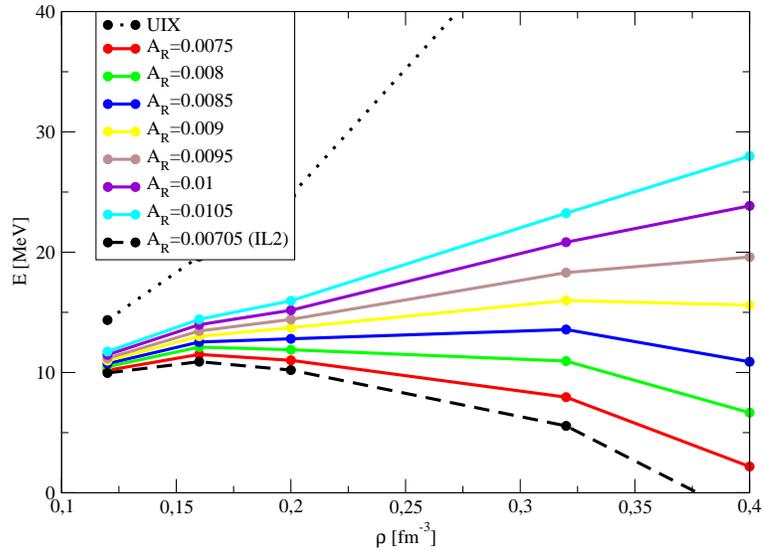}
%\vspace{0.5cm}
\caption{AFDMC equation of state calculated with 66 neutrons interacting with AV8'+IL2 
interaction. Different EOS were obtained by varying the strength of the $A_R$ parameter of IL2.
The AV8'+UIX is also reported for a comparison.
The $A_{3\pi}$ values are expressed in MeV.}
\label{fig:IL2vr}
\end{center}
\end{figure}

The phenomenological $A_R$ part also plays a very important role, but the repulsion given by this term 
is not sufficient to contrast the attraction given by the $3\pi$ exchange term. As it can be seen in 
Fig. \ref{fig:IL2vr} the $A_R$ parameter was increased from the original IL2 value of 0.00705 MeV up to 0.0105 MeV, but the 
EOS is completely different, and still far to be comparable with the equation of state computed with the 
UIX instead.

\section{Superfluidity in low-density neutron matter}
It has been well established that neutron matter is in a superfluid phase in the low-density regime in the 
region of the crust and in the inner part of a neutron star.

In this low-density region neutron superfluidity is expected to be mainly due to the attractive $^1S_0$ channel,
as suggested by the trend of NN phase shifts in each scattering channel. The occurrence of the well known virtual 
neutron-neutron channel determines the possibility of a formation of a pairing condensate at low density\cite{heiselberg00b}.
The physics of $^1S_0$ neutron superfluid is important to be understood in order to understand several phenomena that occur in the 
inner crust of a neutron star which may in turn be related to the formation of glitches\cite{dean03,monrozeau07}.

In recent years several works were performed to estimate the BCS gap for modern NN interaction. Results 
essentially confirm these expectations\cite{lee06,hebeler07}. Several groups are in a close agreement on the gap 
values, converging to an expectation of about 3 MeV at a Fermi momentum close to $k_F\approx$0.8 fm$^{-1}$
\cite{baldo90,khodel96,elgaroy98}. All these calculations consider a bare NN interaction as the pairing 
force. It has been pointed out that the screening by the medium of the interaction could strongly reduce the 
pairing strength in this channel\cite{schulze96,dean03,caiwan03,cao06}.
The many-body calculation of the pairing gap exactly including all the correlations given by the NN interactions is 
of primary relevance, and the AFDMC can be used to accurately perform a calculation of the gap\cite{fabrocini05}.

The AFDMC can be used to describe a superfluid system using a BCS trial wave function as described in Sec. 
\ref{sec:psitrial}. The BCS coefficients appearing in the Pfaffian of paired orbital wave functions can be taken from the BCS amplitudes 
from a correlated basis functions (CBF) calculation\cite{fabrocini05}.
For even N number, all neutrons are paired in the Pfaffian, whereas for odd N values the configuration of the unpaired 
neutron providing the best energy must be found.

The pairing gap energy for odd N values is defined as:
\begin{equation}
\Delta(N)=E(N)-\frac{1}{2}[E(N-1)+E(N+1)] \,,
\end{equation}
and several calculations using different values of N must be performed to verify the eventually gap dependence to 
the number of neutrons N.
The considered Hamiltonian contains the realistic Argonne AV8' NN interaction as well as the Urbana UIX TNI; the 
AV8' finite-size effects as described in Sec. \ref{sec:SNM} and \ref{sec:PNM} were included.

The FP-AFDMC was used to calculate the energy of low-density PNM by simulating 
several systems with N=12$\div$18 and N=62$\div$68 neutrons in a periodic box. The considered Hamiltonian contains
both AV8' NN and UIX TNI interactions. In some case only the NN interaction was employed. 
All the results obtained with FP-AFDMC for several densities are reported in tables \ref{tab:kf02full}, \ref{tab:kf04full}, 
\ref{tab:kf06full} and \ref{tab:kf08full}. In each table is reported the energy per neutron evaluated 
for each considered N. In the case of $k_F$=0.4 fm$^{-1}$ all the calculations were performed for both AV8'
and AV8'+UIX Hamiltonians, while for $f_F$=0.2 fm$^{-1}$, $k_F$=0.6 fm$^{-1}$ and $k_F$=0.8 fm$^{-1}$ full calculations 
were performed only with the full Hamiltonians.

\begin{table}[!h]
\begin{center}
\begin{small}
\begin{tabular}{||c|cc|cc||}
\hline
N   & E/N(AV8'+UIX)   & $\Delta$(AV8'+UIX) \\
\hline
\hline
62  &  0.359(2) &         \\
63  &  0.360(2) &  0.2(2) \\
64  &  0.354(2) &         \\
65  &  0.359(2) &  0.3(2) \\
66  &  0.355(3) &         \\
67  &  0.365(3) &  0.3(3) \\
68  &  0.361(4) &         \\
    &           &  0.3(3) \\
\hline
\end{tabular}
\end{small}
\end{center}
\caption{FP-AFDMC energy per particle calculated with different number N of BCS-paired neutrons with Fermi momentum
$k_F$=0.2 fm$^{-1}$ interacting with AV8'+UIX Hamiltonians in a periodic box.
For odd N the energy gap is reported, and the mean value of $\Delta$ given by N=62$\div$68 is indicated. 
All the energies are expressed in MeV.}
\label{tab:kf02full}
\end{table}

\begin{table}[!h]
\vspace{0.3cm}
\begin{center}
\begin{small}
\begin{tabular}{||c|cc|cc||}
\hline
N   & E/N(AV8'+UIX)   & $\Delta$(AV8'+UIX)& E/N(AV8')   & $\Delta$(AV8') \\
\hline
\hline
12  & 1.294(2) &         & 1.302(2) &         \\ 
13  & 1.407(2) & 1.56(5) & 1.418(1) & 1.62(3) \\
14  & 1.281(3) &         & 1.284(1) &         \\
15  & 1.413(2) & 2.04(6) & 1.420(1) & 2.03(3) \\
16  & 1.273(2) &         & 1.285(1) &         \\
17  & 1.379(2) & 1.79(5) & 1.387(1) & 1.75(3) \\
18  & 1.274(1) &         & 1.282(1) &         \\
    &                       & 1.80(6) &                       & 1.80(3) \\
\hline
62  & 1.230(2) &         & 1.232(2) &        \\
63  & 1.258(2) &  1.5(2) & 1.257(3) & 1.5(3) \\
64  & 1.239(2) &         & 1.237(3) &        \\
65  & 1.261(2) &  1.5(2) & 1.258(3) & 1.0(4) \\
66  & 1.239(2) &         & 1.248(5) &        \\
67  & 1.264(2) &  1.5(2) & 1.262(3) & 1.3(4) \\
68  & 1.245(2) &         & 1.238(1) &        \\
    &          &  1.5(2) &          & 1.3(4) \\
\hline
\end{tabular}
\end{small}
\end{center}
\caption{FP-AFDMC energy per particle calculated with different number N of BCS-paired neutrons with Fermi momentum
$k_F$=0.4 fm$^{-1}$ interacting with AV8'+UIX and AV8' Hamiltonians in a periodic box.
All the energies are expressed in MeV.}
\label{tab:kf04full}
\end{table}

\begin{table}[!ht]
\begin{center}
\begin{small}
\begin{tabular}{||c|cc|cc||}
\hline
N   & E/N(AV8'+UIX)         & $\Delta$(AV8'+UIX)& E/N(AV8')   & $\Delta$(AV8') \\
\hline
\hline
12  & 2.662(2) &         & & \\ 
13  & 2.808(2) & 2.45(4) & & \\
14  & 2.584(2) &         & & \\
15  & 2.822(2) & 2.88(6) & & \\
16  & 2.670(2) &         & & \\
17  & 2.831(2) & 2.4(1)  & & \\
18  & 2.705(7) &         & & \\
    &          & 2.6(1)  & & \\
\hline
62  & 2.553(2) &         & 2.529(3) &        \\
63  & 2.585(2) &  1.7(2) & 2.568(2) & 2.0(3) \\
64  & 2.563(2) &         & 2.546(4) &        \\
65  & 2.604(2) &  2.1(2) & 2.604(2) & 2.2(4) \\
66  & 2.579(2) &         & 2.593(4) &        \\
67  & 2.620(2) &  2.4(2) & 2.612(2) & 1.8(4) \\
68  & 2.587(2) &         & 2.579(2) &        \\
    &          &  2.1(2) &          & 2.0(4) \\
\hline
\end{tabular}
\end{small}
\end{center}
\caption{FP-AFDMC energy per particle calculated with different number N of BCS-paired neutrons with Fermi momentum
$k_F$=0.6 fm$^{-1}$. See table \ref{tab:kf04full} for details.
All the energies are expressed in MeV.}
\label{tab:kf06full}
\end{table}

\begin{table}[ht]
\begin{center}
\begin{small}
\begin{tabular}{||c|cc||}
\hline
N   & E/N(AV8'+UIX)         & $\Delta$(AV8'+UIX) \\
\hline
\hline
12  & 4.651(5) &         \\ 
13  & 4.595(4) & 2.05(9) \\
14  & 4.254(4) &         \\
15  & 4.627(4) & 1.8(1)  \\
16  & 4.730(3) &         \\
17  & 4.934(3) & 2.4(1)  \\
18  & 4.853(3) &         \\
    &          & 2.1(1)  \\
\hline
62  & 4.342(4) &         \\
63  & 4.377(3) & 2.0(4)  \\
64  & 4.350(4) &         \\
65  & 4.360(4) & 2.1(4)  \\
66  & 4.305(3) &         \\
67  & 4.314(3) & 0.6(3)  \\
68  & 4.308(2) &         \\
    &          & 1.6(4)  \\
\hline
\end{tabular}
\end{small}
\end{center}
\caption{FP-AFDMC energy per particle calculated with different number N of BCS-paired neutrons with Fermi momentum
$k_F$=0.8 fm$^{-1}$. 
See table \ref{tab:kf04full} for details.
All the energies are expressed in MeV.}
\label{tab:kf08full}
\end{table}

In order to verify that the system is in a superfluid phase for N=14 and 66 neutrons a calculation using a 
trial wave function of the type of normal PNM was performed.
The energy of the system in the normal phase is compared with the BCS value in table \ref{tab:BCSvsSLAT}. 
At $k_F$=0.2 fm$^{-1}$ and $k_F$=0.6 fm$^{-1}$ the superfluid is well favorite compared to the normal state, and all the energies 
obtained with the BCS are lower respect to the normal trial wave function. It is interesting to note as 
using N=66 at $k_F$=0.4 fm$^{-1}$ the favorite state is again the BCS, but for N=14 the two states are almost
degenerate, being the two energies within error bars.
Instead, at $k_F$=0.8 fm$^{-1}$ the situation is opposite: the BCS state is favorite for N=14, but using N=66 
it seems that the ground-state of the system is the normal phase.

\begin{table}[!h]
\begin{center}
\begin{footnotesize}
\begin{tabular}{||l|cc|cc||}
\hline
Hamiltonian & $k_F$ [fm$^{-1}$] & N  & $E_{BCS}/A$ & $E_n/A$  \\
\hline                                                        
\hline
AV8'+UIX    & 0.2               & 66 & 0.355(3)    & 0.367(2) \\
\hline
AV8'+UIX    & 0.4               & 14 & 1.281(3)    & 1.280(3) \\
AV8'+UIX    & 0.4               & 66 & 1.239(2)    & 1.289(2) \\
\hline                                                        
AV8'+UIX    & 0.6               & 14 & 2.584(2)    & 2.614(4) \\
AV8'+UIX    & 0.6               & 66 & 2.579(2)    & 2.606(4) \\
AV8'        & 0.6               & 66 & 2.593(4)    & 2.621(4) \\
\hline
AV8'+UIX    & 0.8               & 14 & 4.254(4)    & 4.269(4) \\
AV8'+UIX    & 0.8               & 66 & 4.305(3)    & 4.277(7) \\
AV8'        & 0.8               & 66 & 4.261(3)    & 4.218(3) \\
\hline
\end{tabular}
\end{footnotesize}
\end{center}
\caption{Energy per neutron of the system in both normal ($E_n/A$) and BCS ($E_{BCS}/A$) phase for several 
Hamiltonians, densities and N. The estimates of the energy with the system in the normal phase were
performed with the same calculation described in the PNM section.
All the energies are expressed in MeV.}
\label{tab:BCSvsSLAT}
\end{table}

The energy gap was calculated using a different number of neutrons. 
In table \ref{tab:summary} we summarize all the results obtained using FP-AFDMC and reported in the above tables,
and the available CP-AFDMC gap energies taken from Refs. \cite{fabrocini05,illarionov07}.
As it can be seen some finite-size effects are present. With the full AV8'+UIX at each considered density 
the gap energy calculated using N=62$\div$68 is smaller with respect to the N=12$\div$18 case. Also with the AV8' 
at $k_F$=0.4 fm$^{-1}$ this behavior is evident. This fact could indicate some finite-size effect due to the 
volume of the simulation box. Although the potential energy of the system is obtained by considering the effect 
due to inter-particle distances in the neighbor cells, some quantum correlation could still be missing.
However, it has to be noted the opposite behavior given by CP-AFDMC using AV8' at $k_F$=0.6 fm$^{-1}$.

\begin{table}[ht]
\begin{center}
\begin{tabular}{||l|cc|cc||}
\hline
Hamiltonian & $k_F$ [fm$^{-1}$] & N     & $\Delta_{FP}$ & $\Delta_{CP}$ \\
\hline
\hline
AV8'+UIX    & 0.4               & 12$\div$18 & 1.80(6)       &          \\
AV8'+UIX    & 0.4               & 62$\div$68 & 1.5(2)        &          \\
AV8'        & 0.4               & 12$\div$18 & 1.80(3)       & 1.83(7)  \\
AV8'        & 0.4               & 62$\div$68 & 1.3(4)        &          \\
\hline
AV8'+UIX    & 0.6               & 12$\div$18 & 2.6(1)        & 2.73(9)  \\
AV8'+UIX    & 0.6               & 62$\div$68 & 2.1(6)        &          \\
AV8'        & 0.6               & 12$\div$18 &               & 2.46(8)  \\
AV8'        & 0.6               & 50$\div$56 &               & 2.5(5)   \\
AV8'        & 0.6               & 62$\div$66 & 2.0(4)        & 2.6(6)   \\
\hline
AV8'+UIX    & 0.8               & 12$\div$18 & 2.1(1)        &          \\
AV8'+UIX    & 0.8               & 62$\div$68 & 1.6(4)        &          \\
AV8'        & 0.8               & 12$\div$18 &               & 2.2(1)   \\
\hline
\end{tabular}
\end{center}
\caption{Energy gap calculated with FP-AFDMC ($\Delta_{FP}$) and with CP-AFDMC ($\Delta_{CP}$) using 
different Hamiltonians for several densities 
and considering different numbers of neutrons. Some of the CP-AFDMC results are taken from Ref. \cite{fabrocini05} 
and other from \cite{illarionov07}. 
All the energies are expressed in MeV.}
\label{tab:summary}
\end{table}

The gap calculated with FP-AFDMC using N=12$\div$18 is in good agreement with the CP-AFDMC results, while in 
the only comparable case of N=62$\div$68 with AV8' at $k_F$=0.6 fm$^{-1}$ the FP-AFDMC energy gap is 
significantly smaller.

\begin{table}[ht]
\begin{center}
\begin{tabular}{||c|cc||}
\hline
$k_F$ [fm$^{-1}$] & $\Delta$ & $\Delta/E_F$ \\
\hline
\hline
0.2               & 0.3(3)   & 0.55 \\
0.4               & 1.5(2)   & 0.74 \\
0.6               & 2.1(2)   & 0.47 \\
0.8               & 1.6(4)   & 0.20 \\
\hline
\end{tabular}
\end{center}
\caption{Energy gap calculated with FP-AFDMC with N=62$\div$68 using the AV8'+UIX Hamiltonian for several densities.
All the values are expressed in MeV.}
\label{tab:bcstot}
\end{table}

The AFDMC algorithm was applied to microscopically study the $^1S_0$ superfluid gap in low density neutron matter
using realistic Hamiltonian containing a modern NN interaction as well as a TNI. The results are summarized in table \ref{tab:bcstot}. 
The AFDMC allows to 
solve the ground-state of the Hamiltonian by taking into account all the polarization and medium modification 
effects that all other theories neglected or cannot include in the calculation in a realistic way\cite{sedrakian06}.

\chapter{Results: neutron-rich isotopes}
The structure of stable nuclei is generally dominated by shell effects which can be nearly completely explained by the 
single-particle picture of the system. 

However, it has been found experimentally, by means of fragmentation of heavy ion
beams\cite{stanoiu04,thoennessen03,leistenschneider02,reed99}, that neutron rich nuclei show features which seem to go beyond the 
standard shell-model. For example, it was observed that the neutron drip line for O and F isotopes has a 
sudden change. This has been interpreted as a clear sign that many--body effects in this regime cannot 
be treated as a perturbation of the one-particle picture, but become relevant in determining the overall 
structure. Theoretical discussions of the insurgence of a new magic number at N=16 have recently been developed 
by means of mean field calculations\cite{nakada03} and antisymmetrized molecular dynamics\cite{kimura03}.

On the other hand the study of neutron-rich isotopes is an interesting investigation field because their 
structure is particularly relevant to constrain properties of the crust in neutron stars\cite{gupta07},
but the impossibility to use accurate methods such the GFMC to study properties of medium nuclei in 
addition to the complexity of NN and TNI nuclear forces imposes the need to introduce a simpler 
model rather then the full microscopic description of a nucleus with all nucleons as degrees of freedom.

\section{Modelization of neutron-rich nuclei}
A Green's Function Monte Carlo calculation was performed by mimicking oxygen isotopes as neutron 
drops\cite{chang04,pieper05}; in that model all neutrons interact with a realistic interaction (AV18+IL2),
with the addition of an artificial external well providing the attraction necessary to produce 
a bound state of neutrons and representing an average effect of the protons which are not explicitly 
included in the calculation. 
The study of oxygen isotopes is then reduced to the study of the N neutrons in the $^{8+N}$O nucleus.
The chosen external well was a Wood-Saxon potential whose parameters were determined in order to reproduce 
the correct separation energy of $^{17}$O and $^{18}$O isotopes\cite{pieper05}.

The CP-AFDMC was applied to study neutron-rich isotopes of oxygen\cite{gandolfi06} and calcium\cite{gandolfi06b} 
in a any more simple model, where only external neutrons 
are included in the calculations, while all the core is modeled with a single-particle potential acting 
on valence neutrons. The advantage of this model is that the external well is generated in 
a self-consistent way, and the absence of parameters implies no need for any additional calculation.
The study of oxygen isotope excited states were also performed.

The considered Hamiltonian is 
\begin{equation}
H=-\sum_{i=1}^N{\hbar^2\over2m}\nabla_i^2+\sum_{i=1}^NV_{ext}(r_i)+\sum_{i<j}v_{ij}+\sum_{i<j<k}V_{ijk} \,,
\end{equation}
where the NN interaction is the Argonne AV8', the TNI is the Urbana UIX and the N are the interacting 
external neutrons of the $^{16+N}$O and $^{40+N}$Ca isotopes.
The one-body potential $V_{ext}$ describes the core of the closed-shell nucleus
which is replaced by a self-consistent potential.

The trial wave function used in the AFDMC calculation has the same structure of that employed in neutron drop 
calculations.
The radial component were obtained solving the Hartree-Fock (HF) problem with the Skyrme
forces which also provides the external potential well that describes
the closed core. The yielded radial functions are written in the $j,m_j$ 
base, so the single-particle wave function were obtained by coupling the 
spherical harmonics with the spin using the Clebsh-Gordan coefficients.
In  $^n$O $n$=18...22 isotopes neutrons fill only the orbitals in the
$1D_{5/2}$ shell except for the isotope $^{19}$O, where it is necessary 
to put some neutrons also in the $1D_{3/2}$ in order to build the ground state of correct symmetry.
In the $^n$Ca $n$=42...48 isotopes it is assumed that neutrons fill only the orbitals in the
$1F_{7/2}$ and $1F_{5/2}$ shell in order to build the ground state of correct symmetry.

The many-body states are obtained by coupling the single-particle angular momentum by 
constructing eigenstates of total angular momentum $J=j_1+...+j_N$; 
for states with an even number of neutrons, the ground-state 
has $J=0$, while for odd neutron numbers, the ground state has
total angular momentum $J=5/2$ or $J=7/2$.
These states are in general written in terms of a sum of Slater determinants
whose coefficients are determined by the symmetry of the state. Each
determinant is evaluated at the current values of the positions and spin assignments
of the nucleons in the walker $|R,S\rangle$.

The choice of the trial wave function proved to be particularly important for calcium isotopes calculations.
A sufficiently good representation of the ground state of $^{46}$Ca can be obtained by building 
a two-hole state which is complementary to $^{42}$Ca.
For the $^{43}$Ca and $^{45}$Ca completely different trial wave functions are needed; in 
fact the wave function for the first nucleus contains $9$ determinants and the second $35$. 
In principle we could use the $^{43}$Ca wave function as a three-hole state describing 
$^{45}$Ca. However, AFDMC gives in this case an energy a few MeV too high. This is due 
to the approximation used to deal with the sign problem. The two wave functions are 
degenerate in the eigenspace of $J^2$ but they give two different energy, because 
they have a different nodal surface.

The same procedure is used for the construction of the oxygen excited states,
always limiting the choice of the single particle orbitals to the shell $1D_{5/2}$.

In these calculations contributions for the $2S_{1/2}$ orbital were neglected, despite one might 
expect some mixing of this level in the states of isotopes considered.
However, the fact that  AFDMC gives results for the energy differences in very good agreement 
with experimental data indicates that the effects of the mixing are not extremely
significant. 
The inclusion of contributions from the $2S_{1/2}$ orbital also gives rise to
major technical difficulties.
In fact, the component along this orbital tends to be projected 
over the more bound $1S_{1/2}$ state, giving rise to a non-physical density
for the external neutrons. 
This difficulty also prevented from extending this calculations to higher mass
isotopes to the drip line.

An important issue in this calculation is the choice of  the Skyrme force parameters 
to be used to generate the 
effective-HF potentials included in the Hamiltonian for AFDMC calculations.
The use of a self-consistent potential well makes the single--particle
part of the importance function essentially parameter less. In absence of an
efficient variational procedure, it was important to establish that this
choice of orbitals could be in general reasonable, if not optimal. 

Skyrme forces cannot be related directly to the realistic NN interaction
used in the calculation. One must therefore establish a criterion for discriminating
among different choices. As a first step the variational energy of the closed--shell
isotope $^{22}$O was computed on the correlated
Slater determinants obtained using different sets of parameters in 
the Skyrme Hamiltonian, but using the full Hamiltonian with potential
AV8'+UIX. In particular the Skyrme I and Skyrme II\cite{vautherin72} forces were tried.

\begin{figure}[!ht]
\vspace{1.0cm}
\begin{center}
\includegraphics[width=10cm]{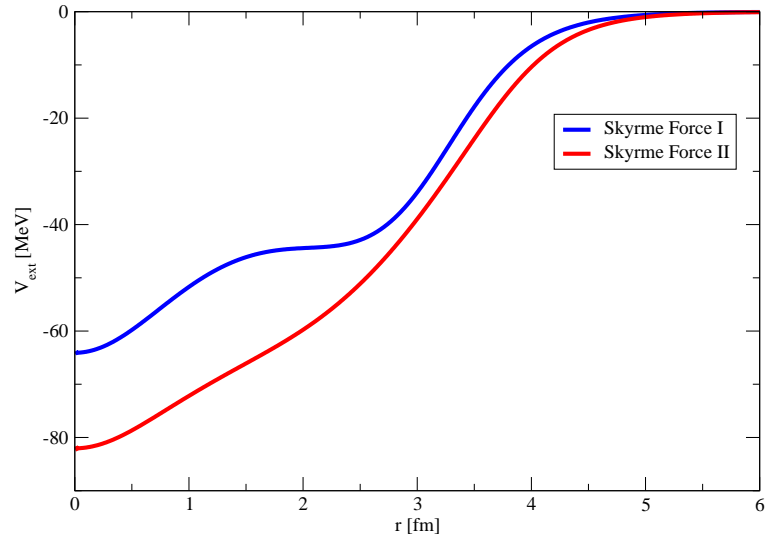}
\vspace{0.5cm}
\caption{Effective single-particle potential generated by HF algorithm, with 
Skyrme's parameter of type Force I (full line) and II\cite{vautherin72} (dashed line).}
\label{figure:HFpot}
\end{center}
\end{figure}

The expectation value of the Hamiltonian considerably differs in the two cases; 
the variational energy is -29.9(1) MeV using Skyrme I and -60.5(1) MeV for Skyrme II, while
the experimental value of the energy for the external neutrons is -34.407 MeV\cite{exp00}.
This is mainly due to the fact that the two sets of Skyrme's parameters
give quite different effective-HF potentials, as plotted in Fig. \ref{figure:HFpot}.  
As a criterion for determining which shell (and which set of orbitals) to
use, it has been chosen to minimize the difference between $\langle H \rangle$ and the 
available experimental estimates of the binding energy of the nucleus. Consequently 
the Skyrme I force was chosen also for calcium isotopes calculation.
However one expects that the choice of the 
self-consistent potential only influences the absolute total binding energy of each isotope. The 
difference between energies of different isotope depends mostly on the choice of the NN and 
TNI interactions that include correlations between the external neutrons. 

\section{Oxygen isotopes}
The AFDMC energies obtained for the isotopes series $^{18}$O-$^{22}$O\cite{gandolfi06} are reported 
in table \ref{table:GSris}, compared with the available experimental values\cite{exp00}. 

\begin{table}[!ht]
\vspace{0.2cm}
\begin{center}
\begin{tabular}{||l|cc||}
\hline  
isotope      & $E_{AFDMC}$ & $E_{exp}$ \\
\hline  
\hline  
E($^{18}$O) & -18.04(2) & -12.188 \\
E($^{19}$O) & -22.4(1)  & -16.145 \\
E($^{20}$O) & -29.36(6) & -23.752 \\
E($^{21}$O) & -33.6(2)  & -27.558 \\
E($^{22}$O) & -40.48(5) & -34.407 \\
\hline  
\end{tabular}
\end{center}
\caption{Ground-state energy calculated with AFDMC. All the energies are 
express in MeV.}
\label{table:GSris}
\end{table}

\begin{figure}[!ht]
\vspace{1.0cm}
\begin{center}
\includegraphics[width=10cm]{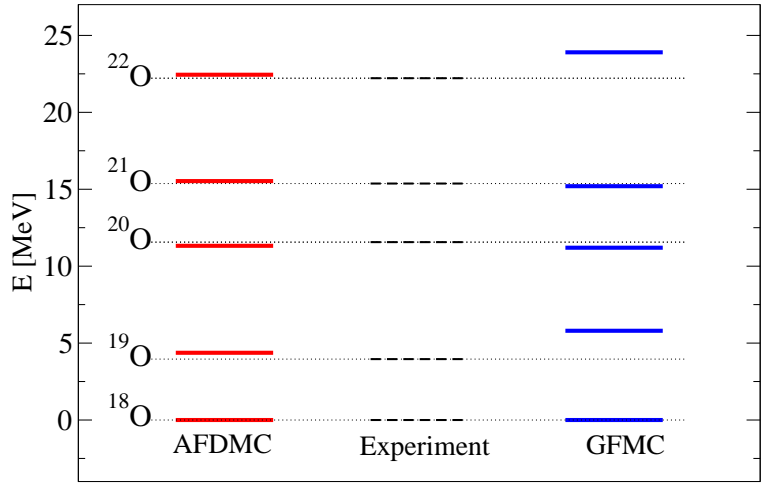}
%\vspace{0.5cm}
\caption{Outline of differences between energies of the isotope series studied.
All the energies are expressed in MeV and all values are referred to
that $^{18}$O. Results are compared with experimental results\cite{exp00} and
with GFMC calculations of Ref. \cite{chang04}.}
\label{figure:GSdiff2}
\end{center}
\end{figure}

As expected, the computed values are quite different from the experiment, although the
relative discrepancy never exceeds 30\%. Of course this is a drawback of the use
of an external potential for including the effects of the filled core of the 
nucleus. In particular the total binding energies are all overestimated.
This reflects the absence of a correct description of the density of neutrons at the
center of the drop, which is underestimated, giving rise to an effective
potential which is too deep for small distances from the center. Moreover
core--polarization effects were completely neglected.   
However, most of the information needed to understand the effects of NN interaction
in the external shell can be obtained looking at energy differences between the 
isotopes considered. In fact, if the intrashell interaction has a dominant effect, the gaps 
should not depend too much on the quality of the external well considered.
In Table \ref{table:GSdiff} and in Figure 
\ref{figure:GSdiff2} are reported
the energy differences for the isotopes considered, compared with the corresponding
differences obtained from the experimental results. 
The AFDMC calculations have also been compared with the GFMC results reported
in the paper by Chang et al.\cite{chang04}.
As it can be seen, in this case the agreement between computed and experimental values 
is excellent. 

\begin{table}[!ht]
\vspace{0.2cm}
\begin{center}
\begin{tabular}{||l|cc||}
\hline  
         & $E_{AFDMC}$ & $E_{exp}$ \\
\hline  
\hline  
E($^{19}$O)-E($^{18}$O) &  -4.4(1)  &  -3.957  \\
E($^{20}$O)-E($^{18}$O) & -11.32(8) & -11.564  \\
E($^{21}$O)-E($^{18}$O) & -15.5(2)  & -15.370  \\
E($^{22}$O)-E($^{18}$O) & -22.44(7) & -22.219  \\
\hline  
\end{tabular}
\end{center}
\caption{Ground-state energy calculated with AFDMC. The 
differences between the studied isotopes are reported. All the energies are 
express in MeV.}
\label{table:GSdiff}
\end{table}

It is possible to notice how the binding energy of $^{22}$O relative to $^{18}$O computed by AFDMC is
almost coincident with the experimental finding, while there is a strong
overestimation in the GFMC calculation. This is quite surprising, being $^{22}$O a closed shell
nucleus. However, it is possible that in this case the choice of the self--consistent HF
potential gives a better description of the outer shell than the all neutrons picture used
in the GFMC calculations. 

\begin{figure}[!h]
\vspace{1.0cm}
\begin{center}
\includegraphics[width=10cm]{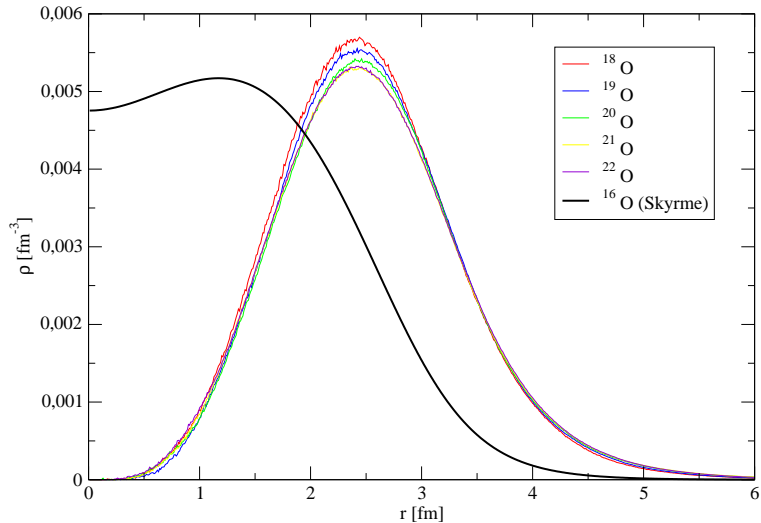}
%\vspace{0.5cm}
\caption{Densities of external neutrons for all isotopes in the ground-state calculated 
with AFDMC and the Skyrme's density of $^{16}$O.}
\label{figure:dens}
\end{center}
\end{figure}

In figure \ref{figure:dens} the AFDMC densities normalized to unity
of the external neutrons for the isotopes considered are displayed.
As it can be seen 
the neutron densities are all quite similar, and very small deviations are present. In fact the density 
profile appears to be larger when there are more neutrons; this can be expected because of the 
repulsive nature of the NN interaction for neutrons.
In the figure it is also displayed the density of $^{16}$O calculated with Skyrme interaction, to make
evident the "halo" effect of the external neutrons. 

By constructing the appropriate wave functions it is possible to study the 
low-lying excited states of a given isotope. Excited states were computed
for the isotopes $^{18}$O, and $^{21}$O. Calculations are limited to 
states which can be built while still remaining within the
$1D_{5/2}$ shell.  
If the considered states are orthogonal to the ground state, 
the projected wave function will be constrained within the same subspace.
The AFDMC results compared with experimental
results of Norum et al.\cite{norum82} for $^{18}$O and of Stanoiu et al.\cite{stanoiu04} for $^{21}$O
are reported in table \ref{table:excdiff}.

\begin{table}[!h]
\begin{center}
\begin{tabular}{||lc|cc||}
\hline  
    & state & $\Delta E_{AFDMC}$ & $\Delta E_{exp}$ \\
\hline  
\hline  
$^{18}$O&&&\\
&$2^+$ & 1.99(9) & 2.0(2) \\
&$4^+$ & 5.08(2) & 3.6(2) \\
\hline
$^{21}$O&&&\\
&$1/2^+$ &1.24(9) &1.218(4)\\
&$3/2^+$ &2.11(9) &2.133(5)\\
\hline  
\end{tabular}
\end{center}
\caption{Excitation energy calculated with AFDMC of the isotopes 
$^{18}$O and $^{21}$O, compared with experimental results (see text). All energies are in MeV.}
\label{table:excdiff}
\end{table}

\begin{figure}[!h]
\vspace{1.0cm}
\begin{center}
\includegraphics[width=10cm]{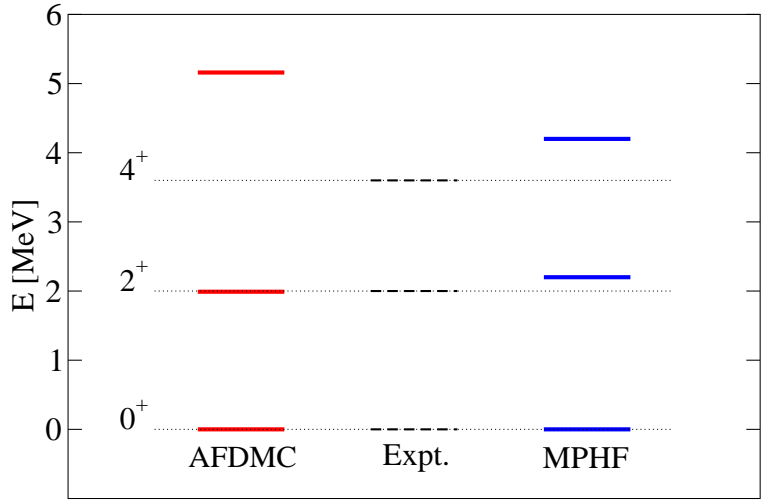}
%\vspace{0.5cm}
\caption{Low-lying excitation energies of the  $^{18}$O isotope, compared with 
experimental values from ref. \cite{norum82}. All energies are in MeV. Results are compared
with MDHF results of Ref. \cite{morrison78}.}
\label{figure:isoexc18}
\end{center}
\end{figure}

In Fig. \ref{figure:isoexc18} it is shown the results for the $^{18}$O  
isotope, compared with experiment and with multideterminant projection Hartree-Fock
(MDHF) results by Morrison et al.\cite{morrison78}.
The excitation energy of the  $2^+$ state turns out to be in excellent
agreement with the experimental findings. The situation for the  
$4^+$ state is much less satisfactory. 
There are essentially two possible sources for this discrepancy. 
In the case of the $J=4$ state, the large angular momentum might 
give the spin--orbit component of the potential a relevant role. 
The presence of spin-orbit terms tends to heavily modify the nodal structure
of the wave function, and this effect can be due to the use of the constrained-path approximation.

\begin{figure}[!h]
\vspace{1.0cm}
\begin{center}
\includegraphics[width=10cm]{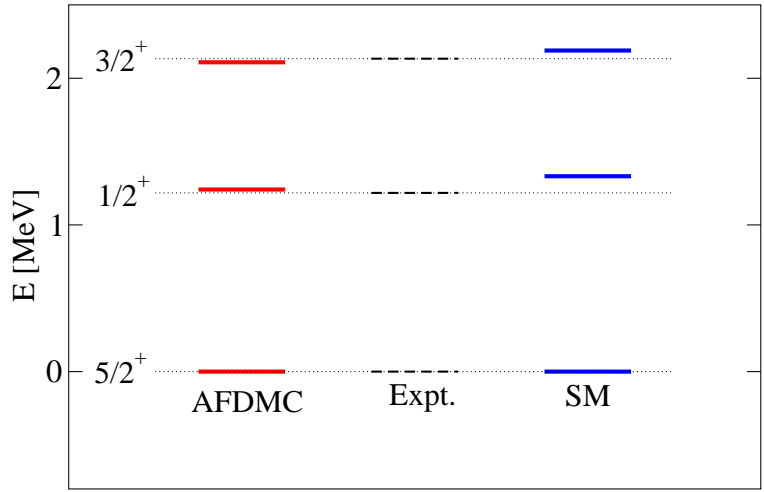}
%\vspace{0.5cm}
\caption{Low-lying excitation energies of the $^{21}$O isotope, compared with 
experimental values from Ref. \cite{stanoiu04}. All energies are in MeV. Results are compared
with shell model results of Ref. \cite{brown01}.}
\label{figure:isoexc21}
\end{center}
\end{figure}

The excitation energies for the $^{21}$O isotope are shown in Fig. \ref{figure:isoexc21}.
We compare our AFDMC results with experiments and with shell model calculations by Brown\cite{brown01}.
In this case the agreement between AFDMC results and experiments is very good for
all the states considered. 

\section{Calcium isotopes}
The constrained-path AFDMC energies obtained for the isotopes series $^{42}$Ca-$^{48}$Ca 
compared with the available experimental values\cite{exp00} are reported in table \ref{tab:caris}.

\begin{table}[!h]
\vspace{0.2cm}
\begin{center}
\begin{tabular}{||l|cc||}
\hline
isotope     & $E_{AFDMC}$ & $E_{exp}$ \\
\hline
\hline
E($^{42}$Ca) & -25.25(7) & -19.85 \\ 
E($^{43}$Ca) & -33.9(4)  & -27.78 \\
E($^{44}$Ca) & -46.3(1)  & -38.91 \\
E($^{45}$Ca) & -52.6(4)  & -46.32 \\ 
E($^{46}$Ca) & -62.9(3)  & -56.72 \\
E($^{47}$Ca) & -70.2(7)  & -63.99 \\
E($^{48}$Ca) & -80.3(8)  & -73.09 \\
\hline
\end{tabular}
\end{center}
\caption{Ground-state energy calculated with AFDMC. All the energies are 
express in MeV. Experimental values are referred to ground-state energy of 
$^{40}$Ca, taken from Ref. \cite{exp00}.}
\label{tab:caris}
\end{table} 

As in the case of oxygen isotopes the absolute binding energies are quite different from the experimental 
results, and the total binding energies are all overestimated.

\begin{figure}[!ht]
\vspace{0.5cm}
\begin{center}
\includegraphics[width=10cm]{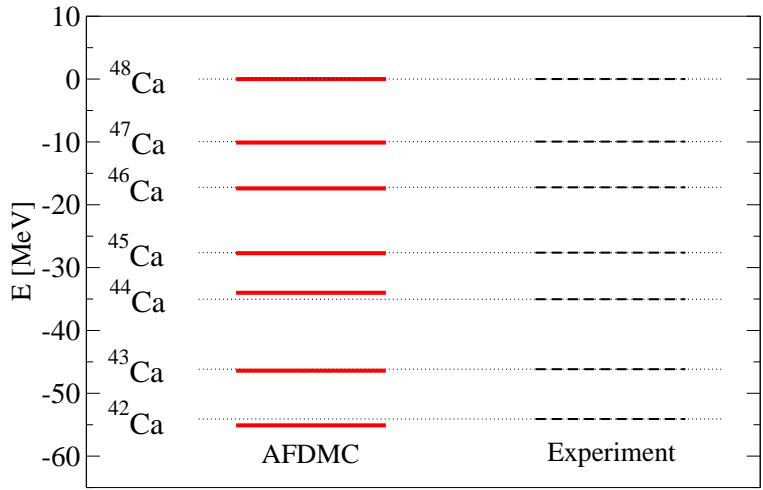}
%\vspace{0.5cm}
\caption{Outline of differences between energies of the isotope series studied.
All the energies are expressed in MeV and all values are referred to $^{48}$Ca.}
\label{fig:cadiff}
\end{center}
\end{figure}

In table \ref{tab:cadiff} and in Fig. \ref{fig:cadiff} the energy differences for the isotopes considered
are reported, compared with the corresponding differences obtained from the experimental results.
As it can be seen, also in this case the agreement between computed and experimental values
is excellent. Small deviations are present only for the two cases of isotopes $^{42}$Ca 
and $^{44}$Ca.

\begin{table}[ht]
\vspace{0.2cm}
\begin{center}
\begin{tabular}{||l|cc||}
\hline
                  & $E_{AFDMC}$ & $E_{exp}$ \\
\hline
\hline
E($^{42}$Ca)-E($^{48}$Ca) & -55.1(8) & -54.09  \\
E($^{43}$Ca)-E($^{48}$Ca) & -46.4(8) & -46.16  \\
E($^{44}$Ca)-E($^{48}$Ca) & -34.0(8) & -35.03  \\
E($^{45}$Ca)-E($^{48}$Ca) & -27.7(8) & -27.62  \\
E($^{46}$Ca)-E($^{48}$Ca) & -17.4(8) & -17.22  \\
E($^{47}$Ca)-E($^{48}$Ca) & -10.1(8) &  -9.95  \\
\hline
\end{tabular}
\end{center}
\caption{Ground-state energy calculated with AFDMC. We report the
differences between the isotopes we had studied. All the energies are
expressed in MeV.}
\label{tab:cadiff}
\end{table}

In this calculation we completely neglected the interaction between shells with different 
orbital angular momentum. This implies that we could verify only the accuracy of NN and TNI interaction 
of nucleons in the $F$ shell. It was shown in other calculations that the Urbana IX TNI seems to be unable 
to correctly describe the correlations between nucleons in different shells\cite{pieper01}.
However, our results show that the energy separation of isotopes with external
neutrons filling the shells with the same orbital angular momentum $L=3$ can be accurately reproduced.

\begin{figure}[!ht]
\vspace{0.5cm}
\begin{center}
\includegraphics[width=10cm]{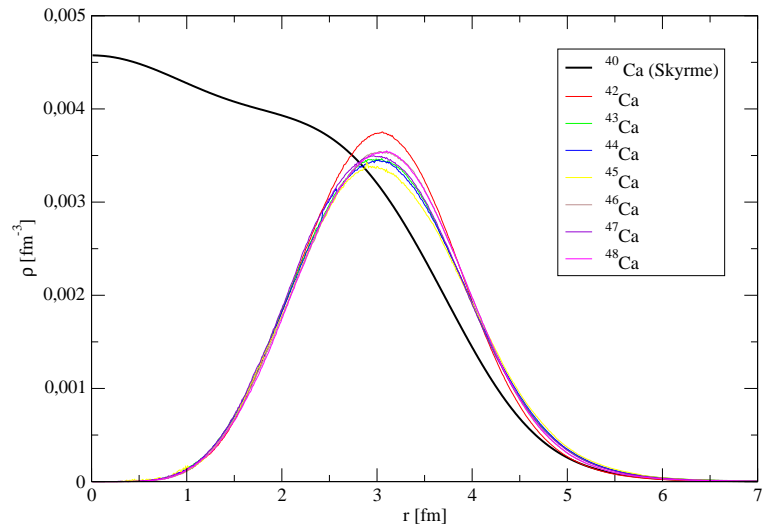}
%\vspace{0.5cm}
\caption{Densities of external neutrons for all isotopes in the ground-state calculated
with AFDMC and the Skyrme density of $^{40}$Ca.}
\label{fig:cadens}
\end{center}
\end{figure}

In figure \ref{fig:cadens} we report the AFDMC densities normalized to unity
of the external neutrons for the isotopes considered in this work.  In the figure 
we also display the density of $^{40}$Ca calculated with the Skyrme I force. As it 
can be seen the neutron densities are all quite similar, and very small deviations 
are present. However external neutrons seem to be very close to the core of $^{40}$Ca 
and because of this one should expect that the interaction between the core with 
external neutrons cannot be a satisfactorily described by a one-body external potential.

\chapter{Conclusions}
In this work I reported and discussed some recent calculations of different nuclear systems. 
The Auxiliary Field Diffusion Monte Carlo method combined with the constrained-path and the fixed-phase 
approximations, needed to control the Fermion sign problem, was described and applied to the study of both finite and infinite systems.
The fixed-phase approximation, introduced and discussed in this thesis, improves all the existing AFDMC results obtained 
with the previously used constrained-path approximation, in particular in dealing with the tensor-$\tau$ and 
spin-orbit interaction.

The fixed-phase constraint allows for a better sampling of the random walks compared to the 
constrained path, leading to an overall lowering of the energies. 
Older calculations showed that CP-AFDMC gives good results for nuclei if a simple $v_4$-like Hamiltonian 
is used, but the imposed constrained cannot correctly build the tensor-$\tau$ correlations.

Instead, the FP-AFDMC has proved to correctly project out from a simple variational wavefunction
the tensor-$\tau$ correlations, needed to include the main NN physics due to the $\pi$-exchange.
It was proved that FP-AFDMC has reached the same level of accuracy of GFMC and the most powerful 
few-body techniques in calculating the binding energies of light nuclei both in the closed- and open-shell configurations. 
The improvement due to the choice of the fixed-phase rather then the constrained-path is particular evident in the ground-state 
energy evaluation of the oxygen compared with other variational calculations.
The FP-AFDMC was applied to simulate nuclei up to A=40, but the first estimates of the CPU time needed for 
the calculations indicate that the method will be practical for much heavier systems. 

Another problem was encountered with CP-AFDMC in dealing with 
the spin-orbit interaction. Some discrepancies were observed in the ground-state calculation of neutron
drops, were a comparison of the spin-orbit splitting with that predicted by GFMC points out some difference.
The problem with the spin-orbit correlations appeared also in pure neutron matter calculations, and was 
explicitly underlined in a comparison between AFDMC and other many-body calculations.
Again, the fixed-phase approximation improves the agreement with GFMC in the prediction of the spin-orbit splitting in neutron drops.

The equation of state of nuclear matter, simulated by means of a periodic system with up to 108 nucleons in a periodic box, was calculated 
with semi-realistic NN potential containing tensor and tensor--$\tau$ forces. 
The FP-AFDMC results are of particular interest and can be used for a benchmark of other many-body theories.

A comparison with other many-body methods, such as the Brueckner-Hartree-Fock and the Fermi Hyper Netted Chain in the Single 
Operator Chain approximation, points out important discrepancies, even in the case with the considered Hamiltonian, that is a simplified form
containing the main physics of the system due to the tensor-$\tau$ term.
However other terms, such the missing NN operators as well as the three-nucleon interaction are
included in a very approximate way in both BHF and FHNC/SOC techniques.

In particular, the BHF seems to better reach convergence with the addition of the three hole-line contribution in
dealing with realistic NN interactions; however the inclusion of the three-nucleon interaction 
(that is essential in nucleonic matter calculations) is absolutely out of control and some work to estimate
its effect should be performed.

The FHNC/SOC approximation is one way to calculate the energy with respect to the wave function provided by the 
Correlated Basis Functions theory. It is a variational estimate of the ground-state energy but 
the uncontrolled introduced approximations violate the variational principle. This violation is not due 
to the theory but to the method used to calculate the energy of the system. 
In particular, the inclusion of the first class of elementary diagrams in the FHNC/SOC calculations revealed 
that their contributions are particular important and not negligible. The effect of other missing terms and of the 
SOC approximation is however unpredictable.

One should note that the FHNC/SOC theory was employed to constrain properties of the Urbana TNI 
on the empirical observations on symmetric nuclear matter coming from experimental data, so a natural 
question concerns the accuracy of the present Urbana TNI. 

The equation of state of neutron matter has been calculated by means of the FP-AFDMC using 
the semi-realistic Argonne AV8' NN interaction and the Urbana UIX TNI. It has to be noted that although 
this Hamiltonian cannot reproduce the ground-state of light nuclei beyond the $s$-shell, it is the most used 
to calculate the equation of state of nuclear and neutron matter and then used to constrain properties of neutron 
stars. 

Modern and more complex TNI were proposed by the Illinois group by fitting the binding energy of light nuclei, 
and tested up to the carbon ground-state.
The AFDMC calculation of neutron drop and neutron matter, using this TNI forces with a semi-realistic NN interaction,
revealed a dramatic dependence of such forces on the density, with the consequence that they cannot be used in a calculation 
to predict neutron star properties. In particular the 3$\pi$-exchange contribution in pure neutron systems 
gives rise to a binding of neutrons for densities below the $\rho_0$.

Hence the obtained results, even if computed with semi-realistic interactions, strongly indicate 
that in the region of densities below the bulk nuclear density most of the physics described by three-body forces is far to be understood
and stress the fact that much work must be addressed to the investigation of TNI.
In addition, many-body forces are likely to play a fundamental role in the understanding of nuclear matter 
properties, particularly at medium and high densities, a region of great interest in nuclear astrophysics 
and heavy ion physics.

The development of the FP-AFDMC opens several important directions for future investigations. First of all, the
addition of the spin-orbit and the TNI forces in the Hamiltonian shall permit to explore all the physics of 
medium-heavy nuclei with fully microscopic calculations.

The first problem is to work out an imaginary time propagator that includes the full AV18 NN interaction. 
Presently, in most of the cases the 
AV18 is evaluated as a perturbation of the AV8'. The assumption is correct by considering that AV8' contains 
most of the contributions of AV18 and was obtained with a reprojection by keeping only the more important terms.
However the operators appearing in the AV18 and not in the AV8' are not included exactly in the GFMC 
calculations. Usually the imaginary-time GFMC evolution is performed with the AV8' and the energy is 
calculated perturbatively with the difference between AV8' and AV18, that usually is very small, of the order of fractions 
of MeV in nuclei. 
This method is used also in the FHNC/SOC calculation where only the 
first NN correlations are included in the variational wave function. However there is not a clear proof that such 
calculation is an upper bound to the true energy, and how good this approximation is especially for higher 
densities.

The AFDMC was implemented using only a NN propagator in a v$_8'$ form with the calculation of the AV18 energy 
in pure neutron matter.
Some simple simulations suggest that this approximation is valid only in the very low-density limit, but not in the general case.
For example for 14 neutrons at $\rho\leq$0.12 fm$^{-3}$ the difference between AV8' and AV18 is less than 2 MeV per neutron and is 
2.7 MeV and 5.1 MeV for densities of 0.16 fm$^{-3}$ and 0.20 fm$^{-3}$ respectively. 
On the other hand the calculation of energy given by the v$_8'$-part of AV18 or AV8' is completely different 
also for low densities. Considering the v$_8'$ propagator only, at $\rho$=0.12 fm$^{-3}$ the energy of 14 neutrons with 
the AV8'+UIX Hamiltonian is 14.12 MeV, while it is 3.60 MeV for AV18+UIX. This means that the extra AV18 terms can not be treated
as correction to the AV8'.
Then some useful indications should concern an accurate study of the role of pure NN interaction in neutron and 
nuclear matter and in nuclei.

Other considerations should be made about three-body forces.
Although at present the Illinois forces cannot be used to calculate neutron star properties, they could
be accurate enough to study properties of heavier nuclei.
However, by knowing the limitation of actual Hamiltonians, at the present the AFDMC can be used to study properties 
of neutron matter, especially in the low-density regime where the superfluid phase, and the TNI effect on 
the pairing gap, can be accurately investigated.

An interesting open question concerns
the range of validity of these effective forces depending on the density of the system. It should be pointed out that
the effect and the role of many-body forces were never explored with other methods. 

The NN interaction is dominant in nuclear matter in the low-density regime but the addition of TNI gives a strong 
contribution to the equation of state, and it is necessary to correctly reproduce the equilibrium density that all modern 
NN potentials all overestimate. 
How could then be thought that higher many-body forces are negligible, or that their contribution is not important? On the other
hand when density increases phenomena at subnuclear scale probably starts to play an important role, and 
an effective interaction model should contain more terms that are small at lower densities.

The explicit inclusion of the TNI in the AFDMC algorithm reveals some problem, because the Hubbard-Stratonovich 
transformation naturally applies to Hamiltonians where only quadratic operators appear. An attempt to write a propagator by introducing 
more auxiliary field is in progress. The main intrinsic difficulty is that an higher number of auxiliary variable to be sampled means 
higher variance in the energy evaluation. 

A very interesting alternative is to consider some fictious excited state of nucleons (by remembering 
that TNI are caused by excitations of nucleons), and a corresponding NN fictious-interaction. 
This new interaction depends on the mass of these excitations, and in the infinite limit of such masses 
this NN potential reduces to usual TNI forces. 
The big improvement is that for intermediate masses it naturally generates many-body forces at infinite-orders.
At the present the AFDMC is being used to test the validity of this idea. The starting point is to verify if such 
new interaction exactly gives the same contribution of the Urbana UIX in the low-density regime (for example 
in nuclei) when the excitation masses go to infinity. The next step should be to increase the density and study the 
effect of many-body forces that naturally appear with a finite excitation mass.

\section*{Some dreams...}
When the mentioned technical question will be addressed, the AFDMC could be used to predict the whole structure of the neutron star, 
with all the phases of the matter at each density.
The inclusion of the full NN interaction as well the TNI is then of primarily importance, being the starting point to predict properties of asymmetric nuclear matter 
with an unprecedented accuracy, and another exciting work to do will be the explicit inclusion of hyperons in the calculation. For this 
purpose it has to be noted that the existing nucleon-hyperon and hyperon-hyperon interactions have the same structure of NN potentials, 
and so easy to include in the AFDMC algorithm. Other interesting problems (like the possibility of neutron near the 
core of the star to form a lattice) as well the role of many-body forces (revealing then a completely new physics in the system) 
should be investigated. In addition, using the AFDMC the full Hamiltonian should be (in principle) computed the ground state of all 
existing nuclei, in order to have a theoretical prediction of the Weizsacker formula directly from NN forces.

Another very interesting achievement is the simulation of a nuclear system starting directly from the QCD. This calculation should be a great challenge. 
At present the nuclear physics can not be directly derived from the QCD without the use of effective potentials derived from EFT (the 
lattice QCD is not considered because limited to hadron physics). 
The idea of begin a calculation involving nucleons that interact with explicit pions degrees of freedom is very exciting and stimulating for the future.
In fact the computation of the equation of state of nuclear matter by including directly the pions in the calculation is absolutely unimaginable
at present.
Some attempt in this direction is also in progress.

\cleardoublepage
\chapter*{Acknowledgements}
First of all I dedicate this work to my son that is coming. I am grateful to my parents Alessandra and Paolo 
(to have sponsored my studies...) and to my wife Serena. They demonstrated a lot of patience with me during my 
years spent at University and the Ph.D. courses.

I am very indebted to my supervisor Francesco Pederiva that encouraged me during my studies, and not only for 
the academic guide, but also for his patience and friendship with me.

I thank all other students in Trento (Lucia Dandrea, Michele Montobbio, Sebastiano Pilati, Marco Cristoforetti, Mario
Nadalini, David Tombolato, Giacomo Ciani, Alberto Ambrosetti, Giacomo Gradenigo, Paolo Armani and Francesco Operetto) 
and all fellow students I met during my studies.

It's a pleasure to thank professor Kevin Schmidt for a lot of fruitful discussions I had with him and also because 
he gave to me the possibility and the hospitality to spend two months in the Arizona State University.
I would like to remember that the work of this thesis started also from a meeting I had with professor Stefano Fantoni in Trieste when I 
started my graduate studies about the nuclear Monte Carlo methods.

I am grateful to Pietro Faccioli, Giuseppina Orlandini, Winfried Leidemann, Marco Traini, Stefano Giorgini and Mal Kalos for useful advices in Trento, 
and to Giampaolo Co', Steven Pieper, Bob Wiringa and Joe Carlson for stimulating discussions during my visits in Lecce, Argonne and 
Los Alamos. At last I thank Alexey Illarionov, Mohamed Bouadani and Cristian Bisconti for their friendship.

\cleardoublepage

\addcontentsline{toc}{chapter}{Bibliography}

\backmatter
\listoftables
\addcontentsline{toc}{chapter}{\numberline{List of Tables}}
\clearpage
\listoffigures
\addcontentsline{toc}{chapter}{\numberline{List of Figures}}
\end{document}